\documentclass[longauth]{aa}

\usepackage{graphicx}
\usepackage{txfonts}
\usepackage{hyperref}

\usepackage{amsmath}
\usepackage{color} 
\usepackage{xcolor} 
\usepackage{makecell} 
\usepackage[para]{threeparttable} 

\newcommand*\pyorbit{\texttt{PyORBIT}}
\newcommand{\thirstee}{\texttt{THIRSTEE}}

\newcommand*\tess{{\it TESS}}

\newcommand{\snr} {\mbox{S/N}}

\newcommand{\prot}{\mbox{P$_{\rm rot}$}}

\newcommand{\teff}{$T_{{\rm eff}}$}
\newcommand{\kms}{\mbox{km\,s$^{-1}$}}
\newcommand{\ms}{\mbox{m s$^{-1}$}}
\newcommand{\gcm}{\mbox{g cm$^{-3}$}}

\newcommand{\logg} {\mbox{log\,{\it g}}}

\newcommand{\mplanet}{\mbox{$M_{\rm p}$}}
\newcommand{\rplanet}{\mbox{$R_{\rm p}$}}
\newcommand{\rhoplanet}{\mbox{$\rho_{\rm p}$}}

\newcommand{\teq}{$T_{{\rm eq}}$}

\newcommand{\mearth}{\mbox{M$_\oplus$}}
\newcommand{\rearth}{\mbox{R$_\oplus$}}

\newcommand{\msun}{\mbox{M$_\odot$}}
\newcommand{\rsun}{\mbox{R$_\odot$}}
\newcommand{\mstar}{\mbox{$M_\star$}}
\newcommand{\rstar}{\mbox{$R_\star$}}

\newcommand{\rhostar}{\mbox{$\rho_\star$}}
\newcommand{\rhosun}{\mbox{$\rho_\odot$}}

\newcommand{\logRHK}{\mbox{$\log {\rm R}^{\prime}_{\rm HK}$}}

\newcommand{\halpha}{\mbox{H$\alpha$}}

\newcommand{\starname}{\mbox{TOI-406}}

\begin{document}

   \title{Characterisation of TOI-406 as a showcase of the THIRSTEE program}

   \subtitle{A two-planet system straddling the M-dwarf density gap}

   \author{G. Lacedelli\inst{\ref{iac}}\,$^{\href{https://orcid.org/0000-0002-4197-7374}{\protect\includegraphics[height=0.19cm]{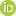}}}$,
          E. Pallé\inst{\ref{iac}, \ref{ull}}\,$^{\href{https://orcid.org/0000-0003-0987-1593}{\protect\includegraphics[height=0.19cm]{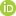}}}$,
          R. Luque\inst{\ref{chicago}\fnmsep\thanks{NHFP Sagan Fellow}}\,$^{\href{https://orcid.org/0000-0002-4671-2957}{\protect\includegraphics[height=0.19cm]{Figures/orcid.jpg}}}$,
          C. Cadieux\inst{\ref{montreal}}\,$^{\href{https://orcid.org/0000-0001-9291-5555}{\protect\includegraphics[height=0.19cm]{Figures/orcid.jpg}}}$,
          J. M. Akana Murphy\inst{\ref{california}\fnmsep\thanks{NSF Graduate Research Fellow}}\,$^{\href{https://orcid.org/0000-0001-8898-8284}{\protect\includegraphics[height=0.19cm]{Figures/orcid.jpg}}}$,
          F. Murgas\inst{\ref{iac},\ref{ull}}\,$^{\href{https://orcid.org/0000-0001-9087-1245}{\protect\includegraphics[height=0.19cm]{Figures/orcid.jpg}}}$,
          M. R. Zapatero Osorio\inst{\ref{CSIC-INTA}}\,$^{\href{https://orcid.org/0000-0001-5664-2852}{\protect\includegraphics[height=0.19cm]{Figures/orcid.jpg}}}$,
          H. M. Tabernero\inst{\ref{computense},\ref{CSIC-INTA}},
          K. A. \ Collins\inst{\ref{cambridge}\, \textsuperscript{$\star \star$}},
          C. N.\ Watkins\inst{\ref{cambridge}}\,$^{\href{https://orcid.org/0000-0001-8621-6731}{\protect\includegraphics[height=0.19cm]{Figures/orcid.jpg}}}$,
          A. L'Heureux\inst{\ref{montreal}}\,$^{\href{https://orcid.org/0009-0005-6135-6769}{\protect\includegraphics[height=0.19cm]{Figures/orcid.jpg}}}$,
          R. Doyon\inst{\ref{montreal},\ref{montreal_observatory}}\,$^{\href{https://orcid.org/0000-0001-5485-4675}{\protect\includegraphics[height=0.19cm]{Figures/orcid.jpg}}}$,
          D. Jankowski\inst{\ref{poland}},
          G. Nowak\inst{\ref{poland}}\,$^{\href{https://orcid.org/0000-0002-7031-7754}{\protect\includegraphics[height=0.19cm]{Figures/orcid.jpg}}}$,
          \'E. Artigau\inst{\ref{montreal}, \ref{montreal_observatory}},
          N. M.\ Batalha\inst{\ref{california}}\,$^{\href{https://orcid.org/0000-0002-7030-9519}{\protect\includegraphics[height=0.19cm]{Figures/orcid.jpg}}}$,
          J. L. \ Bean\inst{\ref{chicago}}\,$^{\href{https://orcid.org/0000-0003-4733-6532}{\protect\includegraphics[height=0.19cm]{Figures/orcid.jpg}}}$,
          F. Bouchy\inst{\ref{geneve}},
          M. Brady\inst{\ref{chicago}}\,$^{\href{https://orcid.org/0000-0003-2404-2427}{\protect\includegraphics[height=0.19cm]{Figures/orcid.jpg}}}$,
          B. L. Canto Martins\inst{\ref{brazil}, \ref{inaf_arcetri}},
          I. Carleo\inst{\ref{inaf}, \ref{iac}}\,$^{\href{https://orcid.org/0000-0002-0810-3747}{\protect\includegraphics[height=0.19cm]{Figures/orcid.jpg}}}$, 
          M. Cointepas\inst{\ref{grenoble}, \ref{geneve}}, 
          D. M. Conti\inst{\ref{aav_observers}}\,$^{\href{https://orcid.org/0000-0003-2239-0567}{\protect\includegraphics[height=0.19cm]{Figures/orcid.jpg}}}$,
          N. J. Cook\inst{\ref{montreal}},
          I. J. M. Crossfield\inst{\ref{kansas}}\,$^{\href{https://orcid.org/0000-0002-1835-1891}{\protect\includegraphics[height=0.19cm]{Figures/orcid.jpg}}}$,
          J. I. Gonz\'alez Hern\'andez\inst{\ref{iac},\ref{ull}}\,$^{\href{https://orcid.org/0000-0002-0264-7356}{\protect\includegraphics[height=0.19cm]{Figures/orcid.jpg}}}$,
          P. Lewin\inst{\ref{lewin}}\,$^{\href{https://orcid.org/0000-0003-0828-6368}{\protect\includegraphics[height=0.19cm]{Figures/orcid.jpg}}}$,
          N. Nari\inst{\ref{las_palmas},\ref{iac},\ref{ull}}\,$^{\href{https://orcid.org/0009-0001-2825-6185}{\protect\includegraphics[height=0.19cm]{Figures/orcid.jpg}}}$,
          L. \,D. Nielsen\inst{\ref{munich},\ref{eso_garching}},
          J.~Orell-Miquel\inst{\ref{ull}, \ref{iac}}\,$^{\href{https://orcid.org/0000-0003-2066-8959}{\protect\includegraphics[height=0.19cm]{Figures/orcid.jpg}}}$,
          L. Parc\inst{\ref{geneve}},
          R. P. Schwarz\inst{\ref{cambridge}}\,$^{\href{https://orcid.org/0000-0001-8227-1020}{\protect\includegraphics[height=0.19cm]{Figures/orcid.jpg}}}$,
          G. Srdoc\inst{\ref{croatia}},
          V. Van Eylen\inst{\ref{ucl}}\,$^{\href{https://orcid.org/0000-0001-5542-8870}{\protect\includegraphics[height=0.19cm]{Figures/orcid.jpg}}}$
          }

   \institute{Instituto de Astrof\'{i}sica de Canarias (IAC), 38205 La Laguna, Tenerife, Spain \label{iac} 
              \email{glacedelli@iac.es}
    \and Departamento de Astrof\'isica, Universidad de La Laguna (ULL), E-38206 La Laguna, Tenerife, Spain \label{ull} 
    \and Department of Astronomy and Astrophysics, University of Chicago, Chicago, IL 60637, USA \label{chicago} 
    \and Institut Trottier de recherche sur les exoplan\`etes, D\'epartement de Physique, Universit\'e de Montr\'eal, Montr\'eal, Qu\'ebec, Canada \label{montreal} 
    \and Department of Astronomy and Astrophysics, University of California, Santa Cruz, CA 95064, USA \label{california} 
    \and Centro de Astrobiología (CSIC-INTA), Crta. Ajalvir km 4, E-28850 Torrejón de Ardoz, Madrid, Spain \label{CSIC-INTA} 
    \and Departamento de Física de la Tierra y Astrofísica \& IPARCOS-UCM (Instituto de Física de Partículas y del Cosmos de la UCM), Facultad de Ciencias Físicas, Universidad Complutense de Madrid, 28040 Madrid, Spain \label{computense} 
    \and  Center for Astrophysics \textbar \ Harvard \& Smithsonian, 60 Garden Street, Cambridge, MA 02138, USA \label{cambridge} 
    \and Observatoire du Mont-Mégantic, Université de Montréal, C.P. 6128, Succ. Centre-ville, Montréal, H3C 3J7, Québec, Canada \label{montreal_observatory}
    \and Institute of Astronomy, Faculty of Physics, Astronomy and Informatics, Nicolaus Copernicus University, Grudzi\c{a}dzka 5, 87-100 Toru\'{n}, Poland \label{poland}
    \and Observatoire de Gen\`eve, D\'epartement d’Astronomie, Universit\'e de Gen\`eve, Chemin Pegasi 51b, 1290 Versoix, Switzerland \label{geneve} 
    \and Departamento de F\'isica Te\'orica e Experimental, Universidade Federal do Rio Grande do Norte, Campus Universit\'ario, Natal, RN, 59072-970, Brazil \label{brazil}
    \and INAF - Osservatorio Astrofisico di Arcetri, Largo E. Fermi 5, Florence, Italy \label{inaf_arcetri}
    \and INAF -- Osservatorio Astrofisico di Torino, Via Osservatorio 20, I-10025, Pino Torinese, Italy \label{inaf} 
    \and University of Grenoble Alpes, CNRS, IPAG, F-38000 Grenoble, France \label{grenoble}
    \and American Association of Variable Star Observers, 185 Alewife Brook Parkway, Suite 410, Cambridge, MA 02138, USA \label{aav_observers}
    \and Department of Physics \& Astronomy, University of Kansas, 1082 Malott, 1251 Wescoe Hall Drive, Lawrence, KS 66045, USA \label{kansas}
    \and The Maury Lewin Astronomical Observatory, Glendora, California, 91741, USA \label{lewin}
    \and Light Bridges S.L., Observatorio del Teide, Carretera del Observatorio, s/n Guimar, 38500, Tenerife, Canarias, Spain \label{las_palmas}
    \and University Observatory, Faculty of Physics, Ludwig-Maximilians-Universit{\"a}t M{\"u}nchen, Scheinerstr. 1, 81679 Munich, Germany \label{munich}
    \and European Southern Observatory (ESO), Karl-Schwarzschild-Str. 2, 85748 Garching bei M\"unchen, Germany \label{eso_garching}
    \and Kotizarovci Observatory, Sarsoni 90, 51216 Viskovo, Croatia \label{croatia}
    \and Mullard Space Science Laboratory, University College London, Surrey, UK \label{ucl}
             }

   \date{Received 13 September 2024; accepted 02 November 2024}

 
  \abstract
   {The exoplanet sub-Neptune population currently poses a conundrum, as to whether small-size planets are volatile-rich cores without an atmosphere, or rocky cores surrounded by a H-He envelope. 
   To test the different hypotheses from an observational point of view, a large sample of small-size planets with precise mass and radius measurements is the first necessary step. On top of that, much more information will likely be needed, including atmospheric characterisation and a demographic perspective on their bulk properties.}
   {We present here the concept and strategy of the \thirstee\ project, which aims to shed light on the composition of the sub-Neptune population across stellar types by increasing their number and improving the accuracy of bulk density measurements, as well as investigating their atmospheres and performing statistical, demographic analysis.
   We report the first results of the program, characterising a new two-planet system around the M-dwarf TOI-406.}
   {We analysed \tess\ and ground-based photometry together with high-precision ESPRESSO and NIRPS/HARPS radial velocities to derive the orbital parameters and investigate the internal composition of the two planets orbiting TOI-406.}
   {TOI-406 hosts two planets with radii and masses of $R_{\rm c} = 1.32 \pm 0.12$~\rearth, $M_{\rm c} = 2.08_{-0.22}^{+0.23}$~\mearth\  and $R_{\rm b} = 2.08_{-0.15}^{+0.16}$~\rearth, $M_{\rm b} = 6.57_{-0.90}^{+1.00}$~\mearth,  orbiting with periods of $3.3$ and $13.2$ days, respectively. The inner planet is consistent with an Earth-like composition, while the external one is compatible with multiple internal composition models, including volatile-rich planets without H/He atmospheres. The two planets are located in two distinct regions in the mass-density diagram, supporting the existence of a density gap among small exoplanets around M dwarfs. With an equilibrium temperature of only \teq~$=368$~K, TOI-406 b stands up as a particularly interesting target for atmospheric characterisation with {\it JWST} in the low-temperature regime.}
  {}

\titlerunning{Characterisation of the TOI-406 system as a showcase of the THIRSTEE program}
\authorrunning{Lacedelli et al.}
   \maketitle
%
\section{Introduction}\label{sec:intro}
The increasing number of discovered exoplanets, more than $5500$ at present, 
now allows for the development of demographic studies based on the general properties of the observed planetary population.
In the last few decades, space-based missions such as {\it Kepler} \citep{Borucki2010} and, more recently, \tess\ \citep{ricker2014}, have identified an emerging population of planets with sizes between the Earth and Neptune, also known as sub-Neptunes. 
This population of small-size planets ($1$~\rearth~$<$~\rplanet~$< 4$~\rearth) is absent in the Solar System, but more than half of all Sun-like stars in the Galaxy host a sub-Neptune interior to $1$~AU \citep[e.g.][]{Batalha2013, marcy2014, Petigura13PNAS}. 
Thus, understanding the origin and nature of the sub-Neptune population is not only essential to inform our understanding of the planetary formation and evolution processes, but it also helps us to picture our Solar System in a global context.

The sub-Neptune population shows a paucity of planets with sizes between $1.5$ and $2.0$~\rearth, known as `radius gap' \citep{Fulton2017, vanEylen2018, hirano2018, berger2020, cloutier2020, Petigura22}, which sparked the development of planet formation and evolution models that could explain this observational feature. 
One of the most common interpretations is attributed to atmospheric evolution, which can reproduce the observed dependency of the gap location as a function of incident stellar irradiation, and the fact that most of the highly-irradiated sub-Neptunes have Earth-like densities \citep[][]{Dai2019}. 
In this scenario, planets below the radius gap (known as super-Earths) are bare rocky cores that have been fully stripped from their primordial hydrogen-dominated atmosphere originally accreted from the protoplanetary disc. 
This atmospheric mass loss is mainly thought to be due to XUV radiation from young stars \citep{OwenWu13, LopezFortney13} or to the heat coming from cooling cores \citep{ginzburg2018, gupta2019}, even though other mechanisms such as giant impacts have also been investigated \citep{Wyatt20}. 
These atmospheric mass loss theories predict that planets above the gap (`gas dwarfs' hereafter - also known as mini-Neptunes) are Earth-like cores formed in the inner regions of the protoplanetary disc that retained their primordial atmospheres without being substantially enriched in volatiles \citep{IkomaHori12,Lee14,LeeChiang16}. 
Models that include in situ gas accretion during the gas-poor phase of disc evolution \citep{Lee22} are also able to reproduce the observations.

An alternative interpretation of the radius distribution of the sub-Neptune population is that super-Earths and mini-Neptunes are instead two different types of planets: the former consisting of a rock-iron mixture (a scaled-up version of Earth) and the latter consisting of a water-rock mixture (a scaled-up version of the Solar System's icy moons), known as `water worlds' \citep{Leger04,Rogers15,Mousis20,DornLichtenberg21}. 
Water worlds have intrinsically larger sizes with respect to super-Earths, due to their lower bulk density. 
In the water world hypothesis, planets formed beyond the ice line (or have been significantly polluted by pebbles or planetesimals coming from that external region), and then migrated to the inner parts of the system \citep{Raymond18,Bitsch19,Izidoro22}. 
While several fine-tuning effects must be introduced to replicate the properties of the gas dwarfs \citep{Lee14, OwenWu16, Liu17, Kite19, Esteves20, Ogihara20}, water worlds are a natural outcome of formation and migration processes independent of the accretion mechanism (planetesimal or pebble) and the initial conditions of the protoplanetary disc \citep{Raymond18,Bitsch19,Venturini20,burn2021,Izidoro22}. 
The same models are also able to reproduce the demographic features that originally favoured the gas dwarfs hypothesis, such as their large abundance \citep{Mayor11, Mulders18}, typical masses \citep{Weiss14, Otegi20}, radii \citep{FultonPetigura18, Petigura22}, and period ratio and multiplicity distributions \citep{Lissauer11, Fabrycky14}. 
Moreover, various recent individual discoveries \citep{Luque21,Luque22,Cadieux22,Piaulet22,Diamond-Lowe22, cherubim2023, osborne2023, Cadieux2024b, Cadieux2024a} have provided increasing evidence that such planets exist.

Recently, using a refined catalogue of small planets (\rplanet$< 4$~\rearth), \cite{luquepalle2022} proposed that for M dwarf hosts, the radius gap is a consequence of interior composition rather than an indicator of atmospheric mass loss.
The authors show that M-dwarf planets seem to be distributed into three main populations: 1) planets that follow a bulk density comparable to the one of the Earth (e.g. rocky planets), 2) planets with cores consisting of a mixture of rock and water-dominated ices in a 50:50 mass proportion (e.g. water worlds), and 3) planets with a significant envelope made of H/He.

In this scenario, the apparent paucity of planets in the $1.5$–$2.0$~\rearth\ radius range is due to the combination of the rocky population reaching at most $10$~\mearth, and the water world population having a minimum mass of $2$–$3$ M$_{\oplus}$. 
The rock-ice proportion in the water world population identified in \cite{luquepalle2022} is not a coincidence, since internal structure models consistent with $50$:$50$ admixtures of rock and water ices \citep{Zeng2019,Mousis20} perfectly match the ratio of rocky and volatile material measured beyond the ice line in protoplanetary discs \citep{Lodders03}~--~even though the amount and storage mechanisms of water in planetary interiors is still debated \citep{DornLichtenberg21,Vazan22,Luo24}. 
Generally, water is not expected to be in condensed form for the known sub-Neptune population (typically close-in; hence, hot), but recent formation and evolution models that account for the effect of supercritical water mixed with H/He still reproduce the radius gap at the observed location \citep{Burn24}.

However, this sub-Neptune population sits in a nexus of interior structure modelling degeneracy.
Often, multiple combinations of interior compositions, including varying percentages of rocky and icy materials, and gases, can match the radius and mass measurements of individual planets \citep{RogersSeager10}, as well as the observed mass-radius trends, as is extensively explored in \cite{parc2024}. 
As an example, \cite{Rogers23} show that the precise bulk densities of M-dwarf planets from \cite{luquepalle2022} can be also reproduced assuming a gas dwarf composition, if using atmospheric mass loss models that include spontaneous mass loss (also known as `boil-off') after disc dispersal \citep{IkomaHori12,OwenWu16}. 
The two scenarios (atmospheric mass loss, and the water-world hypothesis) are thus able to explain the individual and demographic properties of this population, highlighting the need for more detailed studies. This need for a large population of precisely characterised sub-Neptune planets is the subject of this paper, and of the \thirstee\ project.

This paper is organised as follows: in Sect.~\ref{sec:thirstee_strategy} we describe the rational and observational strategy of the \thirstee\  project, and in the rest of the work we present its first result; that is, the analysis and characterisation of a new planetary system around the M dwarf TOI-406 as part of the \thirstee\ sample. 
For the system characterisation, we used \tess\ (Sect.~\ref{sec:tess}) and ground-based photometry together with two radial velocity (RV) datasets coming from ESPRESSO and NIRPS/HARPS, collected under the \thirstee\ project and the NIRPS Guaranteed Time Observations (GTO), respectively (Sect.~\ref{sec:ground_based}). 
With these data, we precisely infer stellar parameters (Sect.~\ref{sec:star}) and mass, radius, and orbital configuration (Sect.~\ref{sec:global_analysis}) for the two planets orbiting TOI-406. 
We discuss our results and the potential of the \thirstee\ program using TOI-406 as a benchmark target within the context of the formation and evolution processes of the sub-Neptune population in Sect.~\ref{sec:discussion}.

\section{THIRSTEE: Rationale and survey strategy}\label{sec:thirstee_strategy}

To investigate the origin and nature of sub-Neptunes, precise and accurate bulk density measurements are needed. Masses and radii are necessary to begin modelling their internal and atmospheric properties at an individual level and enable a demographic understanding of these planets as a population. Radial velocity follow-up of transiting exoplanets is the most extensive approach to measure the masses of these planets since, contrary to transit timing variations, it does not require the systems to be close to orbital resonance in order to characterise them. Despite the influence of stellar activity on multiple timescales, the incredibly precise ephemeris from transit observations (namely, the orbital period and phase of the planet) ease the modelling of the RV data, providing precise and accurate measurements of the planet masses in combination with stellar and instrumental mitigation strategies. The development of extremely precise RV instruments capable of sub-metre-per-second internal precision (e.g. ESPRESSO, MAROON-X, EXPRES, NEID, KPF, NIRPS) enables, in addition, the mass characterisation of transiting sub-Neptunes to be extended from low-mass stars (with typical RV semi-amplitudes of $1-2$~m/s) to Sun-like stars (with typical RV semi-amplitudes of $30-70$~cm/s).

\medskip

To fill this gap, we have started the \thirstee\footnote{{\bf T}racking {\bf H}ydrates {\bf I}n {\bf R}efined {\bf S}ub-neptunes to {\bf T}ackle their {\bf E}mergence and {\bf E}volution.}\ survey. The core of \thirstee\ revolves around the following questions: whether sub-Neptunes are gas dwarfs or water worlds; whether water worlds exist only around M dwarfs or also around Sun-like stars; and, if both exist, what their relative occurrence is and how it depends on other system properties.

\medskip

To tackle these questions, \thirstee\ takes an observational approach and, at the core of the project, is a dedicated RV survey to increase the number and improve the accuracy of bulk density measurements of transiting sub-Neptunes. We are particularly interested in extending the known sample in two important axes that have been largely unexplored: planet equilibrium temperature and host spectral type. The dependence on planet bulk density with host spectral type (as a proxy of stellar mass) contains the most information to connect the observed population with global models of planet formation and evolution \citep[as recently highlighted by, e.g.,][]{venturini2024,ChakrabartyMulders2024}. On the other hand, a sample of sub-Neptunes with ranging equilibrium temperatures and host spectral types is necessary to constrain the chemistry and interior structure of this population via atmospheric characterisation. Aerosol obfuscation has been proposed to explain the attenuated spectral features of sub-Neptune atmospheres in transmission as a function of equilibrium temperature \citep{brande2024}. However, there is a degeneracy between clouds and high atmospheric mean molecular weight in feature amplitude that \textit{HST} could not resolve. The first \textit{JWST} observations of cool sub-Neptunes orbiting M~dwarfs show that aerosols are not needed to explain their feature attenuation \citep{Madhusudhan_2023,bennecke2024,Piaulet2024}. For \textit{JWST} to break this degeneracy within its lifetime, a large sample of transiting sub-Neptunes with excellent observability metrics that spans a large range of planet equilibrium and stellar effective temperatures is needed. Besides, mass measurements are a prerequisite for atmospheric studies of small exoplanets to break degeneracies from atmospheric retrievals \citep{batalha19}. \thirstee\ is designed to fulfil all these necessities and enable an understanding of sub-Neptune properties free of model degeneracies.

\medskip

The \thirstee\ project has secured observing time through open calls in most state-of-the-art spectrographs in both hemispheres, including ESPRESSO (111.24PJ.001, PI: E. Pallé; 112.25F2.001, PI: E. Pallé; 113.26NP.001, PI: E. Pallé), HARPS (111.24PJ.002, PI: E. Pallé), HIRES (2023A\_U156, PI: N. M. Batalha), HARPS-N (CAT23A52, PI: I. Carleo; CAT23B74, PI: I. Carleo; CAT24B20, PI: I. Carleo), CARMENES (24B-3.5-012, PI: E. Pallé), and KPF (2024A\_N158 NASA/Keck Key Strategic Mission Support [KSMS] program, PI: R. Luque). We have done an exhaustive target selection and triage for the program. Here, we summarise the procedure followed to find the targets. We used two sources of targets: 1) confirmed transiting planets from the NASA Exoplanet Archive (NEA) without a mass measurement or with a precision worse than 20\% (and without teams actively working on improving it), and 2) \tess\ candidates vetted and validated by the \tess\ Follow-up Observing Program (TFOP\footnote{\url{https://tess.mit.edu/followup}}; \citealt{collins2019}). For the latter, we removed all candidates: i) without ground-based observations to confirm that the purported transit is on target and not on a nearby star; ii) with reconnaissance spectroscopy revealing spectroscopic binaries, fast-rotating stars, or with high chromospheric activity seen in $\log R'_{\rm HK}$; iii) with high-resolution imaging that shows the presence of a nearby companion that could fall in the fibre aperture of our RV spectrographs (with typical fibre sizes of $\sim 1\arcsec$) or host the transit event; and iv) with ongoing mass characterisation from other RV teams. 

\medskip

For all targets, we applied cuts in brightness\footnote{The lower limit is set to avoid saturating most {\it JWST} observing modes, the upper limit is set by the ability of the RV instruments in our program to reach a photon-limited RV precision below 1.5 m/s in 30\,min exposures.} ($8 < V < 13$\, mag), planet radius ($1.3 < R_{\rm p}/R_\oplus < 2.8$), and prospects for atmospheric characterisation (transmission spectroscopy metric (TSM)~$ > 30$, \citealp{kempton2018,hord2024}). The goal of these cuts is two-fold. First, the radius range ensures that all the planets in the sample span the radius gap \citep{Fulton2017,vanEylen2018} where radius alone is not a proxy of the internal composition of the planet and all populations of rocky planets, rocky cores with H-rich atmospheres, and water-rich cores with or without significant envelopes overlap \citep{spright}. The TSM cut ensures that all targets are suitable for future atmospheric characterisation with {\it JWST} and {\it Ariel} \citep{ARIEL}, which remains the best avenue to understand the origin and nature of sub-Neptune planets and which requires precise planetary masses to break degeneracies from atmospheric retrievals \citep{batalha19}. Both selection cuts critically contribute to community-wide goals pertaining to the most numerous exoplanet population known to date and for which even the most basic questions about their internal and atmospheric properties remain unanswered (see also \citealt{Teske21, Chontos22}; Crossfield et al., in prep; Armstrong et al., in prep.)

\medskip

\begin{figure*}
\centering
  \includegraphics[width=\linewidth]{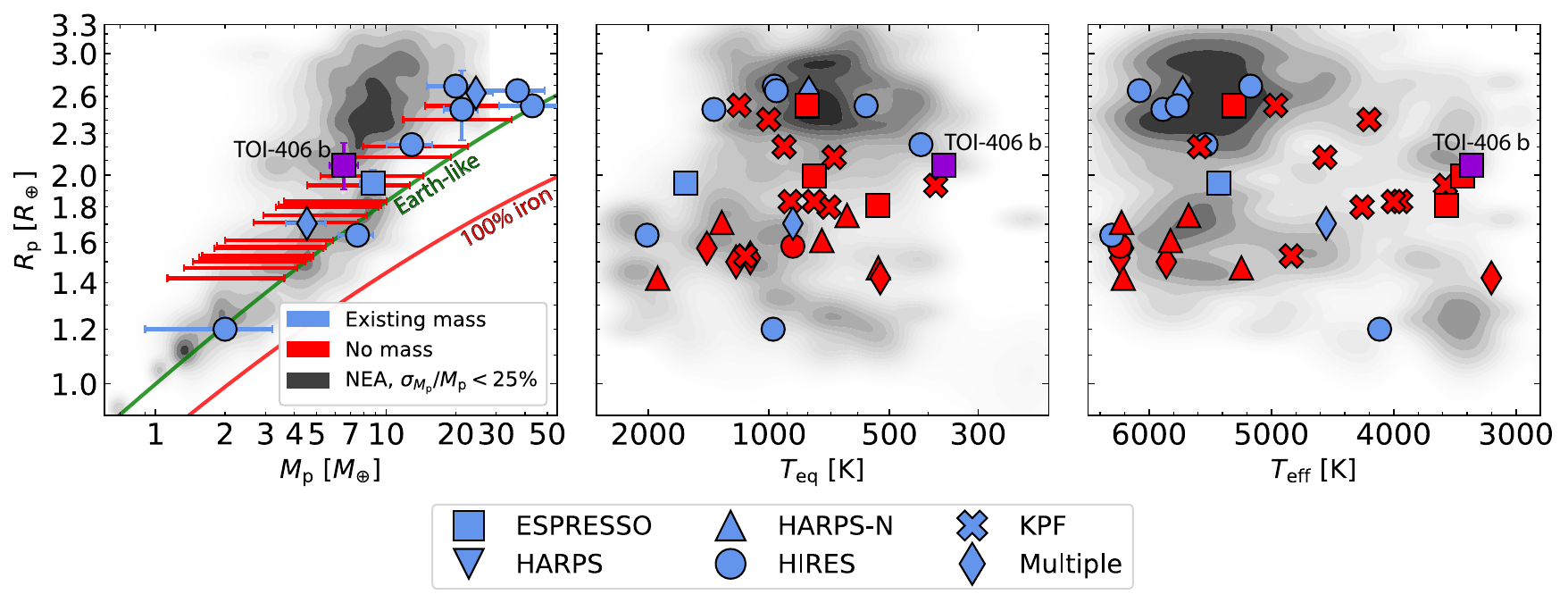}
  \caption{\thirstee\ sample distributed in radius vs mass (left), equilibrium temperature (middle), and stellar effective temperature (right). Blue markers represent planets with an existing mass measurement in the literature and red markers represent those without one. TOI-406 b is shown in purple because this work represents the planet's first mass measurement. Marker symbols correspond to the instruments the \thirstee\ survey is using to measure the mass of each planet. Underlying contours show the relative population of confirmed planets with better than 25\% measurement precision in mass, according to data from the NEA accessed on August 27, 2024. Left: \thirstee\ targets without mass measurements are shown as bars, representing the range of their expected mass given their radius (e.g. \citealt{parviainen2024}). Curves representing an Earth-like and a pure iron composition are shown for reference \citep{Zeng2019}. Middle: Equilibrium temperature taken from the literature where available. If unavailable, it was calculated assuming perfect day-night heat redistribution and zero Bond albedo. Right: For unconfirmed planet candidates, stellar effective temperature taken from the TICv8.
  }
    \label{fig:THIRSTEE_sample}
\end{figure*}

Only 23 confirmed planets and 10 TOI candidates meet all of these criteria and are not currently being actively monitored and analysed by other RV teams\footnote{At the time of target selection and/or proposals submission.}. We looked for additional signs of elevated stellar activity for these targets from their \tess\ light curves and computed if their kinematics were consistent with being part of a young moving group, but no red flags were raised. Our multi-facility strategy reduces the load and the risks of relying on a single instrument or facility to carry out the work, while allowing for the optimisation of the properties of each spectrograph to the necessities of each target (latitude, brightness, spectral type, and expected RV semi-amplitude). 
We list our current sample in Table~\ref{tab:targets}, and we show the position of the targets in the mass-radius, temperature-radius, and stellar temperature-radius diagrams in Fig.~\ref{fig:THIRSTEE_sample}.

\begin{table*}[h!]

\tiny
\caption{The \thirstee\ sample.}
\label{tab:targets}
\begin{threeparttable}
\centering                                      
\begin{tabular}{l c c c c c c c}          
\hline\hline                        
Name & \rplanet\ (\rearth) & \mplanet\ (\mearth)& P (d) & \teq\ (K) & \teff\ (K) & Spectrograph & Reference\\
\hline                                 
HD 119130 b & $2.63^{+0.11}_{-0.10}$ & $24.5 \pm 4.4$ & 16.984 & 795 & 5725& HARPS, HARPS-N, HIRES & \cite{luque2019}\\
K2-155 b & $1.8^{+0.2}_{-0.1}$ & - & 6.342 & 708 & 4258& KPF & \cite{diez_alonso2018}\\
K2-180 b & $2.52^{+0.24}_{-0.10}$ & $11.3 \pm 1.9^{\dag}$ & 8.866 & 802 & 5306& ESPRESSO & \cite{castro-gonzalez_2022}, $^{\dag}$\cite{Korth2019}\\
K2-182 b & $2.69^{+0.07}_{-0.05}$ & $20 \pm 5$ & 4.737 & 969 & 5170& HIRES & \cite{akana_murphy2021}\\
K2-184 b & $1.47^{+0.11}_{-0.05}$ & - & 16.978 & 533 & 5245& HARPS-N & \cite{castro-gonzalez_2022}\\
K2-223 c & $1.5 \pm 0.1$ & - & 4.562 & 1208 & 5859& HARPS-N, ESPRESSO & \cite{adams2021}\\
K2-244 b & $1.75_{-0.10}^{+0.13}$ & - & 21.069 & 638 & 5677& HARPS-N & \cite{livingstone2018}\\
K2-269 b & $1.57 \pm 0.12$ & - & 4.145 & 1429 & 6209& ESPRESSO, HARPS-N & \cite{castro-gonzalez_2022}\\
K2-314 b & $1.95^{+0.09}_{-0.08}$ & $8.57^{+1.09}_{-1.08}$ & 3.595 & 1616 & 5430& ESPRESSO & \cite{hidalgo2020}\\
K2-66 b & $2.49^{+0.34}_{-0.24}$ & $21.3 \pm 3.6$ & 5.070 & 1372 & 5887& HIRES & \cite{sinukoff2017}\\
Kepler-100 d & $1.61 \pm 0.05$ & - & 35.333 & 737 & 5825& HARPS-N & \cite{marcy2014b}\\
Kepler-126 b & $1.52 \pm 0.10$ & - & 10.496 & 1113 & 6239 & HIRES, HARPS-N & \cite{rowe2014}\\
Kepler-126 c & $1.58 \pm 0.13$ & - & 21.870 & 871 & 6239 & HIRES, HARPS-N & \cite{rowe2014}\\
Kepler-129 c & $2.52 \pm 0.07$ & $43^{+13}_{-12}$ & 82.200 & 571 & 5770 & HIRES & \cite{zhang2021}\\
Kepler-145 b & $2.65 \pm 0.08$ & $37.1 \pm 11.6$ & 22.951 & 958 & 6080 & HIRES & \cite{xie2014}\\
Kepler-21 b & $1.639^{+0.019}_{-0.015}$ & $7.5 \pm 1.3$ & 2.786 & 2015 & 6305 & HIRES & \cite{bonomo2023}\\
Kepler-50 b & $1.71^{+0.05}_{-0.10}$ & - & 7.812 & 1311 & 6225 & HARPS-N & \cite{chaplin2013}\\
Kepler-538 b & $2.215^{+0.040}_{-0.034}$ &  $12.9 \pm 2.9$  & 81.738 & 417 & 5534 & HIRES & \cite{bonomo2023}\\
Kepler-65 b & $1.42 \pm 0.03$ &  -  & 2.155 & 1896 & 6211 & HARPS-N & \cite{chaplin2013}\\
TOI-1346.02 & $2.5 \pm 0.3$ &  -  & 1.762 & 1187 & 4965 & KPF & TICv8$^{a}$\\
TOI-1411 b & $1.199^{+0.049}_{-0.045}$ &  $2.0^{+1.2}_{-1.1}$ & 1.452 & 976 & 4115$^{\dag}$ & HIRES & \cite{polanski2024}, $^{\dag}$\cite{MacDougall2023}\\
TOI-1432.01 & $2.20 \pm 0.15$ &  -  & 6.109 & 916 & 5587 & KPF & TICv8\\
TOI-1466.01 & $2.4 \pm 0.8$ &  -  & 1.872 & 1001 & 4201 & KPF & TICv8\\
TOI-1748.01 & $1.5 \pm 0.9$ &  -  & 1.832 & 1144 & 4837 & KPF & TICv8\\
TOI-2079.02 & $1.9 \pm 0.1$ &  -  & 9.317 & 383 & 3577 & KPF & TICv8\\
TOI-2274.01 & $1.8 \pm 0.2$ &  -  & 2.680 & 771 & 3943 & KPF & TICv8\\
TOI-2470.01 & $2.1 \pm 0.2$ &  -  & 7.185 & 687 & 4562 & KPF & TICv8\\
TOI-406 b & $2.08_{-0.15}^{+0.16}$ &  $6.57^{+1.00}_{-0.90}$  & 13.176 & 368 & 3392 & ESPRESSO & This work\\
TOI-4363.01 & $1.8 \pm 0.2$ &  -  & 2.119 & 885 & 4002 & KPF & TICv8\\
TOI-521.01 & $2.0 \pm 0.1$ &  -  & 1.543 & 769 & 3439 & ESPRESSO & TICv8\\
TOI-771 b & $1.422^{+0.108}_{-0.086}$ &  -  & 2.326 & 527 & 3201 & ESPRESSO & \cite{mistry2024}\\
TOI-836 b & $1.704 \pm 0.067$ &  $4.53^{+0.92}_{-0.86}$  & 3.817 & 871 & 4552 & HARPS, HARPS-N & \cite{Hawthorn2023}\\
TOI-912.01 & $1.81 \pm 0.08$ &  -  & 4.678 & 535 & 3566 & ESPRESSO & TICv8\\
\hline
\end{tabular}
\tablefoot{Summary of the current \thirstee\ sample and the spectroscopic facilities used for the follow-up of each target within the project. Different references for radius and mass are indicated in each row with the $\dag$ symbol. \\
\tablefoottext{a}{{\it TESS} Input Catalogue Version 8 \citep{Stassun2018}}}
\end{threeparttable}
\end{table*}

\medskip

The planet masses measured with this program must be both precise --- 5$\sigma$ at least to avoid degeneracies in atmospheric retrievals \citep{batalha19} --- and accurate. Accuracy is critical in this part of the parameter space as, for a given radius, the range of allowed internal compositions of the planet can change the mass by a factor of five\footnote{For example, a $1.8\,R_\oplus$ planet (the most representative example of our sample) can have a mass between 2--10\,$M_\oplus$ \citep{spright}.}, which in turn changes completely the interpretation on the properties of the planet. There are several examples where model mis-specification (Osborne et al., in prep),
the addition of new data revealing new activity and/or planetary signals \citep{rajpaul2021},
the use of homogeneous analyses \citep{Fulton2017,luquepalle2022}, or changes in the stellar parameters \citep{burt2024} have impacted the accuracy of the mass measurement well beyond the reported uncertainties. 

\medskip

To tackle these issues in our survey, we followed the guidelines presented in \cite{murphy2024_cadence}
to choose an observing cadence and number of measurements that minimise the biases typically associated with RV follow-up of transiting planets. Upon the completion of the survey, we shall provide a catalogue of sub-Neptune densities where both planet and stellar properties have been homogeneously derived, and thus minimise any additional source of systematic errors. This catalogue will contribute to current and future efforts by the community to study this poorly understood exoplanet population using atmospheric characterisation with ground- and space-based facilities, and to enable new statistical studies that benefit from the information added by this survey. That is, the increase and extension of the well-characterised sub-Neptune population in planet equilibrium temperature and host spectral type. These axes are two of the most critical ones (together with age) in global models of planet formation and evolution for distinguishing between scenarios that reproduce the observational properties of this population \citep[e.g. see review by][]{bean21}.

\section{TESS photometry}\label{sec:tess}
{\it TESS} observed \starname\ in two-minute cadence mode
in sectors $3$ and $4$, between September 20 and November 15, 2018, and subsequently in sectors $30$ and $31$, between September 22 and November 19, 2020.
\tess\ observations were reduced by the Science Processing Operations Center (SPOC) 
pipeline \citep{jenkins2016,jenkins2020}. After the release of the first two sectors, the SPOC pipeline identified one candidate planetary signal,
with a period of $13.176$ days (TOI-406.01). 
\citealt{hord2024} vetted and statistically validated TOI-406.01 (TOI-406 b hereafter) using \textit{Gaia} DR3 catalogue information, and incorporating reconnaissance photometry and spectroscopy dispositions and ground-based high-resolution imaging
follow-up observations from the TFOP. Subsequently, an additional signal with a period of $3.307$ days was detected as a Community-TOI. This additional signal was further confirmed by the subsequent SPOC analysis of the four available sectors, even though it had a double period of $6.615$ days (TOI-406.02). 

\begin{figure}
\centering
  \includegraphics[width=\linewidth]{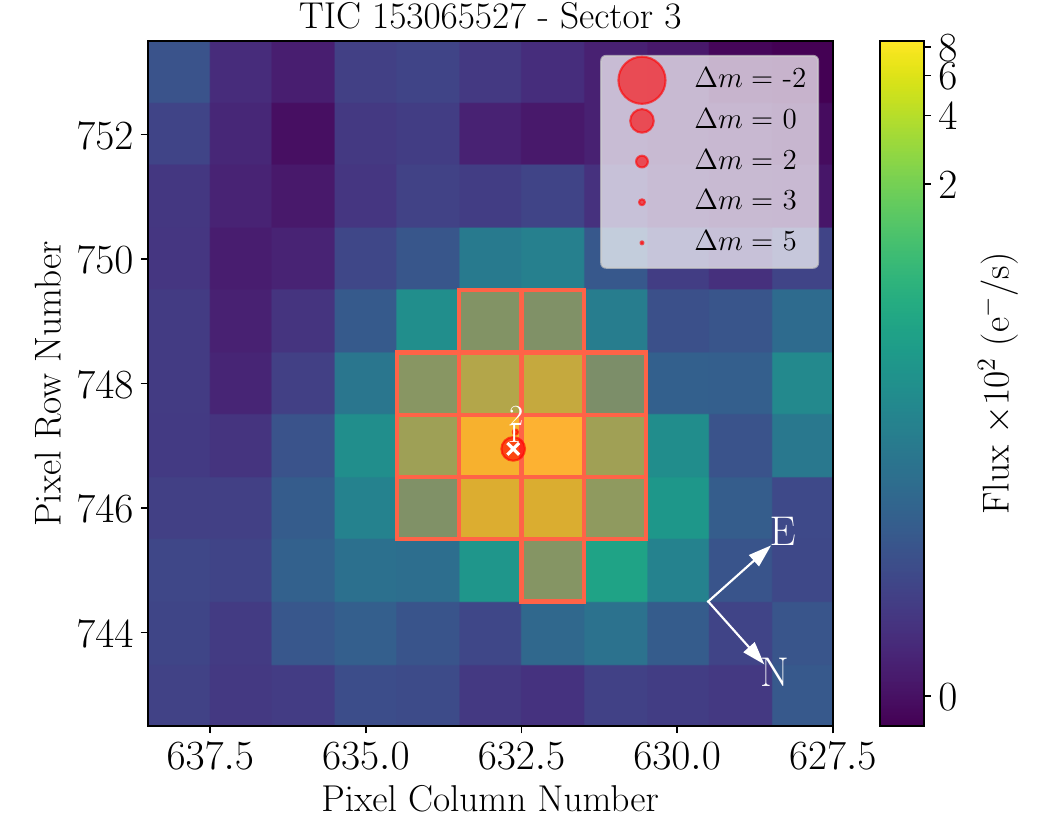}
  \caption{\texttt{tpfplotter} TPF image of \starname\ as in sector 3. The \tess\ aperture used to extract the photometry is shown with red squares. {\it Gaia} DR3 \citep{GaiaColl2023} stars are represented as red circles, with sizes proportional to their magnitude. TOI-406 is positioned at the centre (star 1). Star 2 is a $G=16.887$~mag star whose contribution to the aperture flux was accounted for by the SPOC pipeline through the calculation of the crowding metric and subtraction of the corresponding percentage ($<0.14$~percent in all four sectors) of the median flux level to account for dilution. The TPFs of all sectors are shown in Appendix~\ref{appendix:tpf}.
  }
    \label{fig:tess_tpf}
\end{figure}

In this study, we analysed two-minute cadence Pre-search Data Conditioning Simple Aperture Photometry
\citep[PDCSAP, ][]{smith2012, Stumpe2012, Stumpe2014} light curves downloaded from the
Mikulski Archive for Space Telescopes (MAST)\footnote{
\url{https://mast.stsci.edu/portal/Mashup/Clients/Mast/Portal.html}}.
Figure~\ref{fig:tess_tpf} shows the \tess\ target pixel file (TPF) produced with \texttt{tpfplotter}\footnote{\url{https://github.com/jlillo/tpfplotter}.} \citep{Aller2020}, which displays the field around the
target and highlights the apertures used to extract the light
curves, computed by the SPOC pipeline, which selects for each target and sector the photometric aperture pixels.\\
Before proceeding with the photometric analysis, we removed all points flagged as bad-quality (\texttt{DQUALITY>0}) by the SPOC pipeline\footnote{ \url{https://archive.stsci.edu/missions/tess/doc/EXP-TESS-ARC-ICD-TM-0014.pdf}}, and we rejected the outliers by removing all points above the $3 \sigma$ level for positive outliers and below the $5 \sigma$ level for negative outliers (i.e. lower than the deeper transit). The resulting light curves are shown in Appendix~\ref{appendix:light_curves}.

\section{Ground-based follow-up observations}\label{sec:ground_based}

\cite{hord2024} already statistically validated the planetary nature of TOI-406 b using several ground-based followup observations. Here, we detail the additional photometric and RV observations that we used to measure the masses and densities of the planets in the system.

\subsection{LCOGT photometry}\label{sec:lcogt}

\textit{TESS} has a pixel scale of $\sim 21\arcsec$ pixel$^{-1}$. This can cause multiple stars to be blended in the \textit{TESS} aperture, since photometric apertures generally extend out to about 1 arcminute. In the attempt to identify the real source of the \textit{TESS} transit detection, we performed ground-based follow-up photometry of the field around TOI-406 as part of the TFOP, scheduling the transit observations with the {\tt TESS Transit Finder} tool \citep{jensen2013}.

We observed three transit windows of TOI-406 b on UTC October 30, 2020, August 29, 2021, and November 3, 2021, in Johnson/Cousins $I$ band and Sloan $g'$ (November 3, 2021) from the Las Cumbres Observatory Global Telescope (LCOGT) \citep{brown2013} 1-m network node at Cerro Tololo Inter-American Observatory (CTIO) in Chile. 
We observed a partial and a full transit window on UT February 5, 2019, and October 24, 2022, respectively, in the Sloan $i'$ band from the LCOGT 1\,m network node at South Africa Astronomical Observatory (SAAO) near Sutherland, South Africa.
We also observed another full transit window on UT October 15, 2023, in the Sloan $i'$ band from the LCOGT 1\,m network node at Siding Spring Observatory (SSO) near Coonabarabran, Australia. 
We calibrated the images using the standard LCOGT {\tt BANZAI} pipeline \citep{McCully2018}, and we extracted differential photometry with {\tt AstroImageJ} \citep{Collins:2017} using an aperture radius of $3\farcs 1$ that excluded all of the flux from the nearest known neighbour in the {\it Gaia} DR3 catalogue ({\it Gaia} DR3 4851053994762099840) that is $4\farcs2$ south of TOI-406. 
A $\sim$2.5 ppt event was detected on target in all observations. 
We included all the resulting light curves\footnote{The light curves are available on the {\tt EXOFOP-TESS} webpage: \url{https://exofop.ipac.caltech.edu/tess/target.php?id=153065527}} in our global modelling (Sect.~\ref{sec:joint_fit}). 

The ground-based observations of TOI-406.02 taken with LCOGT were used to rule out unrecognised eclipsing binaries located inside \tess\ photometric aperture that could mimic a planetary signal. However, given the shallow transit depth ($\sim 0.8$~ppt) of TOI-406.02, the transit detection for the existing ground-based LCOGT light curves (not shown) were highly dependent on the detrending parameters and photometric aperture used in the analysis, leading to some ambiguity in the transit evidence. 
Therefore, we did not include these light curves in the global fit.

\subsection{Long-term photometric monitoring}\label{sec:asas_photometry}

\begin{figure*}
\centering
  \includegraphics[width=\linewidth]{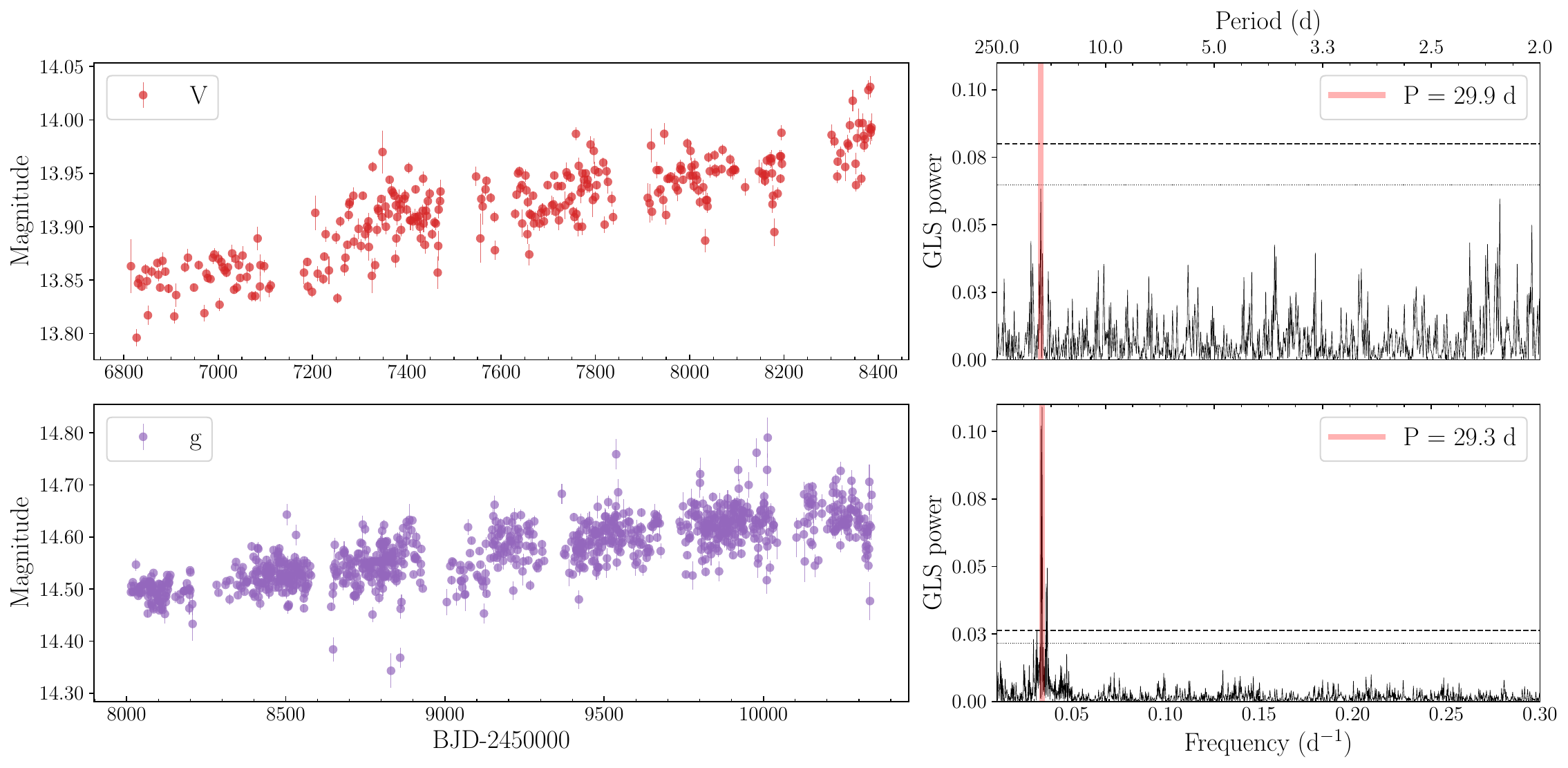}
  \caption{ASAS-SN photometry of TOI-406 in {\it V} and {\it g} filters. Right panel: GLS periodogram of ASAS-SN photometry. The horizontal dashed and dotted lines represent the FAP level of $1$\% and $10$\%, respectively. The vertical red line shows the most significant peak.
  }\label{fig:asas_sn}
\end{figure*}

TOI-406 was observed by the All-Sky Automated Survey for Supernovae (ASAS-SN; \citealt{Shappee2014}, \citealt{Kochanek2017}) between 2014 and 2024.
We retrieved the archival long-term photometry from the ASAS-SN portal\footnote{\url{http://asas-sn.ifa.hawaii.edu/skypatrol/}.} \citep{hart2023}, finding time series both in the {\it V} ($\sim 370$ epochs) and {\it g} ($\sim 1200$ epochs) filters, spanning $\sim 1570$ days and $\sim 2380$ days, respectively (Fig.~\ref{fig:asas_sn}). 
Before proceeding with the analysis, we selected only the observations
flagged as good quality (\texttt{QUALITY=G}), and we performed outlier rejection by doing a $5 \sigma$ clipping, following \citep{hart2023}. 
For the {\it V} filter, the photometry resulted in a median value of $V = 13.92$~mag, with a median uncertainty of $\sigma_{V} = 0.006$ mag and a root mean square (RMS) of $0.04$~mag, while for the {\it g} filter we obtained $g = 14.57$~mag, with a median uncertainty of $\sigma_{g} = 0.009$ mag and an RMS of $0.06$~mag.
The ASAS-SN photometry, even if not appropriate to detect the transit events due to the relatively high uncertainty and cadence (generally $> 2$~d), has been used to investigate the stellar activity and the rotation period of the star (see Sect.~\ref{sec:star_activity}).

\subsection{ESPRESSO spectroscopic observations}\label{sec:spectra}
As part of the \thirstee\ project, we collected $44$ spectra within the 111.24PJ.001 program (PI: E. Pallé) using the Echelle SPectrograph for Rocky Exoplanets and Stable Spectroscopic Observations (ESPRESSO) installed at the Very Large Telescope (VLT) telescope array in the ESO Paranal Observatory, Chile \citep{pepe2021}, with the goal of precisely determining the masses of the candidate planets and searching for additional planetary signals.
The observations started on July 14, 2023, and ended on September 25, 2023. All measurements were gathered using ESPRESSO’s single Unit Telescope (1UT) in high-resolution (HR) mode ($1$~arcsec fibre, $R \sim 140 000$) over a spectral range from $\sim 380$ to $\sim 780$~nm, with $2 \times 1$ detector binning (HR21).
Observations were carried out with fibre B placed on the sky, to measure and subtract the background and possible moon contamination. 
The exposure time was set to $1200$ seconds, which resulted in a median signal-to-noise ratio (\snr) of $23$ at $550$~nm. 
We reduced the data using the ESPRESSO data reduction pipeline (DRS), version 3.0.0. 
We extracted the RV values with the widely used template-matching algorithm \texttt{SERVAL}\footnote{\url{https://github.com/mzechmeister/serval}.} \citep{zechmeister2018}.
The resulting \texttt{SERVAL}-extracted RVs have an RMS of $4.3$~\ms\ and a median individual uncertainty of $0.5$~\ms. The RV data and the associated activity indicators (see Sect.~\ref{sec:star_activity} for more details) are listed in Table~\ref{table:RV_table_ESPRESSO}.

\subsection{NIRPS/HARPS spectroscopic observations}\label{sec:spectra2}
TOI-406 was observed simultaneously from August 25, 2023, to March 16, 2024, with the Near-InfraRed Planet Searcher (NIRPS; \citealt{Bouchy_2017}; \citealt{Wildi_2022}) and the High Accuracy Radial velocity Planet Searcher (HARPS; \citealt{Pepe_2002}), both of which are installed on the ESO 3.6-metre telescope at La Silla, Chile. Both instruments are fibre-fed, stabilised, high-resolution echelle spectrographs, with NIRPS operating in the near-infrared (980--1800\,nm) under an adaptive optics system, and HARPS working in the optical domain (330--690\,nm) in seeing-limited conditions. The observations were conducted as part of the NIRPS-GTO \citep{Artigau_2024} Follow-up of Transiting Planets sub-program (PID: 111.254T.001, 112.25NS.001; PI: Bouchy \& Doyon) in high-efficiency modes; that is, the HE mode for NIRPS ($R\approx70\,000$, 0.9$^{\prime\prime}$ fibre) and the EGGS mode for HARPS ($R\approx80\,000$, 1.4$^{\prime\prime}$ fibre). Over 55 individual nights, we collected two spectra of TOI-406 per night with NIRPS (900\,sec per exposure), plus a single spectra for an additional night, yielding a median SNR of 71.1 per pixel in the middle of $H$ band. 
As NIRPS operated alone for nine nights, the HARPS dataset comprise 47 spectra (a single 1800\,sec exposure per night) with a median SNR of 10.7 per pixel near 550\,nm.
The NIRPS data were reduced with the \texttt{APERO} pipeline (v0.7.290; \citealt{Cook_2022}), the standard DRS for the SPIRou spectrograph \citep{Donati_2020}, which is also compatible with NIRPS. \texttt{APERO} is optimised for near-infrared spectroscopy with many routines specifically developed and refined over the years to address challenges associated with this spectral domain such as telluric contamination, hot pixels, and detector persistence. 
For HARPS, we used the extracted spectra from the HARPS DRS \citep{Lovis_2007} available through the DACE platform\footnote{\href{https://dace.unige.ch/publications/}{\texttt{dace.unige.ch}}}.
The RVs were measured from the processed data using the line-by-line algorithm described in \cite{Artigau_2022}, which is publicly available (v0.69; \texttt{LBL}\footnote{\href{https://github.com/njcuk9999/lbl}{\texttt{github.com/njcuk9999/lbl}}}). 
The \texttt{LBL} package is compatible with both NIRPS and HARPS. The method is conceptually similar to template matching (e.g. \citealt{Anglada-Escude_2012}, \citealt{Astudillo-Defru_2017}), while being more resilient to outlying spectral features (e.g. telluric residuals, cosmic rays, detector defects) as the template fitting is performed line by line, which facilitates the identification and removal of outliers. 
For NIRPS, we used the template of a brighter star with a similar spectral type, GJ~643 (M3.5V), from NIRPS-GTO observations. 
For HARPS, we instead employed the template of GL~699 (M4V) from public data obtained via the ESO archive \citep{Delmotte_2006}. An additional telluric correction was performed for HARPS inside the LBL code by fitting a TAPAS atmospheric model \citep{Bertaux_2014}. 

We removed four outlying epochs (eight measurements) for NIRPS with velocities about 30\,m/s above average. These epochs at barycentric earth radial velocity (BERV) of 12.2--13.1\,km/s could be biased by telluric residuals (imperfect correction) that align with the stellar lines within a line width (that is, with a full width at half maximum (FWHM) of $\sim$ 5\,km/s) given that the systemic velocity of TOI-406 is 15.0\,km/s. All other epochs were taken at BERV ranging from -16.6 and 9.5\,km/s; that is, well outside of one line width of stellar velocity. We also discarded four HARPS measurements with imprecise velocities, each with a low SNR below 3. The final NIRPS (101) and HARPS (43) RVs have a median uncertainty of 4.40 m/s and 3.73 m/s and an RMS of 7.21 m/s and 7.43 m/s, respectively. 
The NIRPS and HARPS RV data and the associated activity indicators (see Sect.~\ref{sec:star_activity} for more details) are listed in Tables~\ref{table:RV_table_NIRPS} and \ref{table:RV_table_HARPS}, respectively. 
\section{The star}\label{sec:star}

\begin{table}[h!]
\small
\caption{Stellar properties of \starname.}
\label{table:star_params}
\begin{threeparttable}[t]
\centering
\begin{tabular}{lll}
\hline\hline
\multicolumn{3}{c}{\starname}\\
\hline
TIC & \multicolumn{2}{l}{153065527} \\
{\it Gaia} DR3 & \multicolumn{2}{l}{4851053999056603904}\\
2MASS & \multicolumn{2}{l}{J03170297-4214323} \\
LP & \multicolumn{2}{l}{994-91}\\[1ex]
\hline
Parameter & Value & Source \\
\hline
RA  (J2000; hh:mm:ss.ss) &  03:17:02.98 & A\\
Dec (J2000; dd:mm:ss.ss) & $-$42:14:32.54 & A\\
$\mu_{\alpha}$ (mas yr$^{-1}$) & $41.649 \pm 0.017$ & A\\
$\mu_{\delta}$ (mas yr$^{-1}$) & $-412.870 \pm 0.021$ & A\\
Parallax (mas) & $32.3545 \pm 0.0178 $& A\\
Distance (pc) & $ 30.91 \pm 0.02 $& B \\
$\gamma$ (\kms) & $14.48 \pm 0.34$ & A \\
U (\kms) & $44.05 \pm 0.31$& F$^a$  \\ 
V (\kms) & $-44.44 \pm 0.29$& F$^a$  \\
W (\kms) & $-0.18 \pm 0.29$& F$^a$  \\
\hline
{\it TESS} (mag) & $11.283 \pm 0.007$& C\\
{\it G} (mag) & $12.544 \pm 0.003$ & A\\
{\it G$_{\rm BP}$} (mag) & $14.050 \pm 0.003$ & A\\
{\it G$_{\rm RP}$} (mag) & $11.341 \pm 0.004$ & A\\
{\it B } (mag) & $15.237 \pm 0.072$ & C\\
{\it V }(mag) & $13.773 \pm	0.038$ & C\\
{\it J }(mag) & $9.695  \pm	0.023$ & D\\
{\it H }(mag) & $9.128  \pm	0.023$ & D\\
{\it K }(mag) & $8.848  \pm	0.021$ & D\\
{\it W1} (mag) & $8.724 \pm	0.023$ & E\\
{\it W2} (mag) & $8.576  \pm	0.020$ & E\\[1ex]
\hline
\teff\ (K) & $ 3392 \pm 44 $ & F\\ 
\logg\  (cgs)  & $ 4.69 \pm 0.04$ & F, {\tiny Spectroscopic} \\
\logg\  (cgs)  & $ 4.82 \pm 0.11$ & F, {\tiny Bolometric} \\ 
$[$M/H$]$ (dex) & $-0.02 \pm 0.04$ & F \\
\logRHK & $ -5.167 \pm  0.017$ & F\\
$P_{\rm rot}$ (d) & $29.1_{-0.5}^{+0.6}$ & F\\
\rstar\ (\rsun) & $  0.410^{+0.029}_{-0.027}$ & F\\
\mstar\ (\msun) & $  0.408^{+0.046}_{-0.042}$ & F \\
$L_{\star}$ ($L_{\odot}$) & $ 0.0193 \pm 0.0003$ & F\\ 
Spectral type & M3 & F\\
\hline\hline
\end{tabular}
\tablebib{
\small
A) {\it Gaia} DR3 \citep{GaiaColl2023}. B) \citet{bailer_jones2021}.
C) {\it TESS} Input Catalogue Version 8 (TICv8, \citealt{Stassun2018}).
D) Two Micron All Sky Survey (2MASS, \citealt{Cutri2003}).
E) {\it Wide-field Infrared Survey Explorer} \citep[{\it AllWISE};][]{cutri_allwise}. F) This work. \\
$^a$ Right-handed, heliocentric galactic space velocity components.}
\end{threeparttable}
\end{table}

\subsection{Stellar parameters}\label{sec:star_properties}
We derived the stellar atmospheric parameters of TOI-406 using {\sc SteParSyn}\footnote{\url{https://github.com/hmtabernero/SteParSyn/}} \citep{tab22} applied to the ESPRESSO spectra. 
We employed a grid of synthetic spectra generated using the Turbospectrum code \citep{ple12} in combination with the BT-Settl stellar atmospheric models \citep{all12}, and atomic and molecular data from the Vienna atomic line database \citep[VALD3, see][]{rya15}. 
We employed a list of \ion{Fe}{i} and \ion{Ti}{i} lines, along with various TiO molecular bands, which have been shown to be well suited for the analysis of M-type stars (see \citealt{mar21} for more detail). 
From our analysis, we obtained the following stellar parameters: $T_{\rm eff}$~$=$~3364~$\pm$~17~K, $\log{g}$~$=$~4.69~$\pm$~0.04 dex, and [M/H]~$=$~$-0.08$~$\pm$~0.08~dex. The first set of error bars refers only to the internal errors, and to estimate a more realistic uncertainty on \teff, we added a systematic component to the errors, following \cite{tayar2022}, obtaining $T_{\rm eff}$~$=$~3364~$\pm$~100~K.

The effective temperature, [M/H], and abundance measurements of several elements were also determined from the NIRPS spectrum using the analysis framework of \citealt{Jahandar_2024}. 
The first step in the analysis consists of fitting both $T_{\rm eff}$  and  [M/H] through  $\chi^2$ minimisation of several spectral regions using ACES stellar models (\citealt{all12}; \citealt{husser2013new}) with a fixed log~$g$ of 5.0, convolved to match the NIRPS resolution.  
The analysis yields \hbox{$T_{\rm eff}= 3399\pm 49$\,K}  and \hbox{[M/H]~$= 0.0 \pm 0.05$\,dex}, in good agreement with the parameters inferred from the ESPRESSO spectrum. 
The values reported in Table~\ref{table:star_params} are the weighted averages of both ESPRESSO and NIRPS measurements; that is, \hbox{$T_{\rm eff} = 3392 \pm 44$\,K} and \hbox{[M/H]~$= -0.02 \pm 0.04$\,dex}.
Finally, a series of fits was performed to a fixed $T_{\rm eff} = 3400$\,K on individual spectral lines of known chemical species to derive their elemental abundances.  
The results are summarised in Table~\ref{table:NIRPS_abundances}.  

\begin{table}[h]
\centering
\small
\caption{TOI-406 stellar abundances measured with NIRPS.}
\begin{tabular}{lccc}
\hline\hline
Element &[X/H] & $\sigma$ & \# of lines \\
\hline
Fe I      &  0.07   &  0.07   & 45      \\
Al I      &  0.27   &  0.12   & 5       \\
Mg I      &  0.32   &  0.06   & 8       \\
Ca I      &  -0.02  &  0.3    & 3       \\
Ti I      &   0.02  &  0.2    & 8       \\
Na I      &   0.17  &  0.18   & 2       \\
K I       &   0.1   &  0.1    & 3       \\
O I$^*$   &  0.00   &  0.01   & 82      \\
\hline
\end{tabular}
\tablefoot{
\tablefoottext{*}{The oxygen abundance was inferred from OH lines.}}
\label{table:NIRPS_abundances}
\end{table}

\begin{figure}
\centering
  \includegraphics[width=\linewidth]{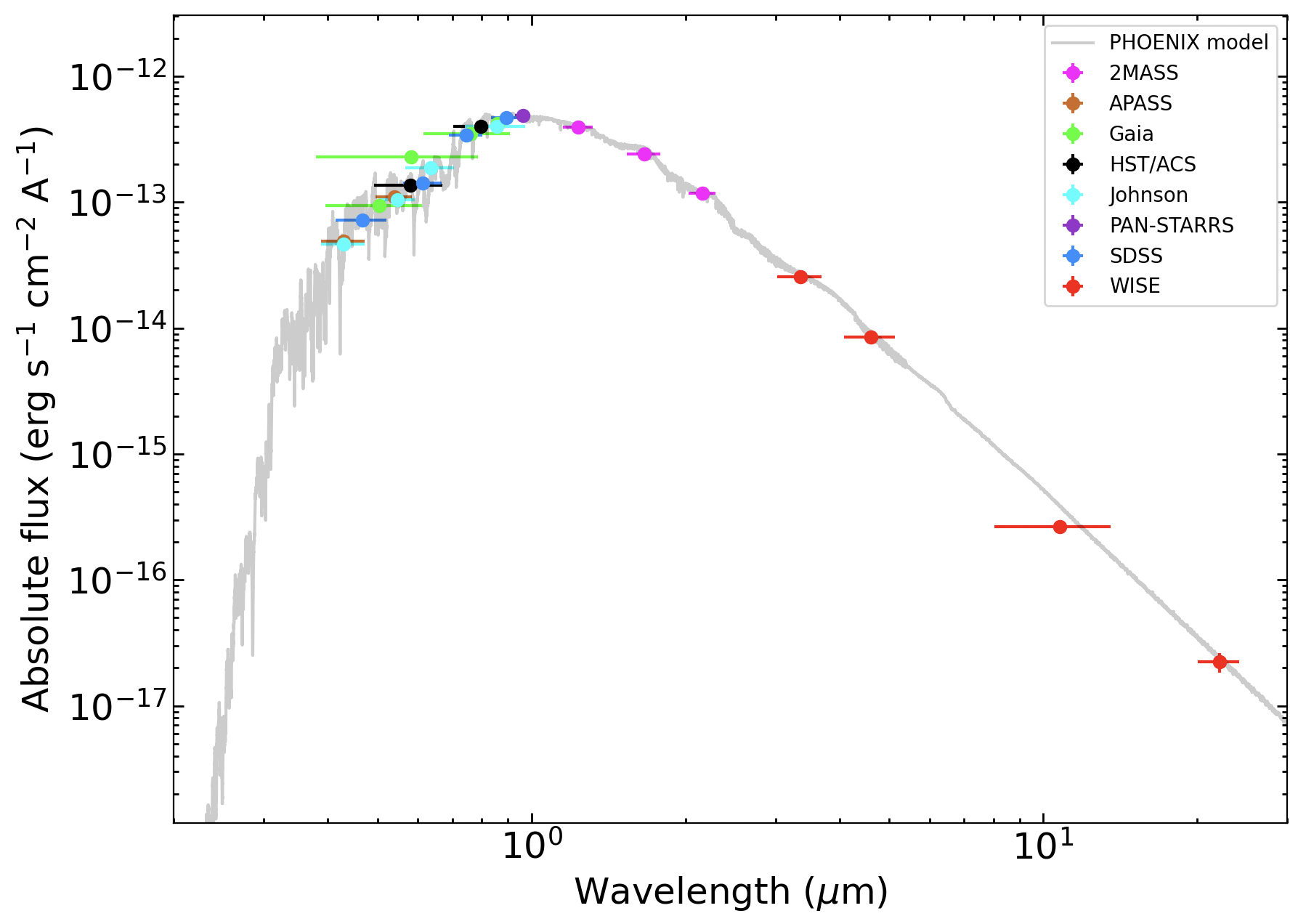}
  \caption{SED for TOI-406. The grey line shows the BT-Settl model for \teff~$=3300$~K, \logg~$= 5.0$~dex, and solar metallicity. 
  }
    \label{fig:sed}
\end{figure}

We determined the photometric spectral energy distribution (SED) of TOI-406 (Fig.~\ref{fig:sed}) by combining all broad-band photometry available from public archives, including 
the American Association of Variable Star Observers Photometric All-Sky Survey \citep[APASS,][]{henden_2014}, 
the Sloan Digital Sky Survey \citep[APASS,][]{york_2000},
the {\it Gaia} DR3 archive \citep{gaia_2016,GaiaColl2023}, 
2MASS \citep{skrutskie_2006}, 
and the Wide-field Infrared Survey Explorer archive \citep[{\sl WISE},][]{wright_2010}.
Observed magnitudes were converted into fluxes by using the zero points given in the Virtual Observatory SED Analyzer tool \citep[VOSA,][]{bayo2008}, and observed fluxes were transformed into absolute fluxes by employing the {\it Gaia} DR3 trigonometric parallax. 
As is shown in Fig.~\ref{fig:sed}, TOI-406's SED is purely photospheric up to $\sim$25 $\mu$m. 
The stellar bolometric luminosity (\hbox{$L_{\star} = 0.0193 \pm 0.0003$~L$_\odot$}) was obtained by integrating the observed SED. The stellar radius ($R_{\star} = 0.410 ^{+0.029}_{-0.027}$~R$_\odot$) was then computed via the Stefan-Boltzmann law, which states $L_{\star} = 4 \, \pi \, R^2 \, \sigma_{\rm SB} \, T_{\rm eff}^4$, where $\sigma_{\rm SB}$ is the Stefan-Boltzmann constant.  
We measured the stellar mass ($M_{\star} = 0.408^{+0.046}_{-0.042}$ M$_\odot$) by using the mass-radius relation of \cite{schweitzer2019} (their Eq.~6).
As a consistency check, we obtained the surface gravity, log\,$g = 4.82 \pm 0.11$ (cm\,s$^{-2}$), using our derived stellar mass and radius, which is compatible at the 1\,$\sigma$ level with the spectroscopic value (log\,$g = 4.69 \pm 0.04$) obtained from the spectral fitting analysis.

Finally, we used the {\it Gaia} DR3 positions, proper motions, parallax, and systemic RV to compute the galactic heliocentric space velocities of \starname\ ($U = 44.05 \pm 0.31$ \kms, $V = -44.44 \pm 0.29 $ \kms, $W = -0.18 \pm 0.29 $ \kms), in the directions of the Galactic centre, Galactic rotation, and north Galactic pole, respectively.
According to the probabilistic approach of \cite{bensby2014}, the galactic kinematic indicates that \starname\ is a thin disc population star. All the adopted astrometric and photometric stellar properties of \starname\ are listed in Table~\ref{table:star_params}.

\subsection{Stellar activity indices and rotation period}\label{sec:star_activity}
We first analysed the periodicity of the \tess\ long-cadence SAP light curve\footnote{We decided to analyse the SAP light curve instead of the PDCSAP for the search of the rotation period as this photometry is less affected by systematic pipeline correction that could over-correct/inject spurious signals \citep{smith2012, Stumpe2012, Stumpe2014}.} and the long-term ASAS-SN photometry to identify the possible rotation period of the star. The generalised Lomb-Scargle (GLS, \citealt{Zechmeister2009}) periodogram of the \tess\ light curve identified a broad peak around $20-30$~d, due to the different results of the 2018 (sectors $3$, $4$) and 2020 (sectors $30$, $31$) data, which, when analysed separately, show a significant peak at $28$ and $21$ days, respectively. 

Both the ASAS-SN series show a periodicity around $\sim 29$~d, especially significant in the {\it g} filter (see Fig.~\ref{fig:asas_sn}. These results are further confirmed by the autocorrelation analysis we performed following the implementation of \citep{angus2021}, from which we identify the main periodicity in the {\it TESS} and ASAS-SN {\it g} light curves at $28.7$ and $28.5$ days, respectively\footnote{The auto-correlation analysis of the ASAS-SN {\it V} time series was inconclusive, even though pointing towards a period of $\sim 21$~days.}.

We also investigated the stellar activity and assessed its impact on the RV measurements using a series of activity indicators (available on CDS). 
For the ESPRESSO dataset, we included in our analysis various absorption line indicators produced by the \texttt{SERVAL} pipeline, such as the Na \textsc{i} doublet ($\lambda \lambda 589.0$~nm, $589.6$~nm), the \halpha\ ($\lambda 656.2$~nm), and the $S$-index (Ca \textsc{ii} H\&K lines, $\lambda\lambda 396.8470$~nm, $393.3664$~nm, computed with \texttt{ACTIN2}\footnote{\url{https://github.com/gomesdasilva/ACTIN2}} \citealt{Gomes2018}, \citealt{Gomes2021}), as well as other \texttt{SERVAL}-derived activity indicators, such as the RV chromatic index (CRX) and the differential line width (dLW; see \citealt{zechmeister2018} and \citealt{jeffers2022} for a detailed description and discussion of the different indicators). 
In addition, we analysed the activity indexes automatically provided by the ESPRESSO DRS \citep{Baranne1996, Pepe2002}, such as the FWHM, the bisector span (BIS) \citep{queloz2001}, and the contrast (Contr) of the cross-correlation function (CCF). For the CCF extraction, we used a {\tt M3} binary mask.

\begin{figure}
\centering
  \includegraphics[width=\linewidth]{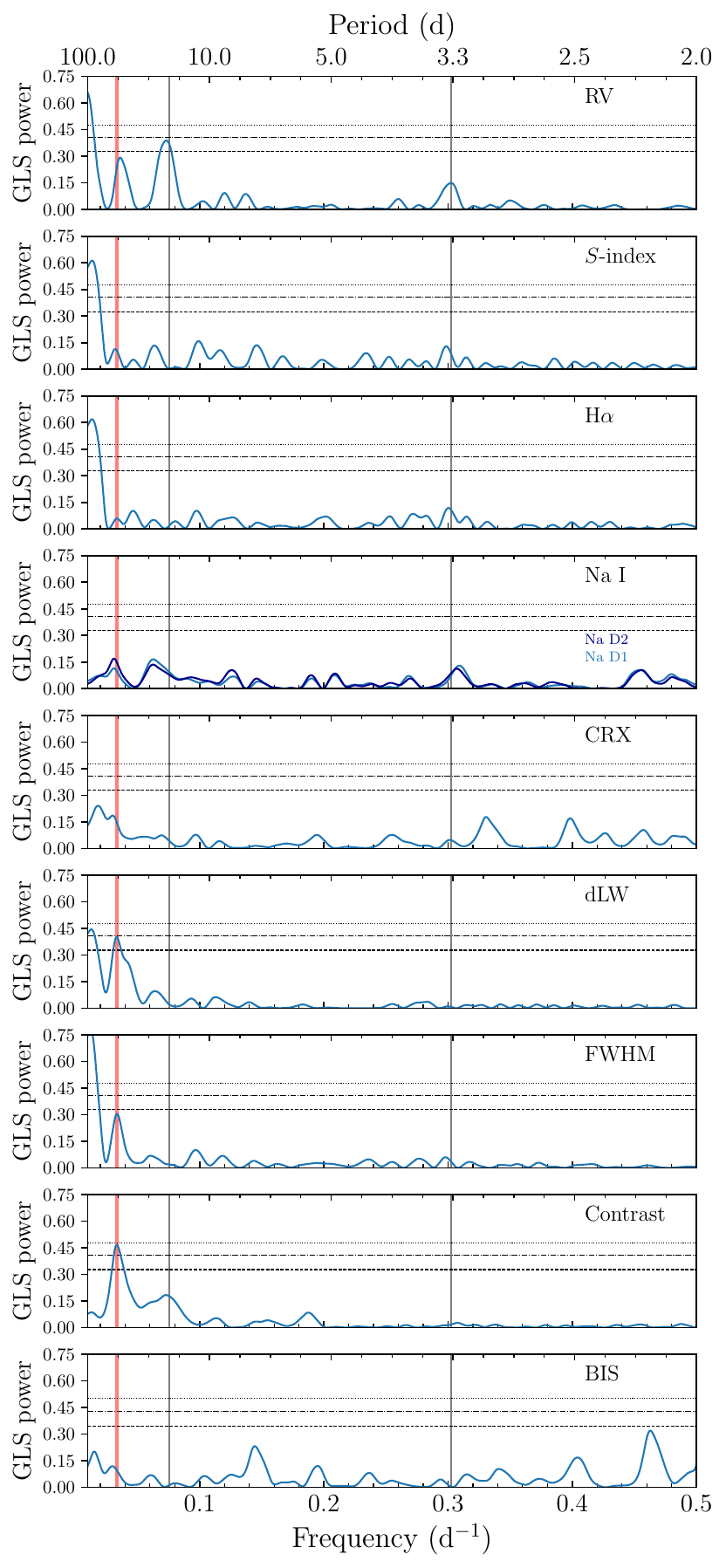}
  \caption{GLS periodogram of the ESPRESSO RVs and the spectroscopic activity indicators. The vertical grey lines indicate the transit-like signals with periods of $13.176$ and $3.307$ days. The vertical red line shows the frequency corresponding to the possible rotational period of the star at $\sim 29$ days. The dashed, dash-dotted, and dotted horizontal lines show the $10\%$, $1$\%, and 0.1\% FAP levels, respectively. 
  }
    \label{fig:GLS_indicators_ESPRESSO}
\end{figure}

\begin{figure*}
\centering
  \includegraphics[width=0.48\linewidth]{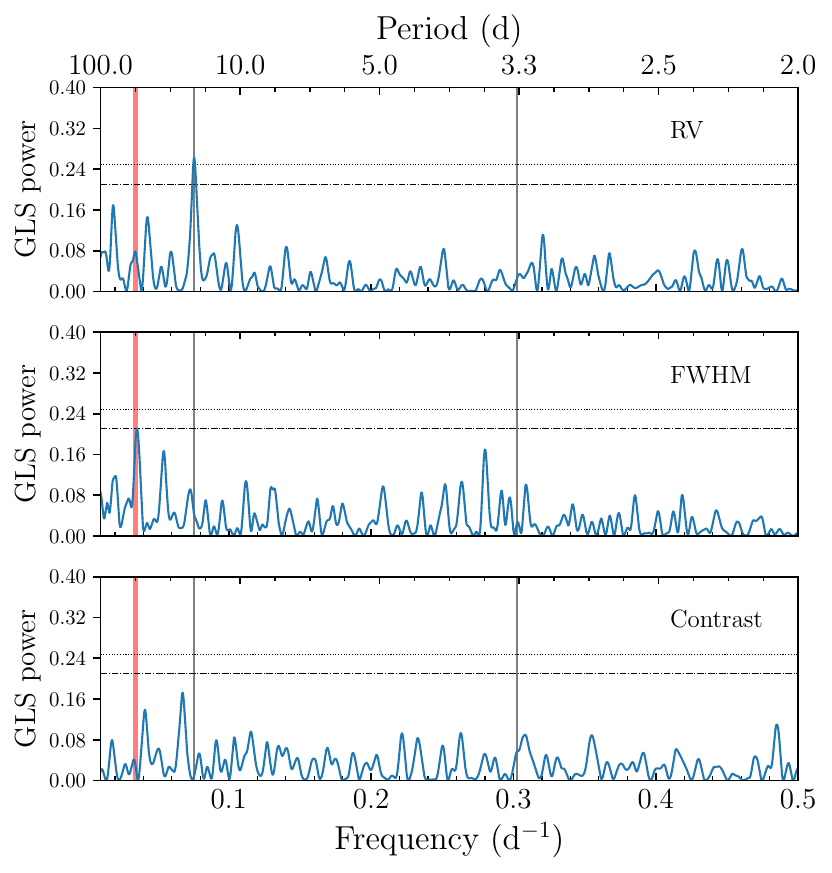}
  \includegraphics[width=0.48\linewidth]{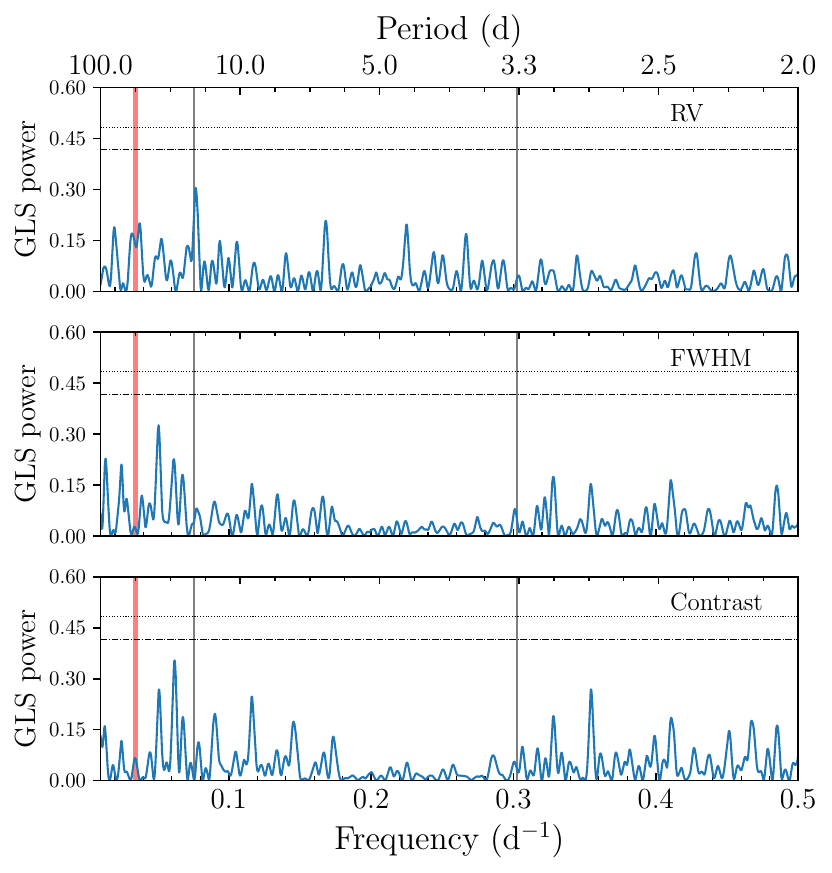}
  \caption{Same as Figure~\ref{fig:GLS_indicators_ESPRESSO}, but for the NIRPS (left) and HARPS (right) datasets. 
  }
    \label{fig:GLS_indicators_HAR_NIR}
\end{figure*}

\begin{figure*}
\centering
  \includegraphics[width=0.48\linewidth]{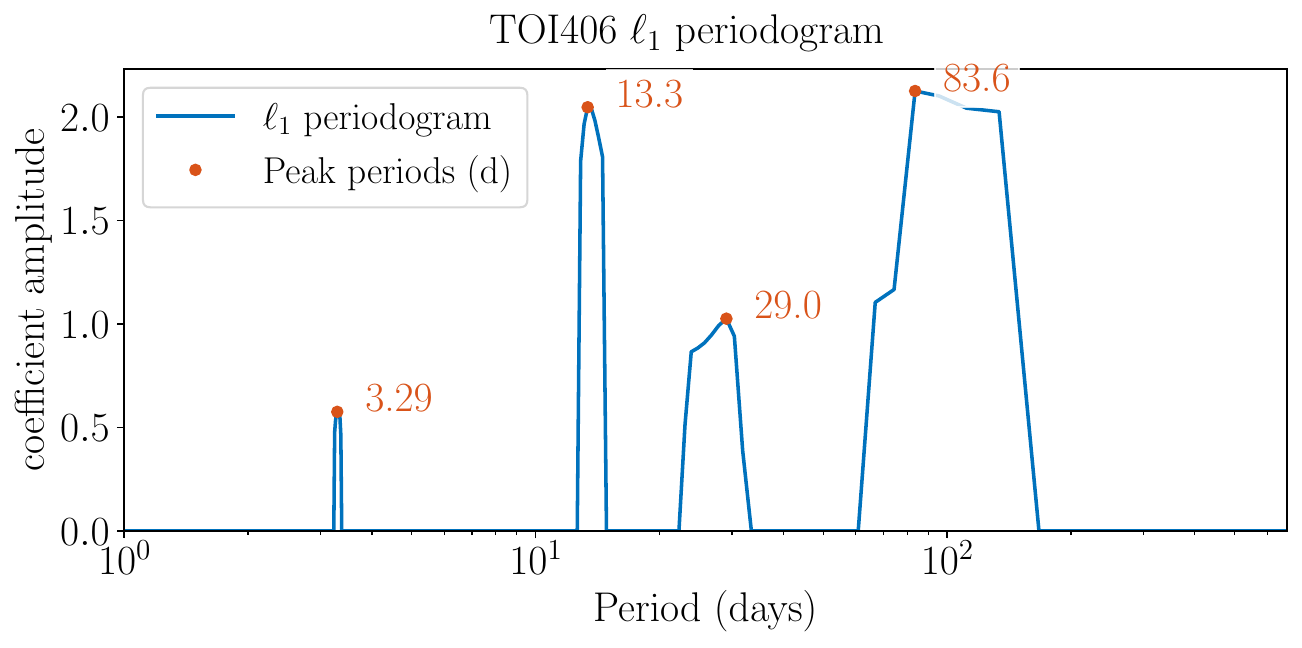}
  \includegraphics[width=0.48\linewidth]{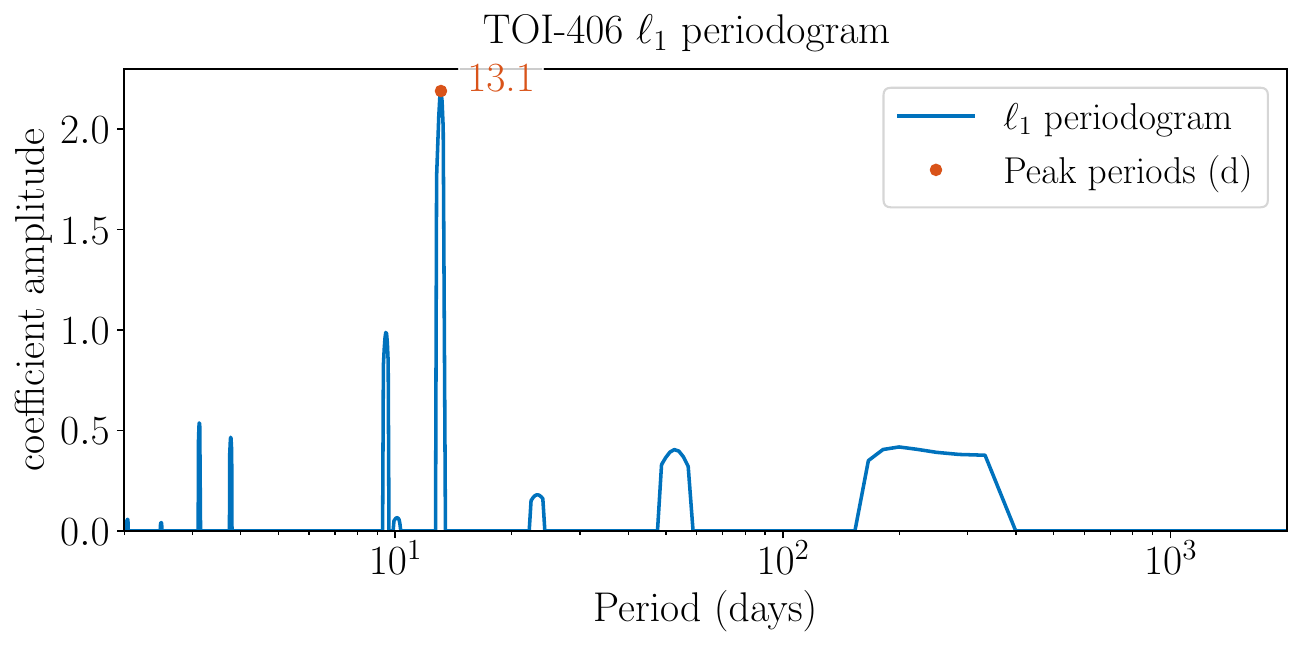}
  \caption{$\ell_1$ periodogram of the ESPRESSO (left) and the combined NIRPS+HARPS datasets (right), computed on a grid of frequencies from 0 to 1 cycle per day. The total time span of the ESPRESSO and NIRPS/HARPS observations is $\sim 72$ and $\sim 204$ days, respectively. Each periodogram highlights in orange the significant peaks (with $\Delta\ln\mathcal{Z} > 10$, where $\mathcal{Z}$ is the Bayesian evidence).
  }
    \label{fig:l1_periodogram}
\end{figure*}

The GLS periodograms of the ESPRESSO RVs and the above-mentioned indexes are shown in Fig.~\ref{fig:GLS_indicators_ESPRESSO}.  
The periodogram of the RVs reveals the presence of a significant peak at $\simeq 13$ days (corresponding to one of the transiting candidates). 
This signal is also clearly identified in the GLS periodogram of the NIRPS RVs, and in the HARPS dataset, despite the lower significance (Fig.~\ref{fig:GLS_indicators_HAR_NIR}). 
Even though not particularly significant, we also identified in the ESPRESSO periodogram the presence of a signal at $\sim 3.3$ days, corresponding to the second planetary candidate (half of the period identified by \tess, see Sect.~\ref{sec:global_analysis}).  
Additionally, the ESPRESSO periodogram shows hints of a signal at $\simeq 29$ days, and a possible long-period signal peaking widely around $80-110$~d. We note that such a signal is longer than the time-span of the ESPRESSO RV observations ($\sim 72$ days), and therefore it cannot be accurately sampled using the current ESPRESSO dataset. The presence of these four significant peaks is further confirmed from our $\ell_1$ periodogram\footnote{\url{https://github.com/nathanchara/l1periodogram}.} analysis (Fig.~\ref{fig:l1_periodogram}). The $\ell_1$ periodogram \citep{hara2017} is designed to search for periodicities in unevenly sampled time series, similarly to the GLS periodogram, but identifying fewer peaks due to aliasing. 
None of these additional signals are clearly identified in the NIRPS/HARPS datasets, 
(see Figs.~\ref{fig:GLS_indicators_HAR_NIR} and \ref{fig:l1_periodogram}), likely due to the higher scatter and internal errors with respect to the ESPRESSO dataset (Sect.~\ref{sec:ground_based}).

Despite most of the ESPRESSO activity indexes not presenting significant periodic signals (false alarm probability, FAP $<10$ \%), the $29$~d peak has a counterpart in the dLW, in the FWHM, and in the contrast of the CCF. 
This periodicity is also identified in the FWHM of the NIRPS dataset (Fig.~\ref{fig:GLS_indicators_HAR_NIR})\footnote{In the case of the HARPS dataset, which is limited with respect to the NIRPS one, such signal is not clearly identified in the FWHM.}.

The peak at $\sim 29$~d corresponds to the period signal identified in the \tess\ and ASAS-SN photometry. According to our activity analysis, we attributed this signal to the rotational period of the star. Moreover, using the \logRHK\ and the activity-rotation calibration for M dwarfs of \cite{suarez_mascareno2016}, we estimate a rotation period of $40_{-12}^{+20}$ days, which is consistent within $1 \sigma$ with the identified signal.

Even though not particularly active, as the low value of the \logRHK\ index indicates (\logRHK $= -5.167 \pm  0.017$; derived from the median $S$-index using the calibrations of \cite{suarez_mascareno2015} and adopting $B - V = 1.464$), from the activity analysis we concluded that the ESPRESSO RV data are affected by stellar activity variations.
Therefore, to consistently model both the planetary and stellar signals, we selected the dLW\footnote{According to the definition \citep{zechmeister2018}, dLW samples the same effect as FWHM, that is, the changes in the absorption line widths of the spectrum related to the changes in the stellar flux caused by the active regions.} series as a proxy of the stellar activity, and in our global analysis we fitted it simultaneously with the RVs (see Sect.~\ref{sec:global_analysis}) in a Gaussian processing (GP) regression framework, which also allowed us to determine the final adopted stellar rotation period (\prot~$=29.1_{-0.5}^{+0.6}$ days).

\section{Data analysis and results}\label{sec:global_analysis}
\subsection{Photometric fit}\label{sec:photometric_fit_only}

Given the ambiguity in the planetary period of the second \tess\ candidate (see Sect.~\ref{sec:tess}), we first performed a photometric fit of the \tess\ light curve only to test the periodicity of TOI-406.02. 
We fitted a two-planet model using the retrieval code \texttt{PyTransit}\footnote{\url{https://github.com/hpparvi/PyTransit}.} \citep{Parviainen2015_Pytransit}, including a GP regression with a Matérn-3/2 kernel to model the short-term periodicity. 
We parametrised the model of each planet using the planetary to stellar radius ratio ($R_\mathrm{p}/$\rstar), mid-transit time ($T_0$), period ($P$), impact parameter ($b$), a common stellar density (\rhostar), two quadratic limb darkening (LD) coefficients, and an average white noise estimate. 
The LD coefficients were constrained using priors calculated with \texttt{PyLDTk}\footnote{\url{https://github.com/hpparvi/ldtk}.} \citep{Husser2013, Parviainen2015}. 
For TOI-406 b, we adopted uniform priors on $T_0$ and $P$ centred around the \tess\ TOI announcement values, with a central time window of $\pm 1$ day and a period range of $P \in [11, 15]$ days. For TOI-406.02, we left the period free to vary between $1$ and $8$ days, and computed a $T_0$ consistent with an orbital period of $\sim$6.6 days and $\sim$3.3 days, while using the same time window range of $\pm 1$ day. We adopted uniform, uninformative priors for the other parameters. 
The best-fitting period for TOI-406.02 from this analysis was $3.307$ days, as is further confirmed by the RV analysis (Sect.~\ref{sec:RV_fit_only}) and by the joint modelling combining all photometric and RV information (Sect.~\ref{sec:joint_fit}). 
All other parameters from the transit fit are in agreement within $1 \sigma$ with the values obtained from the joint modelling and adopted as final values in this work.

\subsection{Radial velocity fit}\label{sec:RV_fit_only}
To prove the feasibility of the \texttt{THIRSTEE} program and check if we can recover the planetary masses using only the \texttt{THIRSTEE} dataset, we performed a RV fit of the ESPRESSO-only data, and a joint fit including ESPRESSO and NIRPS/HARPS GTO data. 
To model the RVs, we used \pyorbit\footnote{\url{https://github.com/LucaMalavolta/PyORBIT}.} \citep{Malavolta2016, Malavolta2018}, a Python package which allows for the modelling of planetary transits and RVs together with stellar activity effects. 
We assumed a two-planet Keplerian model, setting wide boundaries on the semi-amplitude, $K$ ($0.01$-$100$~\ms), for both candidates, and exploring the parameters in logarithmic space. 
We allowed the periods of the two planets to span between $2$ and $20$ days, while we assumed normal priors on $T_0$ based on the values obtained from the photometric fit. 
We assumed the half-Gaussian zero-mean prior of \cite{vanEylen2019} on the orbital eccentricity, $e$, and we adopted the parametrisation of \cite{Eastman2013}, fitting ($\sqrt{e} \sin{\omega}$, $\sqrt{e} \cos{\omega}$) to determine the eccentricity and argument of periastron, $\omega$.
We modelled the stellar activity by including in the fit a GP regression with a quasi-periodic kernel \citep{Grunblatt2015}. 
We modelled simultaneously the RVs and the ESPRESSO dLW time series, which we selected as a proxy for the stellar activity given the presence of a significant periodic signal (Sect.~\ref{sec:star_activity}), in order to better inform the GP \citep{langellier2021, Osborn2021, barragan2023}. 
We used two independent covariance matrices assuming common GP hyper-parameters (stellar rotation period \prot, characteristics decay timescale, $P_\mathrm{dec}$, and coherence scale, {\it w}), except for the amplitude of the covariance matrix, which was fitted separately for the two time series. 
For each RV dataset, we included in the fit a systemic velocity term, and a jitter term to account for any underestimation of the error bars, possible systematics, and short-term stellar activity noise. 
Finally, given the hints of a possible long-period signal in the ESPRESSO RVs (Sect.~\ref{sec:star_activity}), we included in the fit a second-order polynomial trend to take into account the long-term variability (with $c_0$ as the x-intercept, $c_1$ as the linear coefficient, and $c_2$ as the quadratic coefficient).

We performed the RV fit with \texttt{PyORBIT} using the PyDE\footnote{\url{https://github.com/hpparvi/PyDE}} + {\tt emcee} \citep{ForemanMackey2013} set-up, as is detailed in \cite{lacedelli2022} (their Sect. 5), applying the same convergence criteria.
We ran the chains with $2 n_{\mathrm{dim}}$ walkers, where $n_{\mathrm{dim}}$ is the dimensionality of the model, for $200000$ steps, discarding the first $30000$ steps as burn-in and using a thinning factor of $100$.

In both cases (ESPRESSO-only and all available RV data), no signal was identified at $6.6$ days, while the recovered periodicity for the TOI-406.02 candidate was $3.3$ days, with an $8 \sigma$ detection. 
This is in agreement with our photometric analysis (Sect.~\ref{sec:photometric_fit_only}), and further confirms that the true period of the inner candidate (TOI-406 c hereafter) is half the proposed initial period. 
We list the retrieved semi-amplitude of the two planets from the ESPRESSO-only fit and from the fit including all the available RVs in Table~\ref{tab:RV_comparison}, where only the values for TOI-406 b are discrepant but consistent within $1 \sigma$.

\begin{table}[h!]
\caption{RV semi-amplitude of the TOI-406 planets.}
\label{tab:RV_comparison}
\centering                                      
\begin{tabular}{l c c}          
\hline\hline                        
Instrument & $K_{\mathrm b}$ (\ms) & $K_{\mathrm c}$ (\ms)\\
\hline                                 
ESPRESSO & $2.83_{-0.32}^{+0.30}$& $1.60 \pm 0.12$ \\
ESPRESSO+NIRPS+HARPS & $3.20_{-0.51}^{+0.46}$ 
 & $1.66 \pm 0.12$ \\
\hline
\end{tabular}
\tablefoot{Retrieved semi-amplitude of TOI-406 b ($P = 13.17$~d) and TOI-406 c ($P = 3.3$~d) obtained with each RV analysis of the different datasets.}
\end{table}

\subsection{Joint modelling of light curves and RVs}\label{sec:joint_fit}
Finally, we performed the analysis of all the photometric and spectroscopic data using \pyorbit, adopting the model from the RV analysis (Sect.~\ref{sec:RV_fit_only}) and running the fit both in the ESPRESSO-only case and including all the RVs.
We included in the model the \tess\ light curves, and the ground-based observation of TOI-406 b (see Sect.~\ref{sec:ground_based}). 
Within \pyorbit, we modelled the light curves using the {\tt batman} package \citep{Kreidberg2015}, and fitting for each planet the $T_0$, $P$, $R_\mathrm{p}/$\rstar, $b$, a common \rhostar, with a Gaussian prior based on our derived stellar parameters, and a jitter term for each dataset.
We assumed a quadratic LD law on the coefficients, $u_1$, $u_2$, for each dataset, employing the LD parametrisation ($q_1$, $q_2$) introduced by \cite{Kipping2013} and adopting a Gaussian prior with initial values obtained through \texttt{PyLDTk}\footnote{\url{https://github.com/hpparvi/ldtk}.} \citep{Husser2013, Parviainen2015} with a custom large $1 \sigma$ uncertainty of $0.1$ for each coefficient. 
In the modelling of the \tess\ light curves, we incorporated a GP regression with a Matérn-3/2 kernel, to address potential correlated noise.
To remove systematic instrumental effects, for each ground-based light curve we fitted together a quadratic term to detrend the data with respect to airmass. 
If not specified otherwise, we assumed uniform, uninformative priors for all the fitted parameters.

\begin{figure}
\centering
  \includegraphics[width=\linewidth]{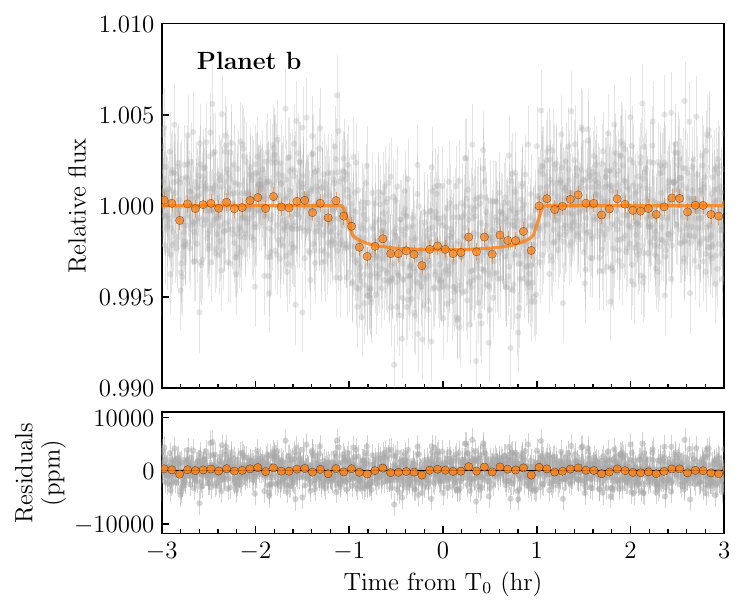}
  \includegraphics[width=\linewidth]{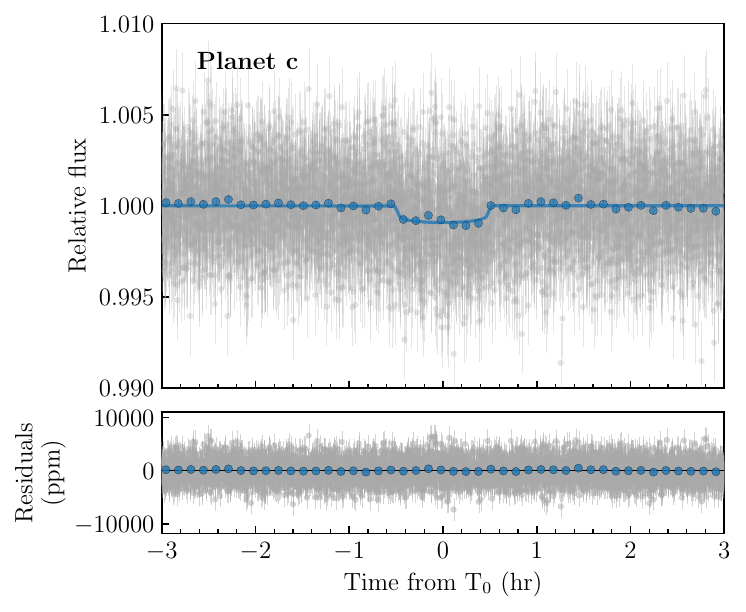}
  \caption{Phase-folded \tess\ light curves of TOI-406 b and c. The best-fitting model is shown as a coloured line, and residuals are plotted in the bottom panels. The coloured dots indicate data points binned over $5$~min. 
  }
    \label{fig:TESS_phase}
\end{figure}

\begin{figure}
\centering
  \includegraphics[width=\linewidth]{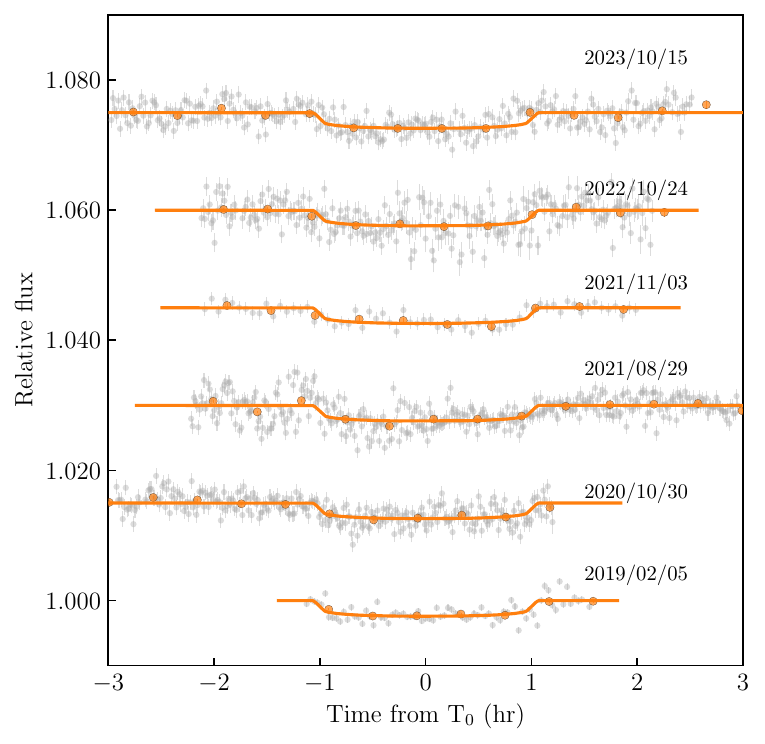}
  \caption{Ground-based individual transit observations of TOI-406 b. Binned data points are shown with coloured dots, and the solid orange line indicates the best-fitting model.
  }
    \label{fig:ground_based}
\end{figure}

\begin{figure*}
\centering
  \includegraphics[width=0.48\linewidth]{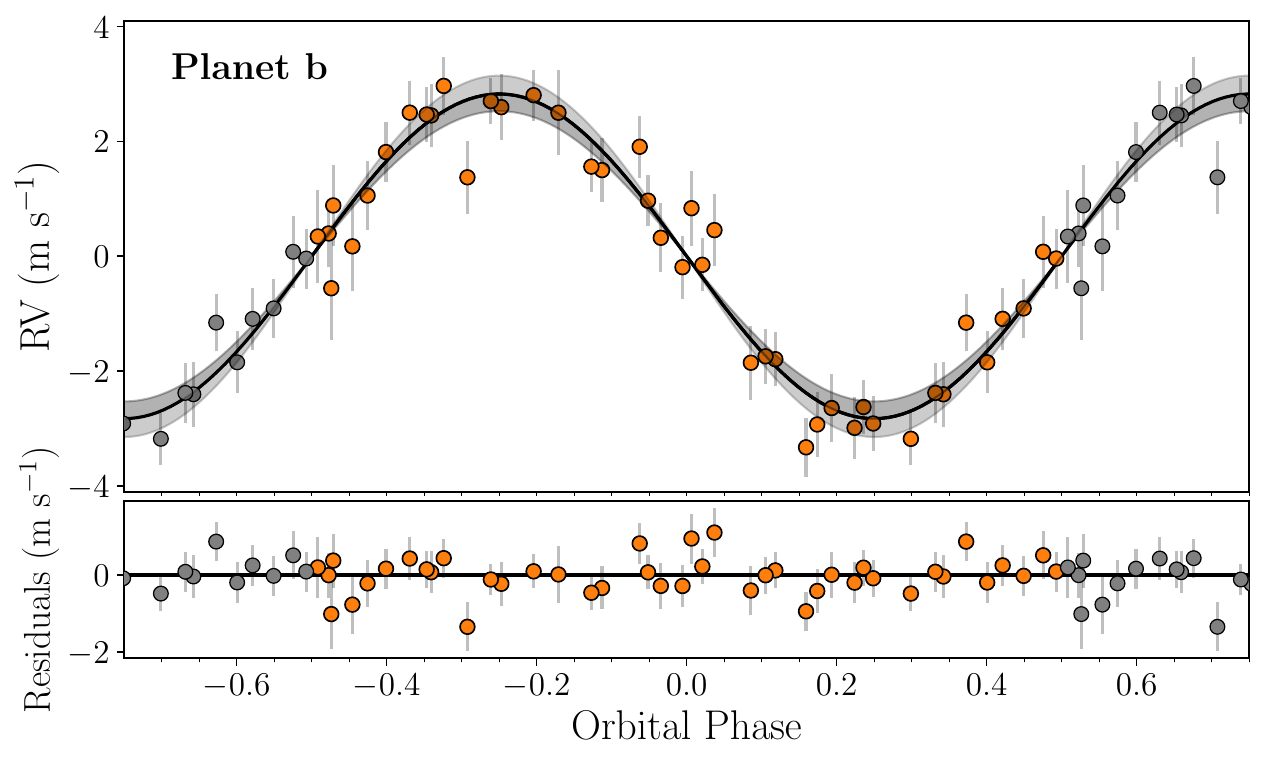}
  \includegraphics[width=0.48\linewidth]{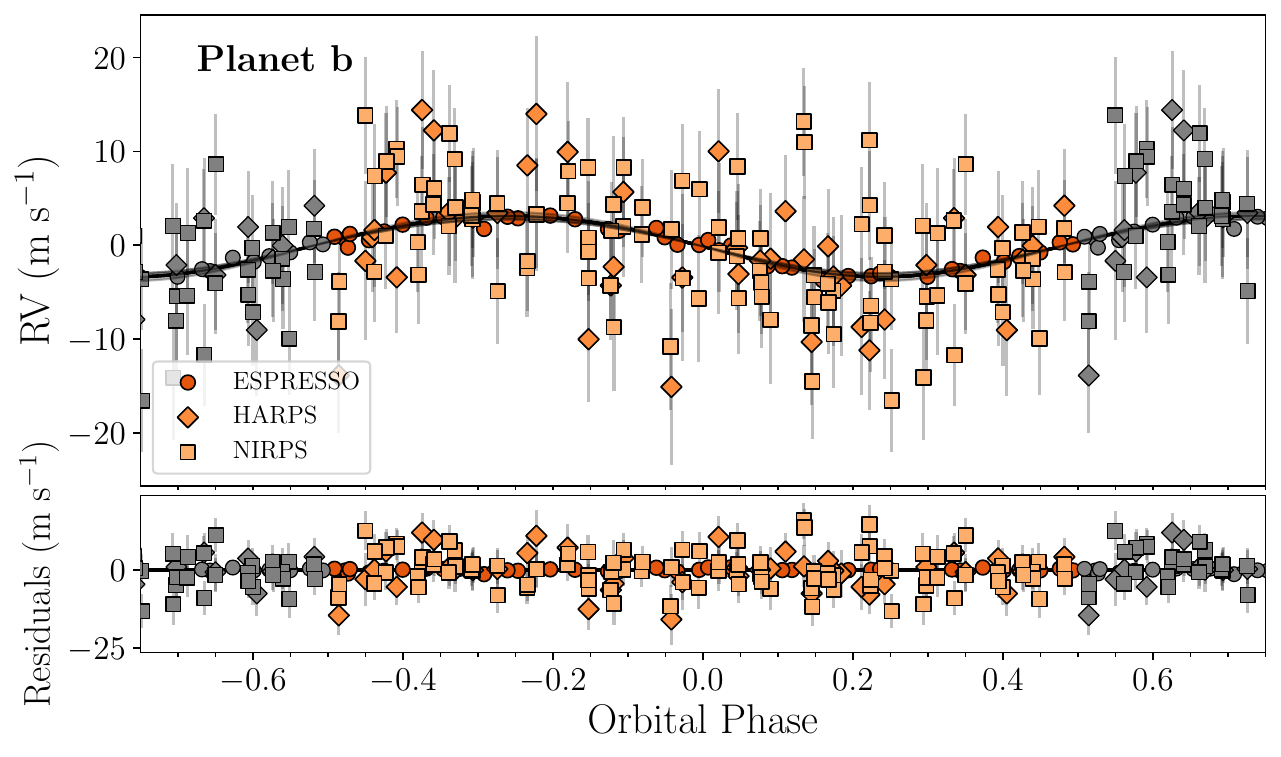}
  \includegraphics[width=0.48\linewidth]{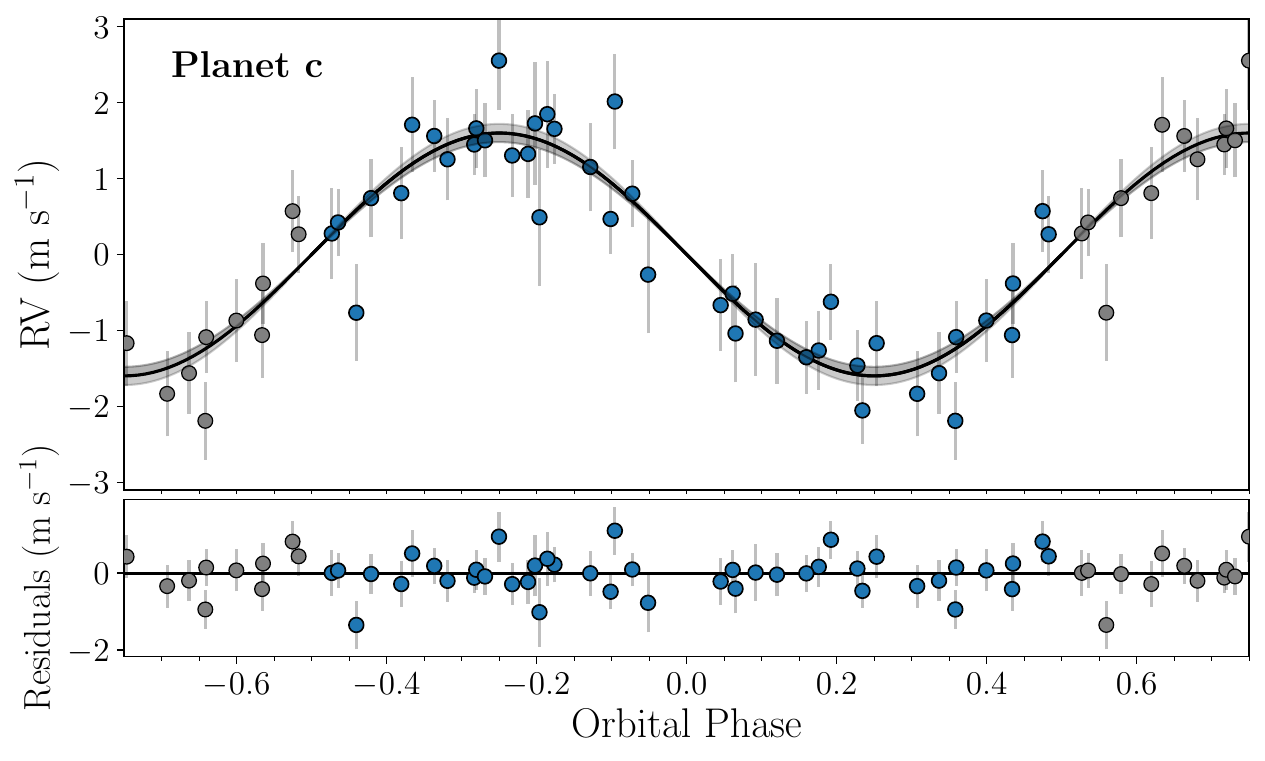}
  \includegraphics[width=0.48\linewidth]{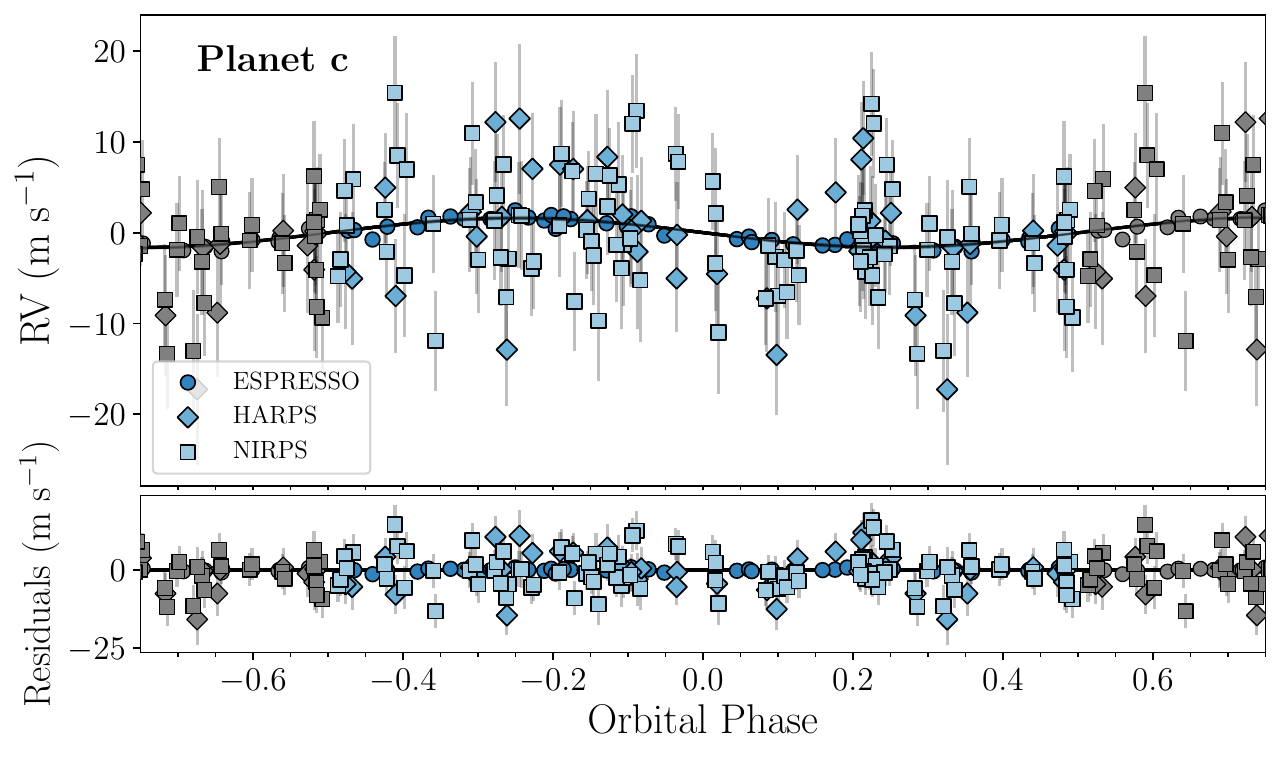}
  \caption{Phase-folded RV fit with residuals from the joint photometric and RV analysis. The left column shows the fit using ESPRESSO data only, while the right one shows the results when including all the available RV data. The shaded area represents the $\pm 1 \sigma$ uncertainties of the model. The reported error bars include the jitter term added in quadrature, and the residuals are shown in the bottom panel. 
  }
    \label{fig:RV_phase}
\end{figure*}

\begin{figure*}
\centering
  \includegraphics[width=0.48\linewidth]{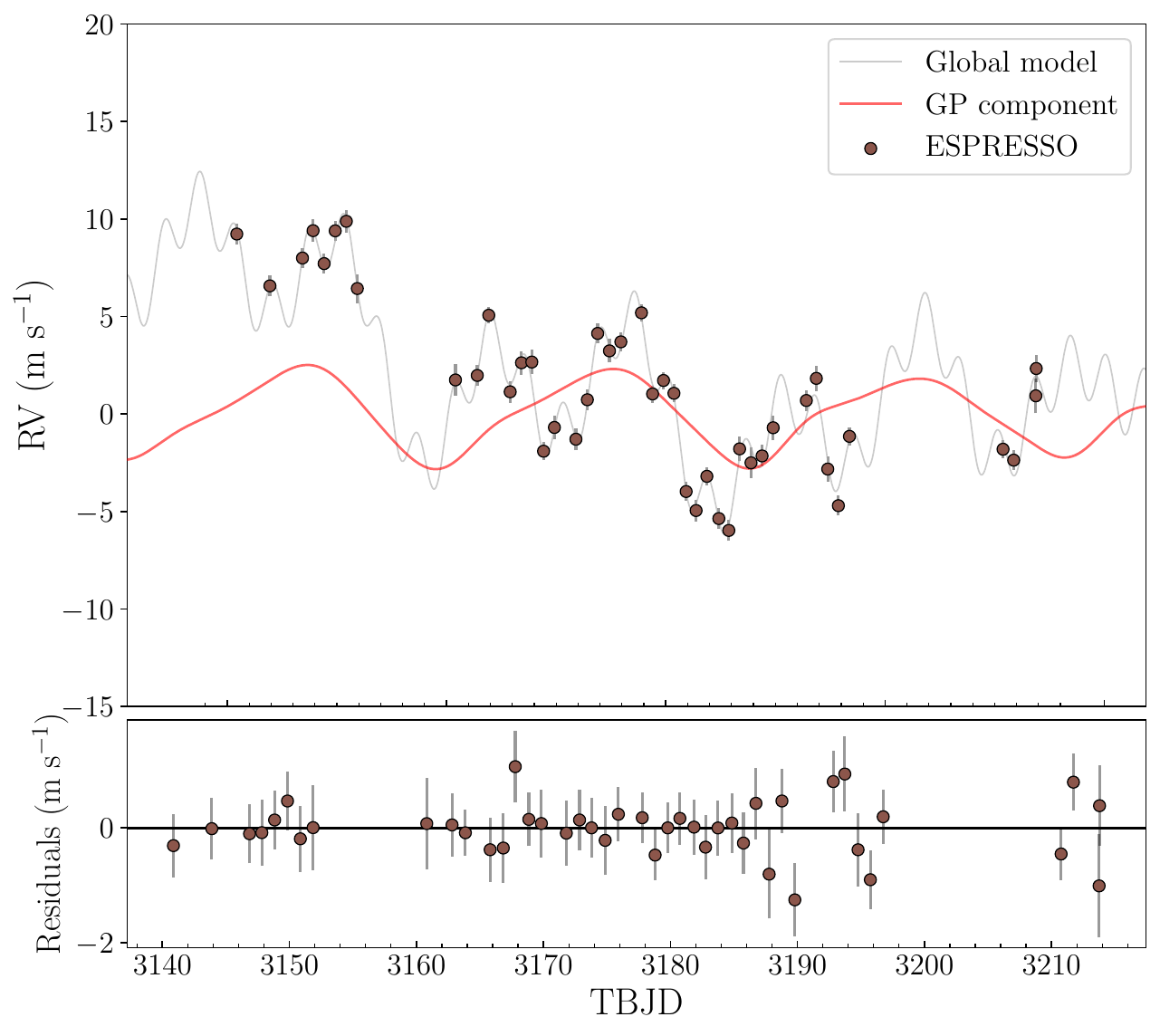}
  \includegraphics[width=0.48\linewidth]{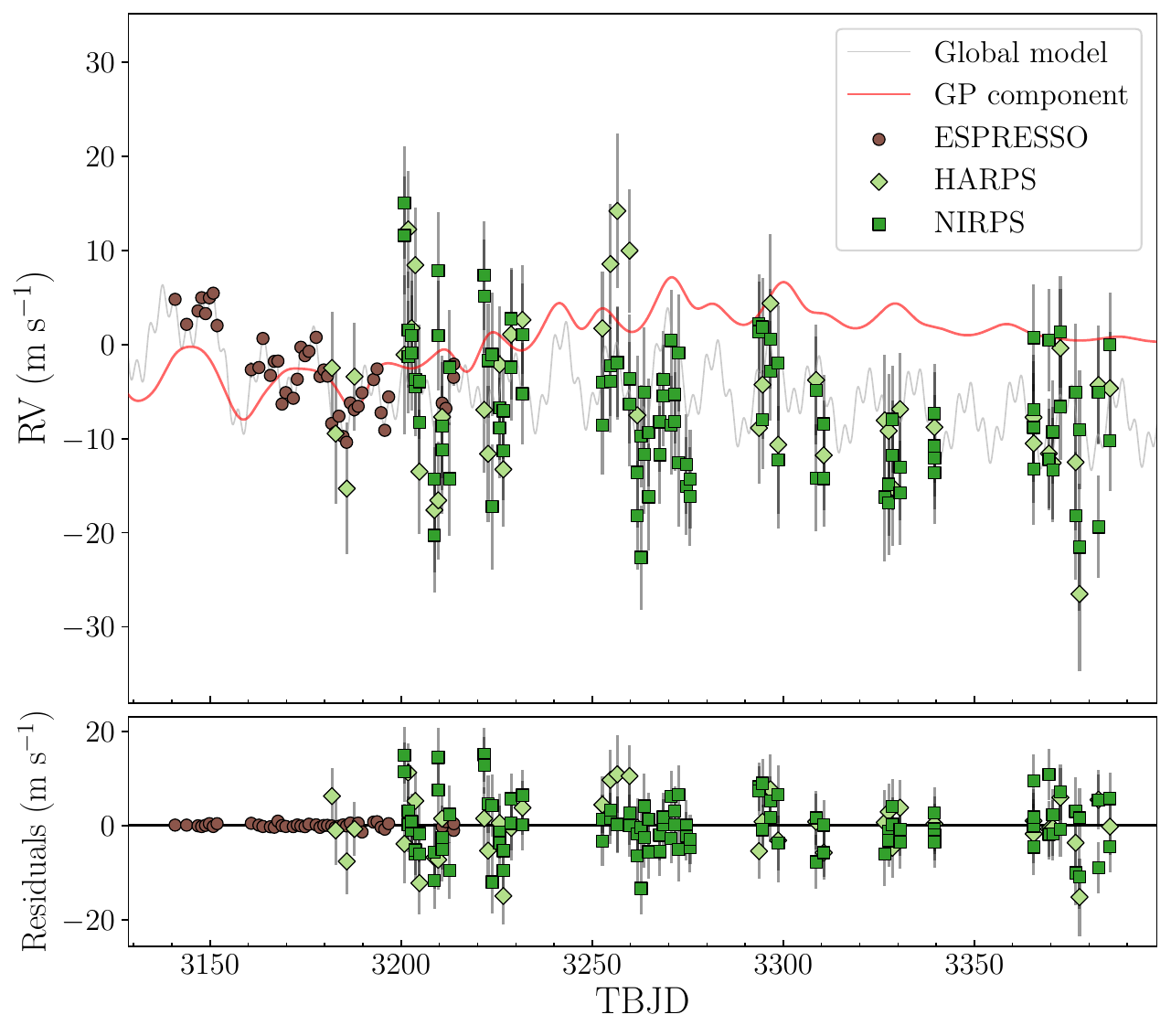}
  \caption{Global RV model from the joint photometric and RV analysis. The left panel shows the fit using ESPRESSO data only, while the right one shows the results when including all the available RV data. Time is expressed in the \tess\ barycentric Julian date (TBJD). The grey line shows the best-fit global model, while the red line highlights the quasi-period GP regression component. The reported error bars include the jitter term added in quadrature, and the residuals are shown in the bottom panel. 
  }
    \label{fig:RV_global}
\end{figure*}

We performed the joint fit both using the ESPRESSO-only dataset, and using ESPRESSO+NIRPS+HARPS. As in the case of the RV fit only, the retrieved semi-amplitudes are virtually the same for TOI-406 c if using the ESPRESSO-only data or all available RVs, but slightly discrepant (but consistent within $1 \sigma$) for TOI-406 b. 
We are unable to discriminate which solution is the most accurate and for completeness we adopted as final values the ones from the joint fit including all the available RVs. 
We list the results of our best-fitting model in Table~\ref{table:joint_parameters}, and we show the transit models, the phase-folded RVs, and the global RV model and in Figs~\ref{fig:TESS_phase}, \ref{fig:ground_based}, \ref{fig:RV_phase}, and \ref{fig:RV_global}, respectively.

\begin{table}[h!]
\caption{Best-fit parameters of the TOI-406 system.}
\label{table:joint_parameters} 
\small
\begin{threeparttable}[t]
\centering
\begin{tabular}{l c c} 
  \hline\hline       
 \multicolumn{3}{c}{Planetary parameters}\\[1ex]
 \hline 
   & TOI-406 b & TOI-406 c \\
  \hline
  $P$ (d) & $13.175682_{-0.000016}^{+0.000017}$ & $3.307441_{-0.000008}^{+0.000010}$ \\
  $T_0$ (TBJD)$^a$  & $1388.570 \pm 0.001$ & $1385.388 \pm 0.002$\\
  $a/$\rstar & $44.6_{-3.1}^{+2.6}$  & $17.6_{-1.3}^{+1.1}$  \\
  $a$ (AU)   & $0.0809\pm 0.0031$ & $0.0322 \pm 0.0013$ \\
  $R_\mathrm{p}/$\rstar & $0.0466 \pm 0.0011$ & $0.0297_{-0.0017}^{+0.0016}$  \\
  \rplanet\ (\rearth) & $2.08_{-0.15}^{+0.16}$  & $1.32 \pm 0.12$ \\
  $b$ & $0.43_{-0.20}^{+0.14}$ & $0.73 \pm 0.06$ \\
  $i$ (deg) & $89.4_{-0.2}^{+0.3}$ & $87.6_{-0.4}^{+0.3}$ \\
  $T_{14}$ (h) & $2.17_{-0.06}^{+0.07}$ & $1.05 \pm 0.05$ \\
  $e$ & $0.056_{-0.034}^{+0.035}$ & $0.032_{-0.023}^{+0.037}$ \\
  $\omega$ (deg)   & $169_{-45}^{+41}$  & $-6_{-118}^{+88}$ \\
  $K$ (\ms)  & $3.25_{-0.38}^{+0.41}$ &$1.63 \pm 0.12$\\
  \mplanet\ (\mearth) & $6.57_{-0.90}^{+1.00}$  &$2.08_{-0.22}^{+0.23}$\\
  \rhoplanet\ (\gcm)  & $4.1 \pm 1.1$ & $4.9 \pm 1.4$\\
  $\rho / \rho_{\oplus , s}^{b}$  & $0.42 \pm 0.11$ & $0.78 \pm 0.22$\\
  $S_{\rm p}$ ($S_{\oplus}$)  & $3.0 \pm 0.5$ & $19 \pm 3$\\
  $T_{\rm eq}^c$ (K)  & $368 \pm 14$ & $584 \pm 22$\\
  $g_{\rm p}^d$ (m s$^{-2}$)  & $15.0 \pm 3.3$ & $11.6 \pm 1.4$ \\[1ex]
 \hline 
 \multicolumn{3}{c}{Common parameters}\\
 \hline
  \rhostar\ (\rhosun) & \multicolumn{2}{c}{$6.7_{-1.4}^{+1.3}$} \\
  $u_{1, \mathrm{TESS}}$ & \multicolumn{2}{c}{$0.15 \pm 0.05$} \\
  $u_{2, \mathrm{TESS}}$ & \multicolumn{2}{c}{$0.45 \pm 0.05$} \\
  $u_{1, \mathrm{i}}$ & \multicolumn{2}{c}{$0.20 \pm 0.08$} \\
  $u_{2, \mathrm{i}}$ & \multicolumn{2}{c}{$0.37 \pm 0.09$} \\
  $u_{1, \mathrm{g}}$ & \multicolumn{2}{c}{$0.24 \pm 0.10$} \\
  $u_{2, \mathrm{g}}$ & \multicolumn{2}{c}{$0.31 \pm 0.10$} \\
  $\sigma_{\rm j, ESPRESSO}^{\rm e}$ (\ms) & \multicolumn{2}{c}{$0.18_{-0.12}^{+0.16}$} \\
  $\sigma_{\rm j, HARPS}^{\rm e}$ (\ms) & \multicolumn{2}{c}{$5.1_{-1.0}^{+1.1}$} \\
  $\sigma_{\rm j, NIRPS}^{\rm e}$ (\ms) & \multicolumn{2}{c}{$3.4 \pm 0.7$} \\
  $\gamma_{\rm ESPRESSO}^{\rm f}$ (\ms) & \multicolumn{2}{c}{$-1.8_{-5.0}^{+5.8}$} \\
  $\gamma_{\rm HARPS}^{\rm f}$ (\ms) & \multicolumn{2}{c}{$14737 \pm 3$} \\
  $\gamma_{\rm NIRPS}^{\rm f}$ (\ms) & \multicolumn{2}{c}{$ 14977 \pm 1$} \\
  $c_{\rm 0}$ (\ms) & \multicolumn{2}{c}{$3275.50$ (fixed)} \\
  $c_{\rm 1}$ (\ms~d$^{-1}$) & \multicolumn{2}{c}{$-0.041 \pm 0.14$} \\
  $c_{\rm 2}$ (\ms~d$^{-2}$) & \multicolumn{2}{c}{$0.0003 \pm 0.0002$} \\[1ex]
   \hline 
 \multicolumn{3}{c}{GP hyper-parameters}\\
 \hline
  \prot\ (d) & \multicolumn{2}{c}{$29.2 \pm 0.5$} \\
  $P_{\rm dec}$ (\ms) & \multicolumn{2}{c}{$62_{-3}^{+6}$} \\
  $w$ (\ms) & \multicolumn{2}{c}{$0.73_{-0.14}^{+0.16}$} \\
  $A^{\rm g}_{\rm ESPRESSO}$ (\ms) & \multicolumn{2}{c}{$6.0_{-1.4}^{+2.0}$} \\
  $A^{\rm g}_{\rm FWHM}$ (\ms) & \multicolumn{2}{c}{$5.8_{-1.8}^{+2.6}$} \\
  $A^{\rm g}_{\rm HARPS}$ (\ms) & \multicolumn{2}{c}{$4.5_{-2.3}^{+2.9}$} \\
  $A^{\rm g}_{\rm NIRPS}$ (\ms) & \multicolumn{2}{c}{$1.4_{-1.0}^{+2.0}$} \\[1ex]
 \hline
\end{tabular}
\tablefoot{
    Parameters were determined from the joint fit (Sect.~\ref{sec:joint_fit}) including all the available RV datasets. \\
    \tablefoottext{a}{\tess\ Barycentric Julian date (BJD$-2457000$).} 
\tablefoottext{b}{Scaled Earth’s bulk density (see Fig.~\ref{fig:density}).} \tablefoottext{c}{$T_{\rm eq} = T_\star \, \biggl(\dfrac{R_\star}{2a}\biggr)^{1/2} \, [f(1-A_{\rm B})]^{1/4}$, assuming $f=1$ and a null Bond albedo ($A_{\rm B} = 0$).}  \tablefoottext{d}{Planetary surface gravity.} 
\tablefoottext{e}{RV jitter term.}
\tablefoottext{f}{RV offset term.} 
\tablefoottext{g}{GP amplitude.}}
\end{threeparttable}
\end{table}

We confirm the planetary nature of the two candidates in the system, with orbital periods of $P_{\rm c} = 3.307$ days (TOI-406 c) and $P_{\rm b} = 13.176$ days (TOI-406 b).
An important outcome of our RV frequency search in Sect.~\ref{sec:star_activity} was the absence of a signal at $\sim 6.6$ days; that is, the transiting candidate TOI-406.02 reported from the SPOC pipeline. 
However, we identified a signal with exactly half the period in the RV periodogram. Such a periodicity was also claimed as a possible alternative period for this candidate in the \tess\ EXOFOP web page.
Our joint photometric and spectroscopic analysis confirms the detection of a planetary signal at $3.307$ days, which we identify with TOI-406 c, and for which we inferred a precise mass ($M_{\rm c} = 2.08_{-0.22}^{+0.23}$~\mearth, from the semi-amplitude $K_{\rm c} = 1.63 \pm 0.12$~\ms) and radius ($R_{\rm c} = 1.32 \pm 0.12$~\rearth), implying an average density of $\rho_{\rm c} = 4.9 \pm 1.4$~\gcm. 

The outer planet, TOI-406 b, has a radius of $R_{\rm b} = 2.08_{-0.15}^{+0.16}$~\rearth, which, combined with the inferred mass of $M_{\rm b} = 6.57_{-0.90}^{+1.00}$~\mearth (from $K_{\rm b} = 3.25_{-0.38}^{+0.41}$~\ms), implies an average density of $\rho_{\rm b} = 4.1 \pm 1.0$~\gcm. With our inferred parameters, planets b and c have equilibrium temperatures of \teq~$=368 \pm 14$~K, and \teq~$=584 \pm 22$~K, respectively. 

From our analysis, we identify a possible long-term trend. However, such a long-term signal could be due both to a planetary companion or to stellar magnetic activity, and additional data of ESPRESSO-like precision covering a longer baseline are needed to understand its nature. This possible additional signal could also be the source of the slight discrepancy in the mass of TOI-406 b when using ESPRESSO-only data or all the available RVs.

The quasi-periodic GP modelling of the RVs and dLW results in the detection of a clear periodicity at \hbox{$P_{\rm rot} = 29.2 \pm 0.5$ days} (see Fig.~\ref{fig:RV_global}). This signal matches the period identified also in the ASAS-SN light curves and (marginally) in the \tess\ photometry (Sect.~\ref{sec:star_activity}). We attribute this signal to the stellar rotational period, and we adopted the value inferred from our joint fit as its final value.

\begin{figure*}
\centering
  \includegraphics[width=0.48\linewidth]{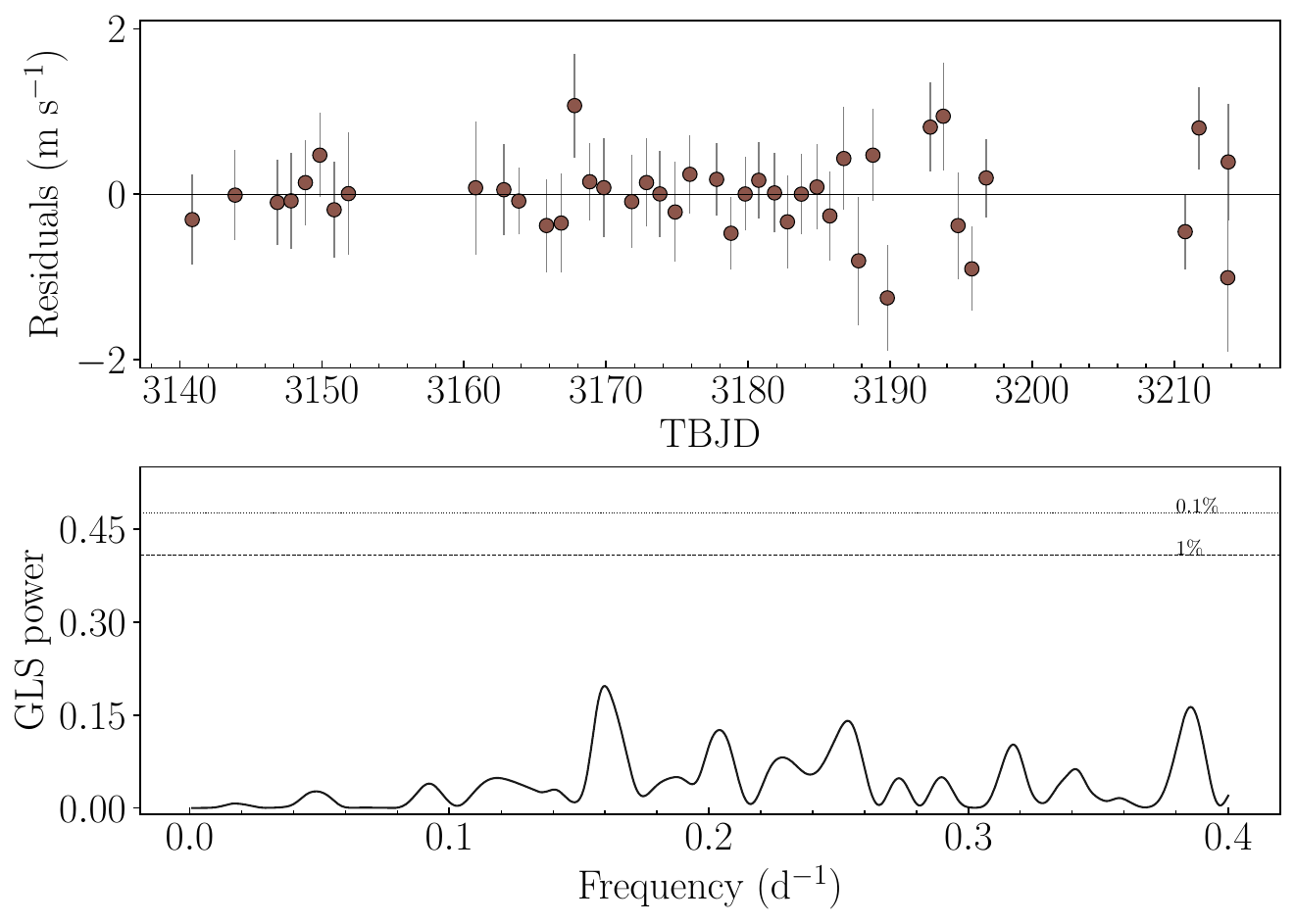}
  \includegraphics[width=0.48\linewidth]{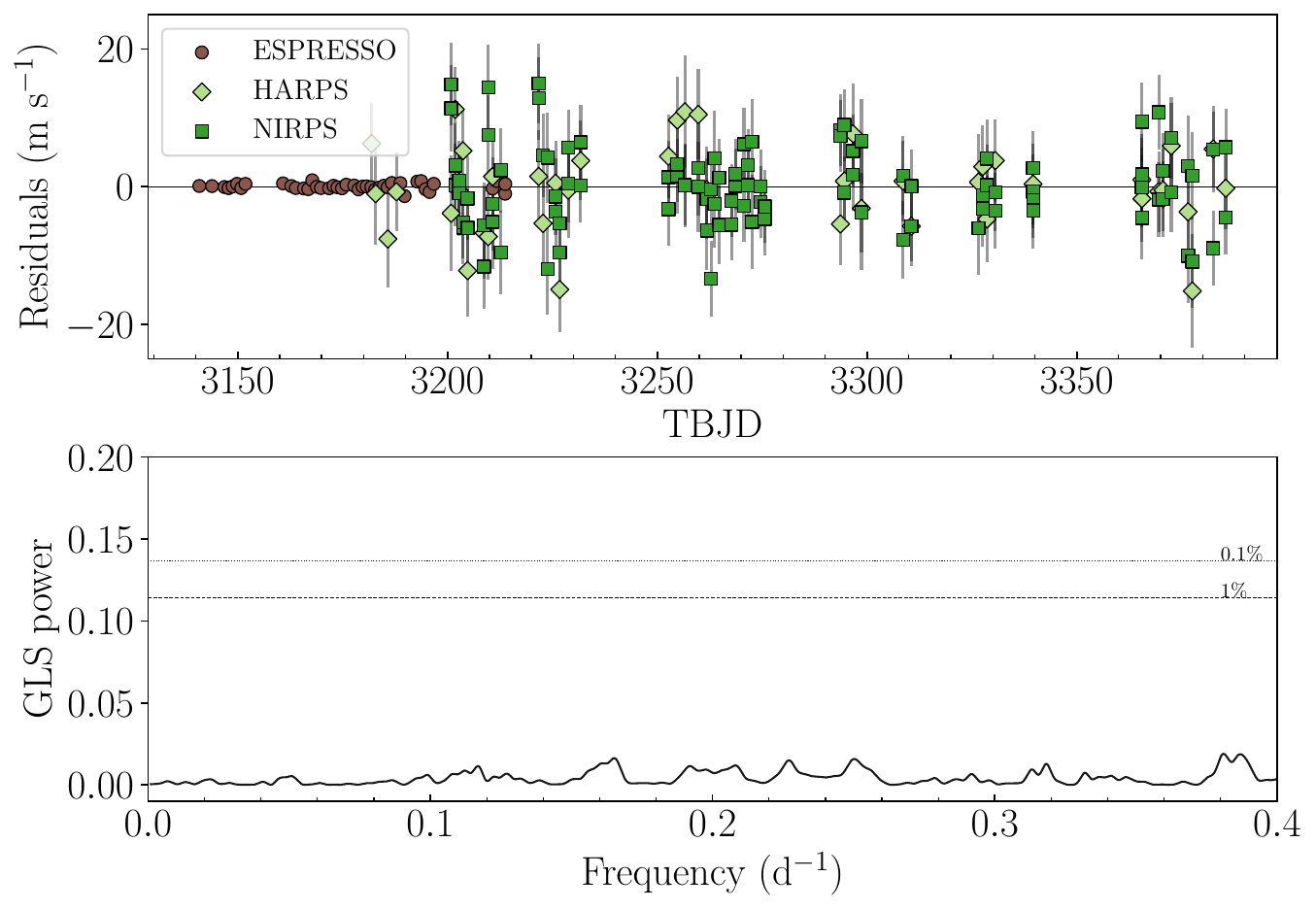}
  \caption{Time series and GLS periodogram of the RV residuals after the joint photometric and spectroscopic fit ( Sect.~\ref{sec:global_analysis}). The left panel shows the results using ESPRESSO data only, while the right one includes all the available RV data.
  In the bottom panel, the dashed and dotted horizontal lines show the $1$ and $0.1$ per cent FAP level, respectively. No significant peaks are present in either periodogram. 
  }
    \label{fig:periodogram_residuals}
\end{figure*}

After the joint analysis, we could not identify any significant additional signals in the periodogram of the RV residuals in either the ESPRESSO-only case or in the case including all the RVs (Fig.~\ref{fig:periodogram_residuals}).
Moreover, we could not identify significant signals in the \tess\ light curve residuals by running a transit search using the transit least squares\footnote{\url{https://github.com/hippke/tls}.} algorithm \citep{hippke2019_TLS}. 
We therefore conclude that with our current dataset we cannot identify additional planetary signals in the system.

\section{Discussion}\label{sec:discussion}

\subsection{TOI-406 planetary properties}\label{sec:planet_properties}

\begin{figure}
\centering
\includegraphics[width=\linewidth]{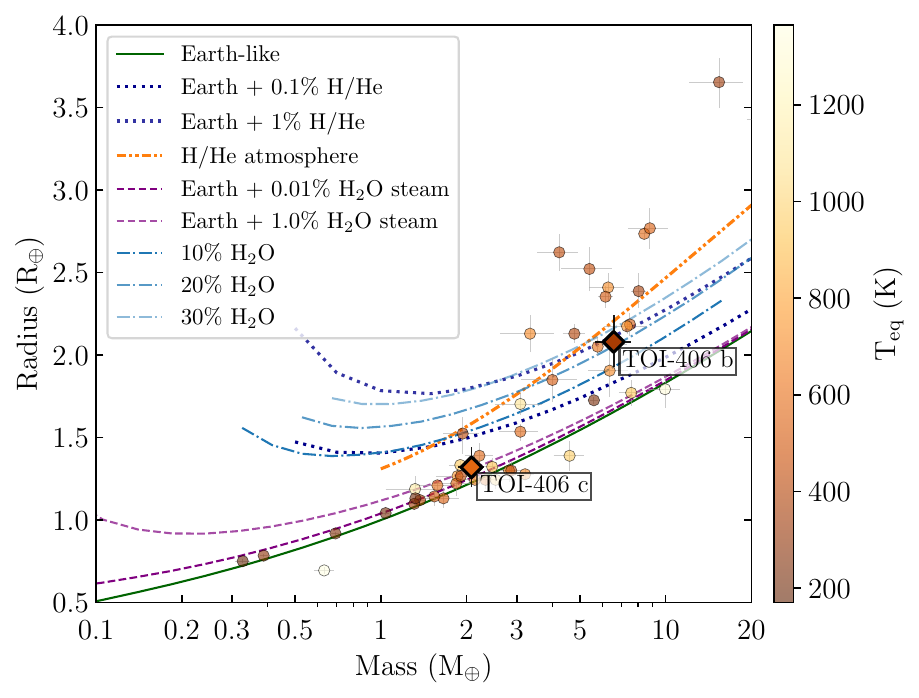}
  \caption{Mass-radius diagram for small transiting planets around M dwarfs, colour-coded according to equilibrium temperature \teq. The TOI-406 planets are labelled, and highlighted with coloured diamonds. The sample was taken from the PlanetS catalogue \citep{parc2024} on July 30, 2024, reporting only planets with mass and radius determinations better than 25\% and 8\%, respectively. The solid green line indicates the theoretical composition models of \citep{Zeng2019} for an Earth-like composition (mass fractions of $32.5$\% iron and $67.5$\% silicates). The dotted dark blue lines correspond to \cite{Zeng2019} theoretical tracks of Earth-like rocky cores with H/He atmospheres by different percentages in mass at $300$~K. The dotted orange line shows \cite{Rogers23} mass-radius distribution derived from photoevaporation models. Dotted purple lines show \cite{turbet2020} models for steam atmospheres, and the dash-dotted blue lines show \cite{aguichine2021} compositional tracks at $400$~K with a varying water mass fraction. 
  }
    \label{fig:MR_diagram}
\end{figure}

One of the goals of the \thirstee\ project is to check whether planets around M dwarfs follow the mass-radius-density trends identified in \cite{luquepalle2022}. Figure~\ref{fig:MR_diagram} shows the position of TOI-406 b and c in the mass-radius diagram of well-characterised small planets around M dwarfs (most recently updated by \citealt{parc2024}). According to the theoretical tracks of \cite{Zeng2019}, TOI-406 c is consistent with an Earth-like composition, even though its density is slightly lower than a pure rocky (Earth-like) planet, making it also consistent with the presence of a small amount of water mass fraction in the form of a supercritical steam atmosphere. 
In fact, rocky planets with a substantial water envelope and more irradiated than the runaway greenhouse limit can develop a supercritical steam atmosphere \citep{turbet2020}, which implies a larger planetary radius, and therefore a slightly lower bulk density. 
The lower density of TOI-406 c with respect to a pure rocky composition could also hint at the presence of a secondary atmosphere, like for example that of 55 Cnc e \citep{Ehrenreich2012, tsiaras2016, keles2022, Hu2024}, TOI-561 b \citep{lacedelli2022, Brinkman2022, patel2023},  and similar planets that make up a significant fraction of the population of rocky planets currently being observed by {\it JWST}.
On the other hand, TOI-406 b lies in a mass-radius region of high degeneracy in terms of interior composition (Fig.~\ref{fig:MR_diagram}). The planet could host a significant amount of water, with a water mass fraction of up to $30$\% according to \cite{aguichine2021} models of irradiated ocean planets, but it could also be consistent with a gas dwarf scenario; that is, a dry Earth-like core surrounded by a substantial H/He envelope, as is inferred from atmospheric mass loss theories \citep{OwenWu13, LopezFortney13, gupta2019, Wu_2019, Rogers23}.

These results are in line with the trend identified by \cite{luquepalle2022} for small planets around M dwarfs, where planets in the $1.4-2$~\rearth\ radius regime tend to fall into two distinct density regions, implying a different internal composition.

\begin{figure}
\centering
  \includegraphics[width=\linewidth]{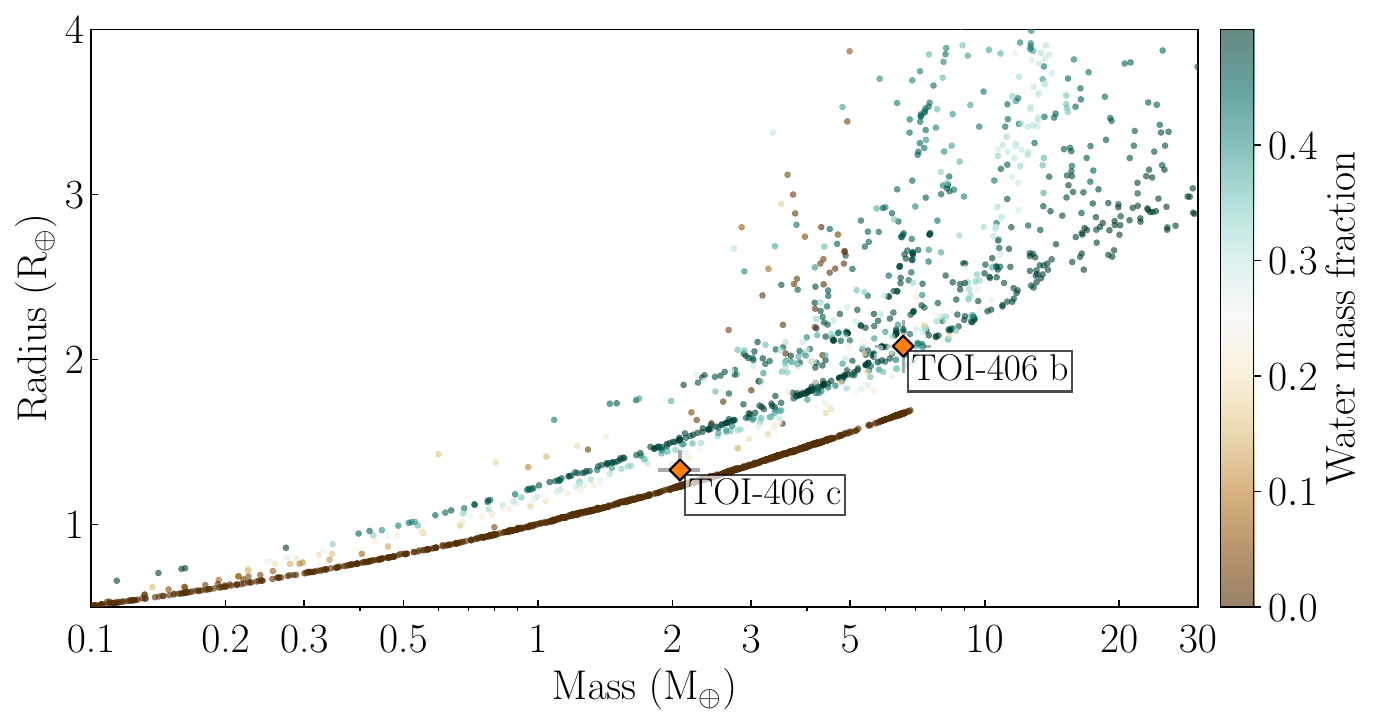}
  \caption{Mass–radius diagram of synthetic planets with a period of less than $100$~d at $2$~Gyr around a $0.4$~\msun\ star from the simulation of \cite{venturini2024}. Planets are colour-coded by water mass fraction. Most of the simulated planets belong to two distinct populations, of either purely rocky cores or planets containing about $50$\% water, plus a few of them have an extended envelope. The TOI-406 planets are labelled and highlighted with orange diamonds. TOI-406 c is consistent with the synthetic population of planets having rocky cores, while TOI-406 b seems to have a substantial water mass fraction. 
  }\label{fig:MR_synthetic}
\end{figure}

\begin{figure*}
\centering
  \includegraphics[width=0.48\linewidth]{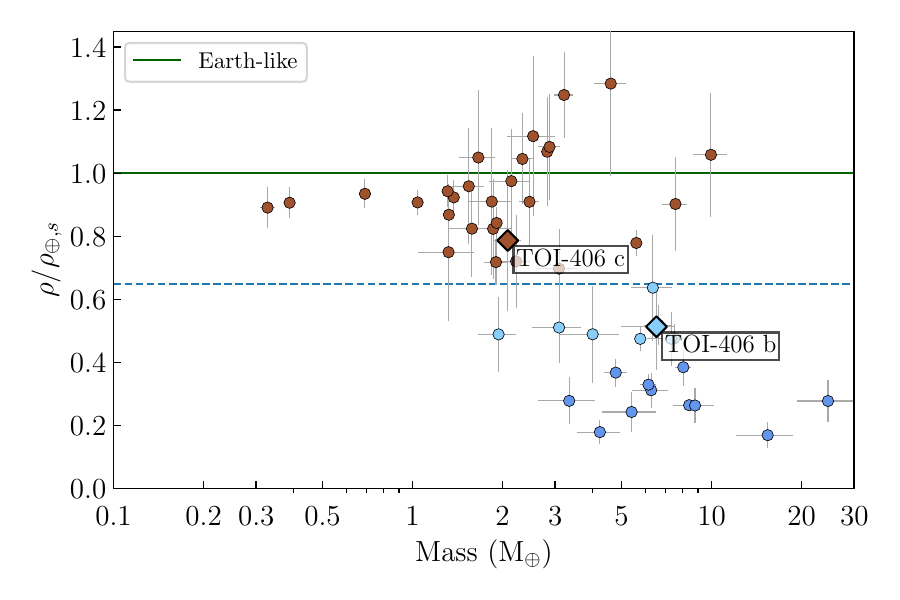}
  \includegraphics[width=0.48\linewidth]{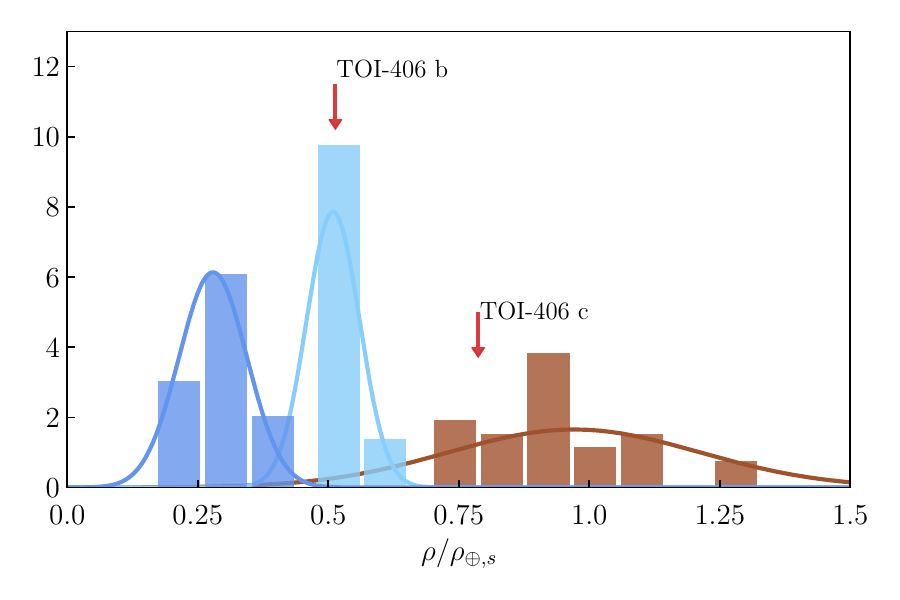}
  \caption{Density properties of the well-characterised small-planet population around M dwarfs. Left: Mass-density diagram for small transiting planets around M dwarfs, adapted from \cite{luquepalle2022}. The TOI-406 planets are labelled, and highlighted with coloured diamonds. The sample is the same as in Fig.~\ref{fig:MR_diagram}. Densities are normalised by \cite{Zeng2019} theoretical model of an Earth-like composition (scaled Earth bulk density, $\rho_{\oplus , s}$, assuming mass fractions of $32.5$~\% iron and $67.5$\% silicates). Planets are colour-coded according to their bulk densities, following \citep{luquepalle2022}: Earth-like densities (brown), possible water worlds (cyan), and sub-Neptunes with H-He envelops (dark blue). The dotted blue line at $0.65$ $\rho_{\oplus , s}$ represents the compositional gap suggested by \cite{luquepalle2022}.
  Right: Normalised histogram of the sample of planets with mass and radius determinations better than 25\% and 8\%, respectively. Density is normalised to the Earth-like model. A Gaussian model fitted to the distribution of each planet population is shown with a solid coloured line. The red arrows show the positions of the TOI-406 planets in the density distributions. 
  }
    \label{fig:density}
\end{figure*}

This outcome is also consistent with recent synthetic planet population simulations.
In the core accretion scenario, planets generally develop into two distinct populations, one composed of a pure rocky core, and the other an ice-rock mixture containing $\sim 50$\% ice. This ice fraction is characteristic of planetary components located beyond the water ice line \citep{Thiabaud2014, marboeuf2014}.
Icy and rocky pebbles have different physical properties, implying that icy cores are born bigger with respect to rocky ones, and this naturally explains the radius valley as an outcome of planet formation and evolution \cite{Venturini20, venturini2024, Burn24}, with most of the mini-Neptunes forming beyond the water ice line, and therefore being water-rich \citep{alibert2013, raymond2018, bitsch2018, brugger2020, izidoro2021, Venturini20}. Figure~\ref{fig:MR_synthetic} shows the position of the two TOI-406 planets with respect to the synthetic population of M dwarfs planets presented in \cite{venturini2024}\footnote{Data are available on \url{https://zenodo.org/records/10719523}.}.
The inner planet, TOI-406 c, is consistent with the population of planets having pure rocky cores, even though a small amount of volatiles could be present, suggesting a formation inside the ice line. 
On the other hand, the more external planet, TOI-406 b, is consistent with the population of planets having a substantial water mass fraction, suggesting a formation beyond the ice line and further inward migration after the disc dispersal. 
Moreover, recent studies show that more complex mechanisms, including the solubility of water in magma, magma ocean dynamics, envelope mixing, interactions between atmosphere and interiors, and redox reactions (i.e. \citealt{DornLichtenberg21, Vazan22, Lichetnberg24, Luo24, Rogers2024}) could play an important role in the determination of the composition of small planets.
However, independently of the model assumption and interpretation of the planetary composition, Fig.~\ref{fig:density} shows that the TOI-406 planets fall into two distinct density populations, as is identified by \cite{luquepalle2022} and supporting the hypothesis of an observational density gap for small exoplanets orbiting M dwarfs. 
This hypothesis is further strengthened by the recent work of \cite{Schulze2024}, which proves the statistical significance of such a density gap using mixture models in a hierarchical framework.

\subsection{Dynamical analysis}\label{sec:dyn_analysis}
We tested the dynamical stability of out best-fitting solution with the the \texttt{REBOUND} package \cite{rein2012, rein2015}, computing the orbits of the system for $100$ Kyr with the \texttt{whfast} integrator (with a fixed time-step of $0.1$~d).
We used the best-fitting parameters in Table~\ref{table:joint_parameters})
to draw ten initial sets of parameters based on a Gaussian distribution, and ran ten distinct simulations.
During the integration, we used the mean exponential growth factor of nearby orbits (MEGNO or Y) indicator \citep{CincottaSimo2000, Rein2016MNRAS.459.2275R} to check the dynamical stability of the solution.
We obtained a MEGNO value of $2$ for all ten runs, proving the stability of this family of solutions.

\begin{figure}[!h]
\centering
\includegraphics[width=0.925\linewidth]{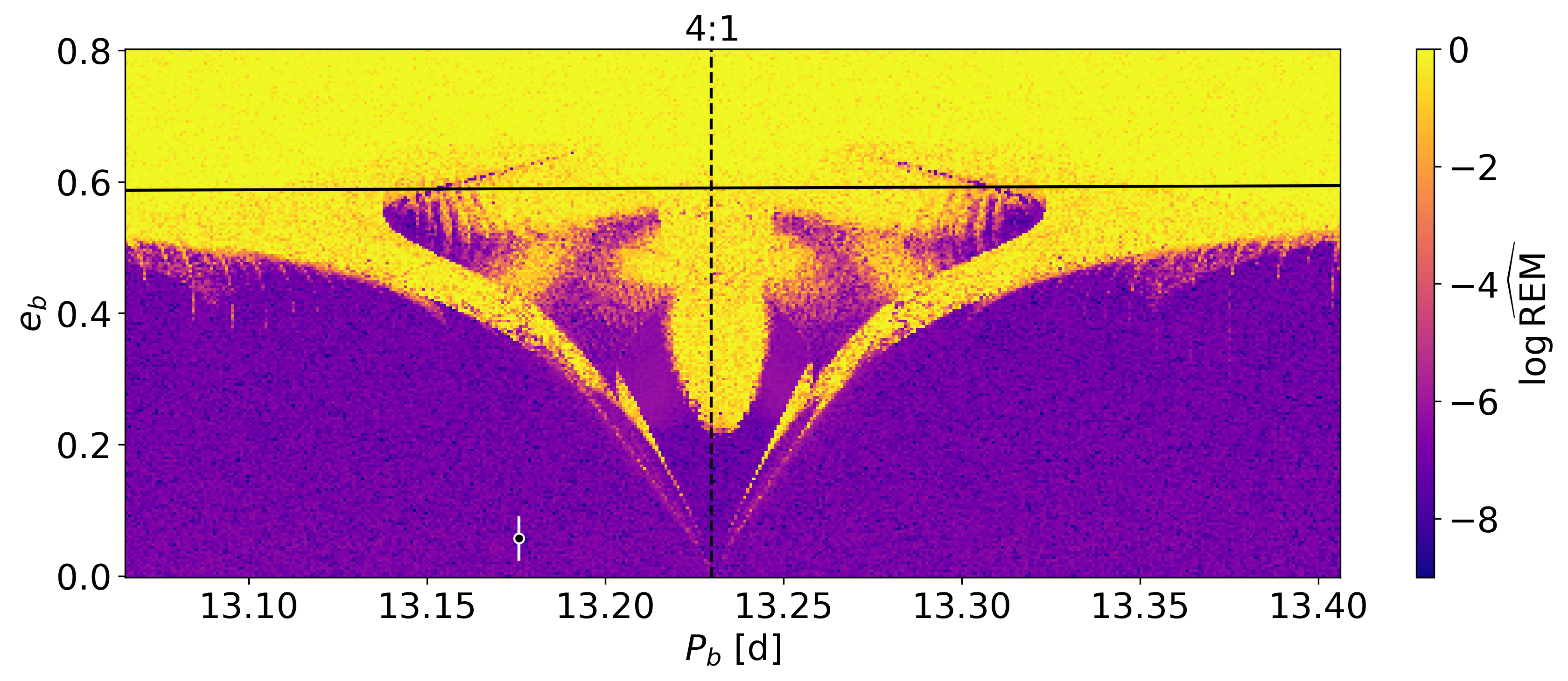}
    \caption{Dynamical map for the solution presented in Table~\ref{table:joint_parameters} for a wide range of orbital periods and eccentricities of the outer planet b. Small values of the fast indicator $\log \mathrm{\widehat{REM}}$ characterise regular (long-term stable) solutions, while chaotic solutions are indicated by $\log \mathrm{\widehat{REM}}$ approaching 0. The black line represents the so-called collision curve of orbits, defined by the condition: $a_b(1 + e_b) = a_c(1 - e_c)$. The vertical dashed black line shows the centre of 4:1 mean-motion resonance. The filled black circle with a white rim indicates the position of the solution with uncertainties for the outer planet b. The resolution for the plot is 401 $\times$ 201 points.
    \label{fig:dynamical_map_REM}}
\end{figure}

To ascertain the dynamical stability of our best-fit solution, we also employed the reversibility error method \citep[REM;][]{Panichi_2017}, which has been demonstrated to be a close analogue of the maximum Lyapunov exponent. In the analysis of multi-body systems, it relies on numerical integration schemes that are time-reversible, in particular symplectic algorithms. This method is based on calculating the difference between the initial state vector and the final state vector, which is obtained by integrating the system of equations at a specific time and returning to the initial epoch. The difference thus defined will depend on the dynamic nature of the system. $\mathrm{\widehat{REM}}=1$ or $\log\mathrm{\widehat{REM}}=0$ means that the difference reaches the size of the orbit. The dynamical stability of the solution with REM was tested using the \verb|whfast| integrator with the 17th order corrector (with a fixed time step of 0.18 d) implemented within the \verb|REBOUND| package for 100 000 orbital periods of the farthest planet. As is illustrated in Fig.~\ref{fig:dynamical_map_REM}, we obtained $\log\mathrm{\widehat{REM}}$ $\leq$ -6, indicating that the solution is stable. Its position in the phase space demonstrates a non-resonant character. The visible resonant structure is at a safe distance, well above the uncertainty value of the orbital period of the outer planet. Like in the case of the MEGNO integrator, we also checked the stability for the family of solutions randomly selected from a Gaussian distribution based on our best-fitting parameters (Table~\ref{table:joint_parameters}). 
We ran 200 simulations for 100 Kyr, and we obtained $\log\mathrm{\widehat{REM}}$ $\leq$ -6 for all of them, confirming that the family of solutions is stable.

\subsection{Prospects for atmospheric characterisation}\label{sec:atmosphere}
\begin{figure}
\centering
  \includegraphics[width=\linewidth]{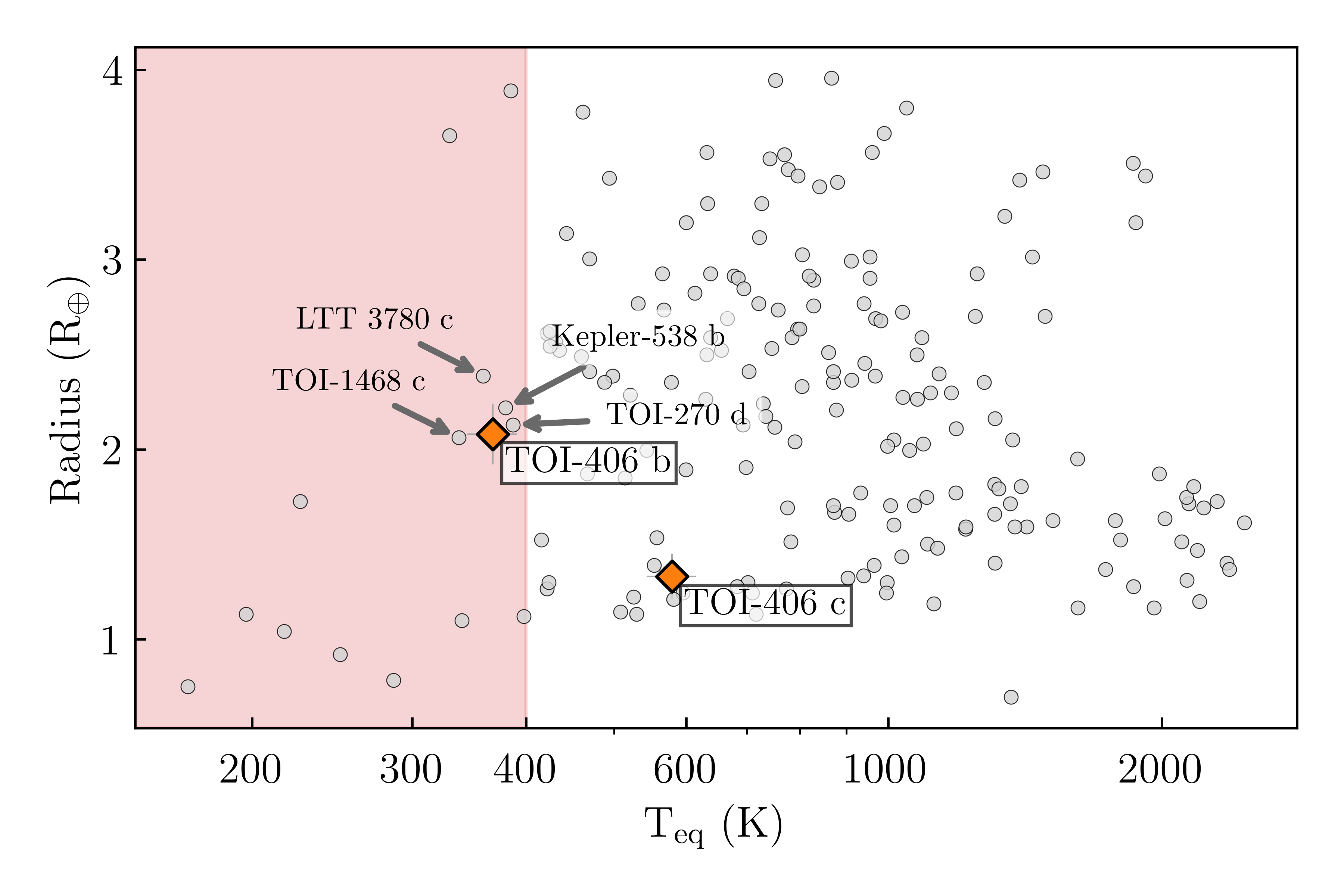}
  \caption{Temperature-radius diagram for small well-characterised exoplanets. The sample was taken from the PlanetS catalogue \citep{parc2024}, selecting only planets with \rplanet~$< 4$~\rearth. TOI-406 b and c are labelled and highlighted with orange diamonds. The low-temperature region (\teq~$<400$~K) is highlighted in red. For comparison, planets with properties similar to TOI-406 b are labelled. }\label{fig:temperature_radius}
\end{figure}

Given the possible presence of a gas layer in both planets and the need for additional constraints to break the degeneracy in planetary composition, we evaluated the feasibility of their atmospheric characterisation with {\it JWST}. To do so, we calculated the TSM and the emission spectroscopy metric (ESM) defined in \citealt{kempton2018}). \\
Using the stellar and planetary parameters and Tables~\ref{table:star_params} and \ref{table:joint_parameters}, TOI-406 c has TSM~$=8.5$ and ESM~$=2.6$, suggesting that its atmospheric characterisation would be a challenging task for {\it JWST}\footnote{For planets with \rplanet~$<1.5$~\rearth, \cite{kempton2018} proposes a threshold of TSM~$>10$ and ESM~$>7.5$.}.
For TOI-406 b, which was the original target we included in our \thirstee\ sample, we obtained TSM~$=43$ and ESM~$=1.0$. 
Considering its radius and temperature, according to the classification of \cite{hord2024}, TOI-406 b lies in the top five best-in-class sample for {\it JWST} transmission spectroscopy follow-up, making it an interesting target with which to investigate the regime of small, low-temperature planets. 
In fact, with \teq~$=368 \pm 14$~K, TOI-406 b adds up to the still-little population of cold planets (\teq~$<400$~K) with small radii (\rplanet~$< 4$~\rearth; Fig.~\ref{fig:temperature_radius}). 
With an incident flux of only $3 \pm 0.7$~$S_{\oplus}$ (P$_{\rm c} = 13.17$~d), the planet lies close to the inner edge of TOI-406’s empirical habitable
zone ($22$~d~$<$~P~$< 92$~ d), as it is defined by \citep{kaltenegger2019}.
As Fig.~\ref{fig:temperature_radius} shows, the planet has very similar properties to TOI-270 d \citep[\rplanet$\sim 2.13$~\rearth, \teq$\sim 350$~K, TSM~$ \sim 124$;][]{Gunther2019}, for which atmospheric features have recently been investigated and detected with {\it HST} and {\it JWST} \citep{Mikal_evans23, Holmberg24, bennecke2024}, demonstrating the potential of this category of planets for atmospheric characterisation, and placing TOI-406 b among the primary targets for future atmospheric follow-up.

\section{Conclusions}\label{sec:conclusions}

In this paper, we present the \thirstee\ project, an observational survey that aims to shed light on the composition of the sub-Neptune population by increasing the number of accurately measured bulk density measurements across all stellar types and planet temperatures. We also aim to investigate their atmospheres to further probe their composition and perform statistical demographic studies of the sub-Neptune population. 

We report the first results of the program, presenting the characterisation of a two-planet system orbiting the M dwarf TOI-406. 
Thanks to ESPRESSO and NIRPS/HARPS RV measurements, we could unveil the orbital architecture of the system, solving the ambiguity surrounding the orbital period of the inner candidate identified by \tess. 
According to our analysis, TOI-406 hosts a $3.3$~d planet (R$_{\rm c} = 1.32 \pm 0.12$~\rearth, M$_{\rm c} = 2.08_{-0.22}^{+0.23}$~\mearth, $\rho_{\rm c} = 4.9 \pm 1.4$~\gcm) consistent with an Earth-like composition, and a temperate (\teq~$=368$~K) $13.2$~d planet (R$_{\rm b} = 2.08_{-0.15}^{+0.16}$~\rearth, M$_{\rm b} =6.57_{-0.90}^{+1.00}$, $\rho_{\rm b} = 4.1 \pm 1.1$~\gcm) compatible with multiple internal composition models, including a water-rich planet scenario. 
The bulk density of the two planets suggests that they fall into two distinct density populations, supporting the evidence for a density gap among the M dwarf population.

One of our important highlights is that TOI-406 b stands out as a particularly interesting target for atmospheric characterisation among low-temperature planets, where cloud formation processes are predicted to be less ubiquitous \citep{yu2021, brande2024}. The planet has very similar properties to TOI-270 d, for which atmospheric features have recently been detected with {\it JWST}, suggesting that it has high potential for atmospheric follow-up. 
Atmospheric characterisation will be essential to investigate the presence of the volatile component, break the degeneracy on the inner structure \citep{RogersSeager10, valencia2013, fortney2013, Hu_2021, tsai2021, Schlichting_2022}, and probe cloud formation processes in sub-Neptunes. Finally, our data suggest the possible presence of a third external planet in addition to TOI-406 b and c, but additional data of ESPRESSO-quality spanning a longer baseline are needed to investigate the nature (or existence) of this signal.

The TOI-406 planetary system shows the potential of the \thirstee\ program, laying a first stone on the path towards creating a reliable and wide observational sample of well-characterised sub-Neptunes. 
Such an observational sample is indeed needed for the comparison with planetary formation and evolution theories, as well as internal composition and atmospheric structure models, with the aim of improving our global picture of the sub-Neptune population.   

\section*{Data availability}\label{sec:data_availability}
Tables \ref{table:RV_table_ESPRESSO}, \ref{table:RV_table_NIRPS} and \ref{table:RV_table_HARPS} are available in electronic form at the CDS via anonymous ftp to cdsarc.u-strasbg.fr (130.79.128.5) or via \url{http://cdsweb.u-strasbg.fr/cgi-bin/qcat?J/A+A/}.

\begin{acknowledgements}
We thank the anonymous referee for the comments that helped improving the quality of this paper. We acknowledge financial support from the Agencia Estatal de Investigaci\'on of the Ministerio de Ciencia e Innovaci\'on MCIN/AEI/10.13039/501100011033 and the ERDF “A way of making Europe” through project PID2021-125627OB-C32, and from the Centre of Excellence “Severo Ochoa” award to the Instituto de Astrofisica de Canarias.
This paper includes data collected by the {\it TESS} mission,
which are publicly available from the Mikulski Archive for Space
Telescopes (MAST). Funding for the {\it TESS} mission is provided 
by the NASA Explorer Program. 
Resources supporting this work were provided by the NASA High-End Computing (HEC) Program through the NASA Advanced Supercomputing (NAS) Division at Ames Research Center for the production of the SPOC data products.
This research has made use of the Exoplanet Follow-up Observation Program (ExoFOP; DOI: 10.26134/ExoFOP5) website, which is operated by the California Institute of Technology, under contract with the National Aeronautics and Space Administration under the Exoplanet Exploration Program. 
Funding for the TESS mission is provided by NASA's Science Mission Directorate. 
This research has made extensive use of the SIMBAD
database, operated at CDS, Strasbourg, France, and NASA’s Astrophysics Data System.
This research has
made use of the NASA Exoplanet Archive, which is
operated by the California Institute of Technology, 
under contract with the National Aeronautics and Space
Administration under the Exoplanet Exploration Program. 
This work has made use of data from the European Space Agency (ESA) mission
{\it Gaia} (\url{https://www.cosmos.esa.int/gaia}), processed by the {\it Gaia}
Data Processing and Analysis Consortium (DPAC,
\url{https://www.cosmos.esa.int/web/gaia/dpac/consortium}). 
Funding for the DPAC
has been provided by national institutions, in particular the institutions
participating in the {\it Gaia} Multilateral Agreement.
This publication makes use of data products from the Two
Micron All Sky Survey, which is a joint project of the
University of Massachusetts and the Infrared Processing and
Analysis Center/California Institute of Technology, funded by
the National Aeronautics and Space Administration and the
National Science Foundation. 
This work makes use of observations from the LCOGT network. Part of the LCOGT telescope time was granted by NOIRLab through the Mid-Scale Innovations Program (MSIP). MSIP is funded by NSF.
This work has been carried out within the framework of the NCCR PlanetS supported by the Swiss National Science Foundation under grant 51NF40\_205606.

R.L. is supported by NASA through the NASA Hubble Fellowship grant HST-HF2-51559.001-A awarded by the Space Telescope Science Institute, which is operated by the Association of Universities for Research in Astronomy, Inc., for NASA, under contract NAS5-26555. R.L. also acknowledges funding from University of La Laguna through the Margarita Salas Fellowship from the Spanish Ministry of Universities ref. UNI/551/2021-May 26, and under the EU Next Generation funds. 
KAC and CNW acknowledge support from the TESS mission via subaward s3449 from MIT. 
J.M.A.M. is supported by the National Science Foundation (NSF) Graduate Research Fellowship Program (GRFP) under Grant No. DGE-1842400. J.M.A.M. and N.M.B. acknowledge support from NASA'S Interdisciplinary Consortia for Astrobiology Research (NNH19ZDA001N-ICAR) under award number 19-ICAR19\_2-0041.
D.J. and G.N. gratefully acknowledges the Centre of Informatics Tricity Academic Supercomputer and networK (CI TASK, Gda´nsk, Poland) for computing resources (grant no. PT01016). G.N. thanks for the research funding from the Ministry of Science and Higher Education programme
the `Excellence Initiative - Research University' conducted at the Centre of Excellence in Astrophysics and Astrochemistry of the Nicolaus Copernicus University in Toru´n, Poland.
NN acknowledges funding from Light Bridges for the Doctoral Thesis `Habitable Earth-like planets with ESPRESSO and NIRPS', in cooperation with the Instituto de Astrofísica de Canarias, and the use of Indefeasible Computer Rights (ICR) being commissioned at the ASTRO POC project in the Island of Tenerife, Canary Islands (Spain). The ICR-ASTRONOMY used for his research was provided by Light Bridges in co-operation with Hewlett Packard Enterprise (HPE).
JIGH acknowledges financial support from the Spanish Ministry of Science and Innovation (MICINN) project PID2020-117493GB-I00.
Research activities of the Board of Observational and Instrumental Astronomy at the Federal University of Rio Grande do Norte are supported by continuous grants from the Brazilian funding agencies CNPq. This study was financed in part by the CAPES-Print program. B.L.C.M. acknowledge CNPq research fellowships grant No. 305804/2022-7.

\end{acknowledgements}

%
%

\bibliographystyle{aa}
\bibliography{bibliography}

\begin{thebibliography}{207}
\expandafter\ifx\csname natexlab\endcsname\relax\def\natexlab#1{#1}\fi

\bibitem[{{Adams} {et~al.}(2021){Adams}, {Jackson}, {Johnson}, {Ciardi},
  {Cochran}, {Endl}, {Everett}, {Furlan}, {Howell}, {Jayanthi}, {MacQueen},
  {Matson}, {Partyka-Worley}, {Schlieder}, {Scott}, {Stanton}, \&
  {Ziegler}}]{adams2021}
{Adams}, E.~R., {Jackson}, B., {Johnson}, S., {et~al.} 2021, \psj, 2, 152

\bibitem[{{Aguichine} {et~al.}(2021){Aguichine}, {Mousis}, {Deleuil}, \&
  {Marcq}}]{aguichine2021}
{Aguichine}, A., {Mousis}, O., {Deleuil}, M., \& {Marcq}, E. 2021, \apj, 914,
  84

\bibitem[{{Akana Murphy} {et~al.}(2021){Akana Murphy}, {Kosiarek}, {Batalha},
  {Gonzales}, {Isaacson}, {Petigura}, {Weiss}, {Grunblatt}, {Ciardi}, {Fulton},
  {Hirsch}, {Behmard}, \& {Rosenthal}}]{akana_murphy2021}
{Akana Murphy}, J.~M., {Kosiarek}, M.~R., {Batalha}, N.~M., {et~al.} 2021, \aj,
  162, 294

\bibitem[{{Akana Murphy} {et~al.}(2024){Akana Murphy}, {Luque}, \&
  {Batalha}}]{murphy2024_cadence}
{Akana Murphy}, J.~M., {Luque}, R., \& {Batalha}, N.~M. 2024, arXiv e-prints,
  arXiv:2411.02521

\bibitem[{{Alibert} {et~al.}(2013){Alibert}, {Carron}, {Fortier}, {Pfyffer},
  {Benz}, {Mordasini}, \& {Swoboda}}]{alibert2013}
{Alibert}, Y., {Carron}, F., {Fortier}, A., {et~al.} 2013, \aap, 558, A109

\bibitem[{{Allard} {et~al.}(2012){Allard}, {Homeier}, \& {Freytag}}]{all12}
{Allard}, F., {Homeier}, D., \& {Freytag}, B. 2012, Philosophical Transactions
  of the Royal Society of London Series A, 370, 2765

\bibitem[{{Aller} {et~al.}(2020){Aller}, {Lillo-Box}, {Jones}, {Miranda}, \&
  {Barcel{\'o} Forteza}}]{Aller2020}
{Aller}, A., {Lillo-Box}, J., {Jones}, D., {Miranda}, L.~F., \& {Barcel{\'o}
  Forteza}, S. 2020, \aap, 635, A128

\bibitem[{{Anglada-Escud{\'e}} \& {Butler}(2012)}]{Anglada-Escude_2012}
{Anglada-Escud{\'e}}, G. \& {Butler}, R.~P. 2012, \apjs, 200, 15

\bibitem[{Angus(2021)}]{angus2021}
Angus, R. 2021, {RuthAngus/starspot: code for measuring stellar rotation
  periods}

\bibitem[{{Artigau} {et~al.}(2024){Artigau}, {Bouchy}, {Doyon}, {Baron},
  {Malo}, {Wildi}, {Pepe}, {Cook}, {Thibault}, {Reshetov}, {Dumusque}, {Lovis},
  {Sosnowska}, {Canto Martins}, {De Medeiros}, {Delfosse}, {Santos}, {Rebolo},
  {Abreu}, {Allain}, {Allart}, {Auger}, {Barros}, {Bazinet}, {Blind}, {Boisse},
  {Bonfils}, {Bourrier}, {Bovay}, {Broeg}, {Brousseau}, {Bruniquel}, {Cabral},
  {Cadieux}, {Carmona}, {Carteret}, {Challita}, {Chazelas}, {Cloutier},
  {Coelho}, {Cointepas}, {Conod}, {Cowan}, {Cristo}, {Gomes da Silva},
  {Dauplaise}, {Lima Gomes}, {Delgado-Mena}, {Ehrenreich}, {Faria}, {Figueira},
  {Forveille}, {Frensch}, {Gagn{\'e}}, {Genest}, {Genolet}, {Gonz{\'a}lez
  Hern{\'a}ndez}, {Gracia Temich}, {Grieves}, {Hernandez}, {Hobson},
  {Hoeijmakers}, {Kerley}, {Krishnamurthy}, {Lafreni{\`e}re}, {Lamontagne},
  {Larue}, {Leaf}, {Le{\~a}o}, {Lim}, {Lo Curto}, {Martins}, {Melo}, {Messias},
  {Mignon}, {Moranta}, {Mordasini}, {Al Moulla}, {Mounzer}, {L'Heureux},
  {Nari}, {Nielsen}, {Osborn}, {Parc}, {Pasquini}, {Passegger}, {Pelletier},
  {Peroux}, {Piaulet}, {Plotnykov}, {Poulin-Girard}, {Rasilla},
  {Saint-Antoine}, {Sarajic}, {Segovia}, {Seidel}, {S{\'e}gransan}, {Silva},
  {Srivastava}, {Stefanov}, {Su{\'a}rez Mascare{\~n}o}, {Sordet}, {Teixeira},
  {Udry}, {Valencia}, {Vall{\'e}e}, {Vandal}, {Vaulato}, {Wade}, {Wardenier},
  {Wehb{\'e}}, {Weisserman}, {Wevers}, \& {Zins}}]{Artigau_2024}
{Artigau}, {\'E}., {Bouchy}, F., {Doyon}, R., {et~al.} 2024, in Society of
  Photo-Optical Instrumentation Engineers (SPIE) Conference Series, Vol. 13096,
  Ground-based and Airborne Instrumentation for Astronomy X, ed. J.~J.
  {Bryant}, K.~{Motohara}, \& J.~R.~D. {Vernet}, 130960C

\bibitem[{{Artigau} {et~al.}(2022){Artigau}, {Cadieux}, {Cook}, {Doyon},
  {Vandal}, {Donati}, {Moutou}, {Delfosse}, {Fouqu{\'e}}, {Martioli}, {Bouchy},
  {Parsons}, {Carmona}, {Dumusque}, {Astudillo-Defru}, {Bonfils}, \&
  {Mignon}}]{Artigau_2022}
{Artigau}, {\'E}., {Cadieux}, C., {Cook}, N.~J., {et~al.} 2022, \aj, 164, 84

\bibitem[{{Astudillo-Defru} {et~al.}(2017){Astudillo-Defru}, {D{\'\i}az},
  {Bonfils}, {Almenara}, {Delisle}, {Bouchy}, {Delfosse}, {Forveille}, {Lovis},
  {Mayor}, {Murgas}, {Pepe}, {Santos}, {S{\'e}gransan}, {Udry}, \&
  {W{\"u}nsche}}]{Astudillo-Defru_2017}
{Astudillo-Defru}, N., {D{\'\i}az}, R.~F., {Bonfils}, X., {et~al.} 2017, \aap,
  605, L11

\bibitem[{{Bailer-Jones} {et~al.}(2021){Bailer-Jones}, {Rybizki}, {Fouesneau},
  {Demleitner}, \& {Andrae}}]{bailer_jones2021}
{Bailer-Jones}, C.~A.~L., {Rybizki}, J., {Fouesneau}, M., {Demleitner}, M., \&
  {Andrae}, R. 2021, VizieR Online Data Catalog, I/352

\bibitem[{{Baranne} {et~al.}(1996){Baranne}, {Queloz}, {Mayor}, {Adrianzyk},
  {Knispel}, {Kohler}, {Lacroix}, {Meunier}, {Rimbaud}, \& {Vin}}]{Baranne1996}
{Baranne}, A., {Queloz}, D., {Mayor}, M., {et~al.} 1996, \aaps, 119, 373

\bibitem[{{Barrag{\'a}n} {et~al.}(2023){Barrag{\'a}n}, {Gillen}, {Aigrain},
  {Meech}, {Klein}, {Nielsen}, {Yu}, {O'Sullivan}, {Nicholson}, \&
  {Lillo-Box}}]{barragan2023}
{Barrag{\'a}n}, O., {Gillen}, E., {Aigrain}, S., {et~al.} 2023, \mnras, 522,
  3458

\bibitem[{{Batalha} {et~al.}(2019){Batalha}, {Lewis}, {Fortney}, {Batalha},
  {Kempton}, {Lewis}, \& {Line}}]{batalha19}
{Batalha}, N.~E., {Lewis}, T., {Fortney}, J.~J., {et~al.} 2019, \apjl, 885, L25

\bibitem[{{Batalha} {et~al.}(2013){Batalha}, {Rowe}, {Bryson}, {Barclay},
  {Burke}, {Caldwell}, {Christiansen}, {Mullally}, {Thompson}, {Brown},
  {Dupree}, {Fabrycky}, {Ford}, {Fortney}, {Gilliland}, {Isaacson}, {Latham},
  {Marcy}, {Quinn}, {Ragozzine}, {Shporer}, {Borucki}, {Ciardi}, {Gautier},
  {Haas}, {Jenkins}, {Koch}, {Lissauer}, {Rapin}, {Basri}, {Boss}, {Buchhave},
  {Carter}, {Charbonneau}, {Christensen-Dalsgaard}, {Clarke}, {Cochran},
  {Demory}, {Desert}, {Devore}, {Doyle}, {Esquerdo}, {Everett}, {Fressin},
  {Geary}, {Girouard}, {Gould}, {Hall}, {Holman}, {Howard}, {Howell},
  {Ibrahim}, {Kinemuchi}, {Kjeldsen}, {Klaus}, {Li}, {Lucas}, {Meibom},
  {Morris}, {Pr{\v{s}}a}, {Quintana}, {Sanderfer}, {Sasselov}, {Seader},
  {Smith}, {Steffen}, {Still}, {Stumpe}, {Tarter}, {Tenenbaum}, {Torres},
  {Twicken}, {Uddin}, {Van Cleve}, {Walkowicz}, \& {Welsh}}]{Batalha2013}
{Batalha}, N.~M., {Rowe}, J.~F., {Bryson}, S.~T., {et~al.} 2013, \apjs, 204, 24

\bibitem[{{Bayo} {et~al.}(2008){Bayo}, {Rodrigo}, {Barrado Y Navascu{\'e}s},
  {Solano}, {Guti{\'e}rrez}, {Morales-Calder{\'o}n}, \& {Allard}}]{bayo2008}
{Bayo}, A., {Rodrigo}, C., {Barrado Y Navascu{\'e}s}, D., {et~al.} 2008, \aap,
  492, 277

\bibitem[{{Bean} {et~al.}(2021){Bean}, {Raymond}, \& {Owen}}]{bean21}
{Bean}, J.~L., {Raymond}, S.~N., \& {Owen}, J.~E. 2021, Journal of Geophysical
  Research (Planets), 126, e06639

\bibitem[{{Benneke} {et~al.}(2024){Benneke}, {Roy}, {Coulombe}, {Radica},
  {Piaulet}, {Ahrer}, {Pierrehumbert}, {Krissansen-Totton}, {Schlichting},
  {Hu}, {Yang}, {Christie}, {Thorngren}, {Young}, {Pelletier}, {Knutson},
  {Miguel}, {Evans-Soma}, {Dorn}, {Gagnebin}, {Fortney}, {Komacek},
  {MacDonald}, {Raul}, {Cloutier}, {Acuna}, {Lafreni{\`e}re}, {Cadieux},
  {Doyon}, {Welbanks}, \& {Allart}}]{bennecke2024}
{Benneke}, B., {Roy}, P.-A., {Coulombe}, L.-P., {et~al.} 2024, arXiv e-prints,
  arXiv:2403.03325

\bibitem[{{Bensby} {et~al.}(2014){Bensby}, {Feltzing}, \& {Oey}}]{bensby2014}
{Bensby}, T., {Feltzing}, S., \& {Oey}, M.~S. 2014, \aap, 562, A71

\bibitem[{{Berger} {et~al.}(2020){Berger}, {Huber}, {Gaidos}, {van Saders}, \&
  {Weiss}}]{berger2020}
{Berger}, T.~A., {Huber}, D., {Gaidos}, E., {van Saders}, J.~L., \& {Weiss},
  L.~M. 2020, \aj, 160, 108

\bibitem[{{Bertaux} {et~al.}(2014){Bertaux}, {Lallement}, {Ferron}, {Boonne},
  \& {Bodichon}}]{Bertaux_2014}
{Bertaux}, J.~L., {Lallement}, R., {Ferron}, S., {Boonne}, C., \& {Bodichon},
  R. 2014, \aap, 564, A46

\bibitem[{{Bitsch} {et~al.}(2018){Bitsch}, {Morbidelli}, {Johansen}, {Lega},
  {Lambrechts}, \& {Crida}}]{bitsch2018}
{Bitsch}, B., {Morbidelli}, A., {Johansen}, A., {et~al.} 2018, \aap, 612, A30

\bibitem[{{Bitsch} {et~al.}(2019){Bitsch}, {Raymond}, \& {Izidoro}}]{Bitsch19}
{Bitsch}, B., {Raymond}, S.~N., \& {Izidoro}, A. 2019, \aap, 624, A109

\bibitem[{{Bonomo} {et~al.}(2023){Bonomo}, {Dumusque}, {Massa}, {Mortier},
  {Bongiolatti}, {Malavolta}, {Sozzetti}, {Buchhave}, {Damasso}, {Haywood},
  {Morbidelli}, {Latham}, {Molinari}, {Pepe}, {Poretti}, {Udry}, {Affer},
  {Boschin}, {Charbonneau}, {Cosentino}, {Cretignier}, {Ghedina}, {Lega},
  {L{\'o}pez-Morales}, {Margini}, {Mart{\'\i}nez Fiorenzano}, {Mayor},
  {Micela}, {Pedani}, {Pinamonti}, {Rice}, {Sasselov}, {Tronsgaard}, \&
  {Vanderburg}}]{bonomo2023}
{Bonomo}, A.~S., {Dumusque}, X., {Massa}, A., {et~al.} 2023, \aap, 677, A33

\bibitem[{Borucki {et~al.}(2010)Borucki, Koch, Basri, Batalha, Brown, Caldwell,
  Caldwell, Christensen-Dalsgaard, Cochran, DeVore, Dunham, Dupree, Gautier,
  Geary, Gilliland, Gould, Howell, Jenkins, Kondo, Latham, Marcy, Meibom,
  Kjeldsen, Lissauer, Monet, Morrison, Sasselov, Tarter, Boss, Brownlee, Owen,
  Buzasi, Charbonneau, Doyle, Fortney, Ford, Holman, Seager, Steffen, Welsh,
  Rowe, Anderson, Buchhave, Ciardi, Walkowicz, Sherry, Horch, Isaacson,
  Everett, Fischer, Torres, Johnson, Endl, MacQueen, Bryson, Dotson, Haas,
  Kolodziejczak, Van~Cleve, Chandrasekaran, Twicken, Quintana, Clarke, Allen,
  Li, Wu, Tenenbaum, Verner, Bruhweiler, Barnes, \& Prsa}]{Borucki2010}
Borucki, W.~J., Koch, D., Basri, G., {et~al.} 2010, Science, 327, 977

\bibitem[{{Bouchy} {et~al.}(2017){Bouchy}, {Doyon}, {Artigau}, {Melo},
  {Hernandez}, {Wildi}, {Delfosse}, {Lovis}, {Figueira}, {Canto Martins},
  {Gonz{\'a}lez Hern{\'a}ndez}, {Thibault}, {Reshetov}, {Pepe}, {Santos}, {de
  Medeiros}, {Rebolo}, {Abreu}, {Adibekyan}, {Bandy}, {Benz}, {Blind},
  {Bohlender}, {Boisse}, {Bovay}, {Broeg}, {Brousseau}, {Cabral}, {Chazelas},
  {Cloutier}, {Coelho}, {Conod}, {Cumming}, {Delabre}, {Genolet}, {Hagelberg},
  {Jayawardhana}, {K{\"a}ufl}, {Lafreni{\`e}re}, {de Castro Le{\~a}o}, {Malo},
  {de Medeiros Martins}, {Matthews}, {Metchev}, {Oshagh}, {Ouellet}, {Parro},
  {Rasilla Pi{\~n}eiro}, {Santos}, {Sarajlic}, {Segovia}, {Sordet}, {Udry},
  {Valencia}, {Vall{\'e}e}, {Venn}, {Wade}, \& {Saddlemyer}}]{Bouchy_2017}
{Bouchy}, F., {Doyon}, R., {Artigau}, {\'E}., {et~al.} 2017, The Messenger,
  169, 21

\bibitem[{{Brande} {et~al.}(2024){Brande}, {Crossfield}, {Kreidberg}, {Morley},
  {Barman}, {Benneke}, {Christiansen}, {Dragomir}, {Fortney}, {Greene},
  {Hardegree-Ullman}, {Howard}, {Knutson}, {Lothringer}, \&
  {Mikal-Evans}}]{brande2024}
{Brande}, J., {Crossfield}, I. J.~M., {Kreidberg}, L., {et~al.} 2024, \apjl,
  961, L23

\bibitem[{{Brinkman} {et~al.}(2023){Brinkman}, {Weiss}, {Dai}, {Huber}, {Kite},
  {Valencia}, {Bean}, {Beard}, {Behmard}, {Blunt}, {Brady}, {Fulton},
  {Giacalone}, {Howard}, {Isaacson}, {Kasper}, {Lubin}, {MacDougall}, {Akana
  Murphy}, {Plotnykov}, {Polanski}, {Rice}, {Seifahrt}, {Stef{\'a}nsson}, \&
  {St{\"u}rmer}}]{Brinkman2022}
{Brinkman}, C.~L., {Weiss}, L.~M., {Dai}, F., {et~al.} 2023, \aj, 165, 88

\bibitem[{{Brown} {et~al.}(2013){Brown}, {Baliber}, {Bianco}, {Bowman},
  {Burleson}, {Conway}, {Crellin}, {Depagne}, {De Vera}, {Dilday}, {Dragomir},
  {Dubberley}, {Eastman}, {Elphick}, {Falarski}, {Foale}, {Ford}, {Fulton},
  {Garza}, {Gomez}, {Graham}, {Greene}, {Haldeman}, {Hawkins}, {Haworth},
  {Haynes}, {Hidas}, {Hjelstrom}, {Howell}, {Hygelund}, {Lister}, {Lobdill},
  {Martinez}, {Mullins}, {Norbury}, {Parrent}, {Paulson}, {Petry}, {Pickles},
  {Posner}, {Rosing}, {Ross}, {Sand}, {Saunders}, {Shobbrook}, {Shporer},
  {Street}, {Thomas}, {Tsapras}, {Tufts}, {Valenti}, {Vander Horst}, {Walker},
  {White}, \& {Willis}}]{brown2013}
{Brown}, T.~M., {Baliber}, N., {Bianco}, F.~B., {et~al.} 2013, \pasp, 125, 1031

\bibitem[{{Br{\"u}gger} {et~al.}(2020){Br{\"u}gger}, {Burn}, {Coleman},
  {Alibert}, \& {Benz}}]{brugger2020}
{Br{\"u}gger}, N., {Burn}, R., {Coleman}, G.~A.~L., {Alibert}, Y., \& {Benz},
  W. 2020, \aap, 640, A21

\bibitem[{{Burn} {et~al.}(2024){Burn}, {Mordasini}, {Mishra}, {Haldemann},
  {Venturini}, {Emsenhuber}, \& {Henning}}]{Burn24}
{Burn}, R., {Mordasini}, C., {Mishra}, L., {et~al.} 2024, Nature Astronomy, 8,
  463

\bibitem[{{Burn} {et~al.}(2021){Burn}, {Schlecker}, {Mordasini}, {Emsenhuber},
  {Alibert}, {Henning}, {Klahr}, \& {Benz}}]{burn2021}
{Burn}, R., {Schlecker}, M., {Mordasini}, C., {et~al.} 2021, \aap, 656, A72

\bibitem[{{Burt} {et~al.}(2024){Burt}, {Hooton}, {Mamajek}, {Barrag{\'a}n},
  {Millholland}, {Fairnington}, {Fisher}, {Halverson}, {Huang}, {Brady},
  {Seifahrt}, {Gaidos}, {Luque}, {Kasper}, \& {Bean}}]{burt2024}
{Burt}, J.~A., {Hooton}, M.~J., {Mamajek}, E.~E., {et~al.} 2024, \apjl, 971,
  L12

\bibitem[{{Cadieux} {et~al.}(2024{\natexlab{a}}){Cadieux}, {Doyon},
  {MacDonald}, {Turbet}, {Artigau}, {Lim}, {Radica}, {Fauchez}, {Salhi},
  {Dang}, {Albert}, {Coulombe}, {Cowan}, {Lafreni{\`e}re}, {L'Heureux},
  {Piaulet-Ghorayeb}, {Benneke}, {Cloutier}, {Charnay}, {Cook},
  {Fournier-Tondreau}, {Plotnykov}, \& {Valencia}}]{Cadieux2024b}
{Cadieux}, C., {Doyon}, R., {MacDonald}, R.~J., {et~al.} 2024{\natexlab{a}},
  \apjl, 970, L2

\bibitem[{{Cadieux} {et~al.}(2022){Cadieux}, {Doyon}, {Plotnykov},
  {H{\'e}brard}, {Jahandar}, {Artigau}, {Valencia}, {Cook}, {Martioli},
  {Vandal}, {Donati}, {Cloutier}, {Narita}, {Fukui}, {Hirano}, {Bouchy},
  {Cowan}, {Gonzales}, {Ciardi}, {Stassun}, {Arnold}, {Benneke}, {Boisse},
  {Bonfils}, {Carmona}, {Cort{\'e}s-Zuleta}, {Delfosse}, {Forveille},
  {Fouqu{\'e}}, {Gomes da Silva}, {Jenkins}, {Kiefer}, {K{\'o}sp{\'a}l},
  {Lafreni{\`e}re}, {Martins}, {Moutou}, {do Nascimento}, {Ould-Elhkim},
  {Pelletier}, {Twicken}, {Bouma}, {Cartwright}, {Darveau-Bernier}, {Grankin},
  {Ikoma}, {Kagetani}, {Kawauchi}, {Kodama}, {Kotani}, {Latham}, {Menou},
  {Ricker}, {Seager}, {Tamura}, {Vanderspek}, \& {Watanabe}}]{Cadieux22}
{Cadieux}, C., {Doyon}, R., {Plotnykov}, M., {et~al.} 2022, \aj, 164, 96

\bibitem[{{Cadieux} {et~al.}(2024{\natexlab{b}}){Cadieux}, {Plotnykov},
  {Doyon}, {Valencia}, {Jahandar}, {Dang}, {Turbet}, {Fauchez}, {Cloutier},
  {Cherubim}, {Artigau}, {Cook}, {Edwards}, {Hallatt}, {Charnay}, {Bouchy},
  {Allart}, {Mignon}, {Baron}, {Barros}, {Benneke}, {Canto Martins}, {Cowan},
  {De Medeiros}, {Delfosse}, {Delgado-Mena}, {Dumusque}, {Ehrenreich},
  {Frensch}, {Gonz{\'a}lez Hern{\'a}ndez}, {Hara}, {Lafreni{\`e}re}, {Lo
  Curto}, {Malo}, {Melo}, {Mounzer}, {Passeger}, {Pepe}, {Poulin-Girard},
  {Santos}, {Sosnowska}, {Su{\'a}rez Mascare{\~n}o}, {Thibault}, {Vaulato},
  {Wade}, \& {Wildi}}]{Cadieux2024a}
{Cadieux}, C., {Plotnykov}, M., {Doyon}, R., {et~al.} 2024{\natexlab{b}},
  \apjl, 960, L3

\bibitem[{{Castro-Gonz{\'a}lez} {et~al.}(2022){Castro-Gonz{\'a}lez}, {D{\'\i}ez
  Alonso}, {Men{\'e}ndez Blanco}, {Livingston}, {de Leon}, {Lillo-Box},
  {Korth}, {Fern{\'a}ndez Men{\'e}ndez}, {Recio}, {Izquierdo-Ruiz}, {Coya
  Lozano}, {Garc{\'\i}a de la Cuesta}, {G{\'o}mez Hern{\'a}ndez}, {Vidal
  Blanco}, {Hevia D{\'\i}az}, {Pardo Silva}, {P{\'e}rez Acevedo}, {Polancos
  Ruiz}, {Padilla Tijer{\'\i}n}, {V{\'a}zquez Garc{\'\i}a}, {Su{\'a}rez
  G{\'o}mez}, {Garc{\'\i}a Riesgo}, {Gonz{\'a}lez Guti{\'e}rrez}, {Bonavera},
  {Gonz{\'a}lez-Nuevo}, {Rodr{\'\i}guez Pereira}, {S{\'a}nchez Lasheras},
  {S{\'a}nchez Rodr{\'\i}guez}, {Mu{\~n}iz}, {Santos Rodr{\'\i}guez}, \& {de
  Cos Juez}}]{castro-gonzalez_2022}
{Castro-Gonz{\'a}lez}, A., {D{\'\i}ez Alonso}, E., {Men{\'e}ndez Blanco}, J.,
  {et~al.} 2022, \mnras, 509, 1075

\bibitem[{{Chakrabarty} \& {Mulders}(2024)}]{ChakrabartyMulders2024}
{Chakrabarty}, A. \& {Mulders}, G.~D. 2024, \apj, 966, 185

\bibitem[{{Chaplin} {et~al.}(2013){Chaplin}, {Sanchis-Ojeda}, {Campante},
  {Handberg}, {Stello}, {Winn}, {Basu}, {Christensen-Dalsgaard}, {Davies},
  {Metcalfe}, {Buchhave}, {Fischer}, {Bedding}, {Cochran}, {Elsworth},
  {Gilliland}, {Hekker}, {Huber}, {Isaacson}, {Karoff}, {Kawaler}, {Kjeldsen},
  {Latham}, {Lund}, {Lundkvist}, {Marcy}, {Miglio}, {Barclay}, \&
  {Lissauer}}]{chaplin2013}
{Chaplin}, W.~J., {Sanchis-Ojeda}, R., {Campante}, T.~L., {et~al.} 2013, \apj,
  766, 101

\bibitem[{{Cherubim} {et~al.}(2023){Cherubim}, {Cloutier}, {Charbonneau},
  {Stockdale}, {Stassun}, {Schwarz}, {Safonov}, {Mortier}, {Lewin}, {Latham},
  {Horne}, {Haywood}, {Gonzales}, {Goliguzova}, {Collins}, {Ciardi}, {Bieryla},
  {Belinski}, {Wohler}, {Watson}, {Vanderspek}, {Udry}, {Sozzetti},
  {S{\'e}gransan}, {Sasselov}, {Ricker}, {Rice}, {Poretti}, {Piotto}, {Pepe},
  {Molinari}, {Micela}, {Mayor}, {Lovis}, {L{\'o}pez-Morales}, {Jenkins},
  {Essack}, {Dumusque}, {Doty}, {Col{\'o}n}, {Cameron}, \&
  {Buchhave}}]{cherubim2023}
{Cherubim}, C., {Cloutier}, R., {Charbonneau}, D., {et~al.} 2023, \aj, 165, 167

\bibitem[{{Chontos} {et~al.}(2022){Chontos}, {Murphy}, {MacDougall},
  {Fetherolf}, {Van Zandt}, {Rubenzahl}, {Beard}, {Huber}, {Batalha},
  {Crossfield}, {Dressing}, {Fulton}, {Howard}, {Isaacson}, {Kane}, {Petigura},
  {Robertson}, {Roy}, {Weiss}, {Behmard}, {Dai}, {Dalba}, {Giacalone}, {Hill},
  {Lubin}, {Mayo}, {Mo{\v{c}}nik}, {Polanski}, {Rosenthal}, {Scarsdale},
  {Turtelboom}, {Ricker}, {Vanderspek}, {Latham}, {Seager}, {Winn}, {Jenkins},
  {Quinn}, {Guerrero}, {Collins}, {Ciardi}, {Shporer}, {Goeke}, {Levine},
  {Ting}, {Bieryla}, {Collins}, {Kielkopf}, {Barkaoui}, {Benni},
  {Esparza-Borges}, {Conti}, {Hooton}, {Kagetani}, {Laloum}, {Marino},
  {Massey}, {Murgas}, {Papini}, {Schwarz}, {Srdoc}, {Stockdale}, {Wang},
  {Wittrock}, \& {Zou}}]{Chontos22}
{Chontos}, A., {Murphy}, J. M.~A., {MacDougall}, M.~G., {et~al.} 2022, \aj,
  163, 297

\bibitem[{{Cincotta} \& {Sim{\'o}}(2000)}]{CincottaSimo2000}
{Cincotta}, P.~M. \& {Sim{\'o}}, C. 2000, \aaps, 147, 205

\bibitem[{{Cloutier} {et~al.}(2020){Cloutier}, {Rodriguez}, {Irwin},
  {Charbonneau}, {Stassun}, {Mortier}, {Latham}, {Isaacson}, {Howard}, {Udry},
  {Wilson}, {Watson}, {Pinamonti}, {Lienhard}, {Giacobbe}, {Guerra}, {Collins},
  {Beiryla}, {Esquerdo}, {Matthews}, {Matson}, {Howell}, {Furlan},
  {Crossfield}, {Winters}, {Nava}, {Ment}, {Lopez}, {Ricker}, {Vanderspek},
  {Seager}, {Jenkins}, {Ting}, {Tenenbaum}, {Sozzetti}, {Sha}, {S{\'e}gransan},
  {Schlieder}, {Sasselov}, {Roy}, {Robertson}, {Rice}, {Poretti}, {Piotto},
  {Phillips}, {Pepper}, {Pepe}, {Molinari}, {Mocnik}, {Micela}, {Mayor},
  {Martinez Fiorenzano}, {Mallia}, {Lubin}, {Lovis}, {L{\'o}pez-Morales},
  {Kosiarek}, {Kielkopf}, {Kane}, {Jensen}, {Isopi}, {Huber}, {Hill},
  {Harutyunyan}, {Gonzales}, {Giacalone}, {Ghedina}, {Ercolino}, {Dumusque},
  {Dressing}, {Damasso}, {Dalba}, {Cosentino}, {Conti}, {Col{\'o}n}, {Collins},
  {Cameron}, {Ciardi}, {Christiansen}, {Chontos}, {Cecconi}, {Caldwell},
  {Burke}, {Buchhave}, {Beichman}, {Behmard}, {Beard}, \& {Akana
  Murphy}}]{cloutier2020}
{Cloutier}, R., {Rodriguez}, J.~E., {Irwin}, J., {et~al.} 2020, \aj, 160, 22

\bibitem[{{Collins}(2019)}]{collins2019}
{Collins}, K. 2019, in American Astronomical Society Meeting Abstracts, Vol.
  233, American Astronomical Society Meeting Abstracts \#233, 140.05

\bibitem[{{Collins} {et~al.}(2017){Collins}, {Kielkopf}, {Stassun}, \&
  {Hessman}}]{Collins:2017}
{Collins}, K.~A., {Kielkopf}, J.~F., {Stassun}, K.~G., \& {Hessman}, F.~V.
  2017, \aj, 153, 77

\bibitem[{{Cook} {et~al.}(2022){Cook}, {Artigau}, {Doyon}, {Hobson},
  {Martioli}, {Bouchy}, {Moutou}, {Carmona}, {Usher}, {Fouqu{\'e}}, {Arnold},
  {Delfosse}, {Boisse}, {Cadieux}, {Vandal}, {Donati}, \&
  {Desli{\`e}res}}]{Cook_2022}
{Cook}, N.~J., {Artigau}, {\'E}., {Doyon}, R., {et~al.} 2022, \pasp, 134,
  114509

\bibitem[{{Cutri} {et~al.}(2021){Cutri}, {Wright}, {Conrow},
  {et~al.}}]{cutri_allwise}
{Cutri}, R., {Wright}, E.~L., {Conrow}, T., {et~al.} 2021, AllWISE Data Release

\bibitem[{{Cutri} {et~al.}(2003){Cutri}, {Skrutskie}, {van Dyk}, {Beichman},
  {Carpenter}, {Chester}, {Cambresy}, {Evans}, {Fowler}, {Gizis}, {Howard},
  {Huchra}, {Jarrett}, {Kopan}, {Kirkpatrick}, {Light}, {Marsh}, {McCallon},
  {Schneider}, {Stiening}, {Sykes}, {Weinberg}, {Wheaton}, {Wheelock}, \&
  {Zacarias}}]{Cutri2003}
{Cutri}, R.~M., {Skrutskie}, M.~F., {van Dyk}, S., {et~al.} 2003, VizieR Online
  Data Catalog, 2246

\bibitem[{{Dai} {et~al.}(2019){Dai}, {Masuda}, {Winn}, \& {Zeng}}]{Dai2019}
{Dai}, F., {Masuda}, K., {Winn}, J.~N., \& {Zeng}, L. 2019, \apj, 883, 79

\bibitem[{{Delmotte} {et~al.}(2006){Delmotte}, {Dolensky}, {Padovani},
  {Retzlaff}, {Rit{\'e}}, {Rosati}, {Slijkhuis}, {Wicenec}, {Fernique}, \&
  {Micol}}]{Delmotte_2006}
{Delmotte}, N., {Dolensky}, M., {Padovani}, P., {et~al.} 2006, in Astronomical
  Society of the Pacific Conference Series, Vol. 351, Astronomical Data
  Analysis Software and Systems XV, ed. C.~{Gabriel}, C.~{Arviset}, D.~{Ponz},
  \& S.~{Enrique}, 690

\bibitem[{{Diamond-Lowe} {et~al.}(2022){Diamond-Lowe}, {Kreidberg}, {Harman},
  {Kempton}, {Rogers}, {Joyce}, {Eastman}, {King}, {Kopparapu}, {Youngblood},
  {Kosiarek}, {Livingston}, {Hardegree-Ullman}, \&
  {Crossfield}}]{Diamond-Lowe22}
{Diamond-Lowe}, H., {Kreidberg}, L., {Harman}, C.~E., {et~al.} 2022, \aj, 164,
  172

\bibitem[{{D{\'\i}ez Alonso} {et~al.}(2018){D{\'\i}ez Alonso}, {Su{\'a}rez
  G{\'o}mez}, {Gonz{\'a}lez Hern{\'a}ndez}, {Su{\'a}rez Mascare{\~n}o},
  {Gonz{\'a}lez Guti{\'e}rrez}, {Velasco}, {Toledo-Padr{\'o}n}, {de Cos Juez},
  \& {Rebolo}}]{diez_alonso2018}
{D{\'\i}ez Alonso}, E., {Su{\'a}rez G{\'o}mez}, S.~L., {Gonz{\'a}lez
  Hern{\'a}ndez}, J.~I., {et~al.} 2018, \mnras, 476, L50

\bibitem[{{Donati} {et~al.}(2020){Donati}, {Kouach}, {Moutou}, {Doyon},
  {Delfosse}, {Artigau}, {Baratchart}, {Lacombe}, {Barrick}, {H{\'e}brard},
  {Bouchy}, {Saddlemyer}, {Par{\`e}s}, {Rabou}, {Micheau}, {Dolon}, {Reshetov},
  {Challita}, {Carmona}, {Striebig}, {Thibault}, {Martioli}, {Cook},
  {Fouqu{\'e}}, {Vermeulen}, {Wang}, {Arnold}, {Pepe}, {Boisse}, {Figueira},
  {Bouvier}, {Ray}, {Feugeade}, {Morin}, {Alencar}, {Hobson}, {Castilho},
  {Udry}, {Santos}, {Hernandez}, {Benedict}, {Vall{\'e}e}, {Gallou}, {Dupieux},
  {Larrieu}, {Perruchot}, {Sottile}, {Moreau}, {Usher}, {Baril}, {Wildi},
  {Chazelas}, {Malo}, {Bonfils}, {Loop}, {Kerley}, {Wevers}, {Dunn}, {Pazder},
  {Macdonald}, {Dubois}, {Carri{\'e}}, {Valentin}, {Henault}, {Yan}, \&
  {Steinmetz}}]{Donati_2020}
{Donati}, J.~F., {Kouach}, D., {Moutou}, C., {et~al.} 2020, \mnras, 498, 5684

\bibitem[{{Dorn} \& {Lichtenberg}(2021)}]{DornLichtenberg21}
{Dorn}, C. \& {Lichtenberg}, T. 2021, \apjl, 922, L4

\bibitem[{{Eastman} {et~al.}(2013){Eastman}, {Gaudi}, \& {Agol}}]{Eastman2013}
{Eastman}, J., {Gaudi}, B.~S., \& {Agol}, E. 2013, \pasp, 125, 83

\bibitem[{{Ehrenreich} {et~al.}(2012){Ehrenreich}, {Bourrier}, {Bonfils},
  {Lecavelier des Etangs}, {H{\'e}brard}, {Sing}, {Wheatley}, {Vidal-Madjar},
  {Delfosse}, {Udry}, {Forveille}, \& {Moutou}}]{Ehrenreich2012}
{Ehrenreich}, D., {Bourrier}, V., {Bonfils}, X., {et~al.} 2012, \aap, 547, A18

\bibitem[{{Esteves} {et~al.}(2020){Esteves}, {Izidoro}, {Raymond}, \&
  {Bitsch}}]{Esteves20}
{Esteves}, L., {Izidoro}, A., {Raymond}, S.~N., \& {Bitsch}, B. 2020, \mnras,
  497, 2493

\bibitem[{{Fabrycky} {et~al.}(2014){Fabrycky}, {Lissauer}, {Ragozzine}, {Rowe},
  {Steffen}, {Agol}, {Barclay}, {Batalha}, {Borucki}, {Ciardi}, {Ford},
  {Gautier}, {Geary}, {Holman}, {Jenkins}, {Li}, {Morehead}, {Morris},
  {Shporer}, {Smith}, {Still}, \& {Van Cleve}}]{Fabrycky14}
{Fabrycky}, D.~C., {Lissauer}, J.~J., {Ragozzine}, D., {et~al.} 2014, \apj,
  790, 146

\bibitem[{{Foreman-Mackey} {et~al.}(2013){Foreman-Mackey}, {Hogg}, {Lang}, \&
  {Goodman}}]{ForemanMackey2013}
{Foreman-Mackey}, D., {Hogg}, D.~W., {Lang}, D., \& {Goodman}, J. 2013, \pasp,
  125, 306

\bibitem[{{Fortney} {et~al.}(2013){Fortney}, {Mordasini}, {Nettelmann},
  {Kempton}, {Greene}, \& {Zahnle}}]{fortney2013}
{Fortney}, J.~J., {Mordasini}, C., {Nettelmann}, N., {et~al.} 2013, \apj, 775,
  80

\bibitem[{{Fulton} \& {Petigura}(2018)}]{FultonPetigura18}
{Fulton}, B.~J. \& {Petigura}, E.~A. 2018, \aj, 156, 264

\bibitem[{{Fulton} {et~al.}(2017){Fulton}, {Petigura}, {Howard}, {Isaacson},
  {Marcy}, {Cargile}, {Hebb}, {Weiss}, {Johnson}, {Morton}, {Sinukoff},
  {Crossfield}, \& {Hirsch}}]{Fulton2017}
{Fulton}, B.~J., {Petigura}, E.~A., {Howard}, A.~W., {et~al.} 2017, \aj, 154,
  109

\bibitem[{{Gaia Collaboration} {et~al.}(2016){Gaia Collaboration}, {Prusti},
  {de Bruijne}, {Brown}, {Vallenari}, {Babusiaux}, {Bailer-Jones}, {Bastian},
  {Biermann}, {Evans}, {Eyer}, {Jansen}, {Jordi}, {Klioner}, {Lammers},
  {Lindegren}, {Luri}, {Mignard}, {Milligan}, {Panem}, {Poinsignon},
  {Pourbaix}, {Randich}, {Sarri}, {Sartoretti}, {Siddiqui}, {Soubiran},
  {Valette}, {van Leeuwen}, {Walton}, {Aerts}, {Arenou}, {Cropper}, {Drimmel},
  {H{\o}g}, {Katz}, {Lattanzi}, {O'Mullane}, {Grebel}, {Holland}, {Huc},
  {Passot}, {Bramante}, {Cacciari}, {Casta{\~n}eda}, {Chaoul}, {Cheek}, {De
  Angeli}, {Fabricius}, {Guerra}, {Hern{\'a}ndez}, {Jean-Antoine-Piccolo},
  {Masana}, {Messineo}, {Mowlavi}, {Nienartowicz}, {Ord{\'o}{\~n}ez-Blanco},
  {Panuzzo}, {Portell}, {Richards}, {Riello}, {Seabroke}, {Tanga},
  {Th{\'e}venin}, {Torra}, {Els}, {Gracia-Abril}, {Comoretto},
  {Garcia-Reinaldos}, {Lock}, {Mercier}, {Altmann}, {Andrae}, {Astraatmadja},
  {Bellas-Velidis}, {Benson}, {Berthier}, {Blomme}, {Busso}, {Carry},
  {Cellino}, {Clementini}, {Cowell}, {Creevey}, {Cuypers}, {Davidson}, {De
  Ridder}, {de Torres}, {Delchambre}, {Dell'Oro}, {Ducourant}, {Fr{\'e}mat},
  {Garc{\'\i}a-Torres}, {Gosset}, {Halbwachs}, {Hambly}, {Harrison}, {Hauser},
  {Hestroffer}, {Hodgkin}, {Huckle}, {Hutton}, {Jasniewicz}, {Jordan},
  {Kontizas}, {Korn}, {Lanzafame}, {Manteiga}, {Moitinho}, {Muinonen},
  {Osinde}, {Pancino}, {Pauwels}, {Petit}, {Recio-Blanco}, {Robin}, {Sarro},
  {Siopis}, {Smith}, {Smith}, {Sozzetti}, {Thuillot}, {van Reeven}, {Viala},
  {Abbas}, {Abreu Aramburu}, {Accart}, {Aguado}, {Allan}, {Allasia},
  {Altavilla}, {{\'A}lvarez}, {Alves}, {Anderson}, {Andrei}, {Anglada Varela},
  {Antiche}, {Antoja}, {Ant{\'o}n}, {Arcay}, {Atzei}, {Ayache}, {Bach},
  {Baker}, {Balaguer-N{\'u}{\~n}ez}, {Barache}, {Barata}, {Barbier}, {Barblan},
  {Baroni}, {Barrado y Navascu{\'e}s}, {Barros}, {Barstow}, {Becciani},
  {Bellazzini}, {Bellei}, {Bello Garc{\'\i}a}, {Belokurov}, {Bendjoya},
  {Berihuete}, {Bianchi}, {Bienaym{\'e}}, {Billebaud}, {Blagorodnova},
  {Blanco-Cuaresma}, {Boch}, {Bombrun}, {Borrachero}, {Bouquillon}, {Bourda},
  {Bouy}, {Bragaglia}, {Breddels}, {Brouillet}, {Br{\"u}semeister},
  {Bucciarelli}, {Budnik}, {Burgess}, {Burgon}, {Burlacu}, {Busonero}, {Buzzi},
  {Caffau}, {Cambras}, {Campbell}, {Cancelliere}, {Cantat-Gaudin}, {Carlucci},
  {Carrasco}, {Castellani}, {Charlot}, {Charnas}, {Charvet}, {Chassat},
  {Chiavassa}, {Clotet}, {Cocozza}, {Collins}, {Collins}, {Costigan}, {Crifo},
  {Cross}, {Crosta}, {Crowley}, {Dafonte}, {Damerdji}, {Dapergolas}, {David},
  {David}, {De Cat}, {de Felice}, {de Laverny}, {De Luise}, {De March}, {de
  Martino}, {de Souza}, {Debosscher}, {del Pozo}, {Delbo}, {Delgado},
  {Delgado}, {di Marco}, {Di Matteo}, {Diakite}, {Distefano}, {Dolding}, {Dos
  Anjos}, {Drazinos}, {Dur{\'a}n}, {Dzigan}, {Ecale}, {Edvardsson}, {Enke},
  {Erdmann}, {Escolar}, {Espina}, {Evans}, {Eynard Bontemps}, {Fabre},
  {Fabrizio}, {Faigler}, {Falc{\~a}o}, {Farr{\`a}s Casas}, {Faye}, {Federici},
  {Fedorets}, {Fern{\'a}ndez-Hern{\'a}ndez}, {Fernique}, {Fienga}, {Figueras},
  {Filippi}, {Findeisen}, {Fonti}, {Fouesneau}, {Fraile}, {Fraser}, {Fuchs},
  {Furnell}, {Gai}, {Galleti}, {Galluccio}, {Garabato}, {Garc{\'\i}a-Sedano},
  {Gar{\'e}}, {Garofalo}, {Garralda}, {Gavras}, {Gerssen}, {Geyer}, {Gilmore},
  {Girona}, {Giuffrida}, {Gomes}, {Gonz{\'a}lez-Marcos},
  {Gonz{\'a}lez-N{\'u}{\~n}ez}, {Gonz{\'a}lez-Vidal}, {Granvik}, {Guerrier},
  {Guillout}, {Guiraud}, {G{\'u}rpide}, {Guti{\'e}rrez-S{\'a}nchez}, {Guy},
  {Haigron}, {Hatzidimitriou}, {Haywood}, {Heiter}, {Helmi}, {Hobbs},
  {Hofmann}, {Holl}, {Holland}, {Hunt}, {Hypki}, {Icardi}, {Irwin}, {Jevardat
  de Fombelle}, {Jofr{\'e}}, {Jonker}, {Jorissen}, {Julbe}, {Karampelas},
  {Kochoska}, {Kohley}, {Kolenberg}, {Kontizas}, {Koposov}, {Kordopatis},
  {Koubsky}, {Kowalczyk}, {Krone-Martins}, {Kudryashova}, {Kull}, {Bachchan},
  {Lacoste-Seris}, {Lanza}, {Lavigne}, {Le Poncin-Lafitte}, {Lebreton},
  {Lebzelter}, {Leccia}, {Leclerc}, {Lecoeur-Taibi}, {Lemaitre}, {Lenhardt},
  {Leroux}, {Liao}, {Licata}, {Lindstr{\o}m}, {Lister}, {Livanou}, {Lobel},
  {L{\"o}ffler}, {L{\'o}pez}, {Lopez-Lozano}, {Lorenz}, {Loureiro},
  {MacDonald}, {Magalh{\~a}es Fernandes}, {Managau}, {Mann}, {Mantelet},
  {Marchal}, {Marchant}, {Marconi}, {Marie}, {Marinoni}, {Marrese},
  {Marschalk{\'o}}, {Marshall}, {Mart{\'\i}n-Fleitas}, {Martino}, {Mary},
  {Matijevi{\v{c}}}, {Mazeh}, {McMillan}, {Messina}, {Mestre}, {Michalik},
  {Millar}, {Miranda}, {Molina}, {Molinaro}, {Molinaro}, {Moln{\'a}r},
  {Moniez}, {Montegriffo}, {Monteiro}, {Mor}, {Mora}, {Morbidelli}, {Morel},
  {Morgenthaler}, {Morley}, {Morris}, {Mulone}, {Muraveva}, {Musella},
  {Narbonne}, {Nelemans}, {Nicastro}, {Noval}, {Ord{\'e}novic},
  {Ordieres-Mer{\'e}}, {Osborne}, {Pagani}, {Pagano}, {Pailler}, {Palacin},
  {Palaversa}, {Parsons}, {Paulsen}, {Pecoraro}, {Pedrosa}, {Pentik{\"a}inen},
  {Pereira}, {Pichon}, {Piersimoni}, {Pineau}, {Plachy}, {Plum}, {Poujoulet},
  {Pr{\v{s}}a}, {Pulone}, {Ragaini}, {Rago}, {Rambaux}, {Ramos-Lerate},
  {Ranalli}, {Rauw}, {Read}, {Regibo}, {Renk}, {Reyl{\'e}}, {Ribeiro},
  {Rimoldini}, {Ripepi}, {Riva}, {Rixon}, {Roelens}, {Romero-G{\'o}mez},
  {Rowell}, {Royer}, {Rudolph}, {Ruiz-Dern}, {Sadowski}, {Sagrist{\`a}
  Sell{\'e}s}, {Sahlmann}, {Salgado}, {Salguero}, {Sarasso}, {Savietto},
  {Schnorhk}, {Schultheis}, {Sciacca}, {Segol}, {Segovia}, {Segransan},
  {Serpell}, {Shih}, {Smareglia}, {Smart}, {Smith}, {Solano}, {Solitro},
  {Sordo}, {Soria Nieto}, {Souchay}, {Spagna}, {Spoto}, {Stampa}, {Steele},
  {Steidelm{\"u}ller}, {Stephenson}, {Stoev}, {Suess}, {S{\"u}veges}, {Surdej},
  {Szabados}, {Szegedi-Elek}, {Tapiador}, {Taris}, {Tauran}, {Taylor},
  {Teixeira}, {Terrett}, {Tingley}, {Trager}, {Turon}, {Ulla}, {Utrilla},
  {Valentini}, {van Elteren}, {Van Hemelryck}, {van Leeuwen}, {Varadi},
  {Vecchiato}, {Veljanoski}, {Via}, {Vicente}, {Vogt}, {Voss}, {Votruba},
  {Voutsinas}, {Walmsley}, {Weiler}, {Weingrill}, {Werner}, {Wevers},
  {Whitehead}, {Wyrzykowski}, {Yoldas}, {{\v{Z}}erjal}, {Zucker}, {Zurbach},
  {Zwitter}, {Alecu}, {Allen}, {Allende Prieto}, {Amorim},
  {Anglada-Escud{\'e}}, {Arsenijevic}, {Azaz}, {Balm}, {Beck}, {Bernstein},
  {Bigot}, {Bijaoui}, {Blasco}, {Bonfigli}, {Bono}, {Boudreault}, {Bressan},
  {Brown}, {Brunet}, {Bunclark}, {Buonanno}, {Butkevich}, {Carret}, {Carrion},
  {Chemin}, {Ch{\'e}reau}, {Corcione}, {Darmigny}, {de Boer}, {de Teodoro}, {de
  Zeeuw}, {Delle Luche}, {Domingues}, {Dubath}, {Fodor}, {Fr{\'e}zouls},
  {Fries}, {Fustes}, {Fyfe}, {Gallardo}, {Gallegos}, {Gardiol}, {Gebran},
  {Gomboc}, {G{\'o}mez}, {Grux}, {Gueguen}, {Heyrovsky}, {Hoar}, {Iannicola},
  {Isasi Parache}, {Janotto}, {Joliet}, {Jonckheere}, {Keil}, {Kim},
  {Klagyivik}, {Klar}, {Knude}, {Kochukhov}, {Kolka}, {Kos}, {Kutka}, {Lainey},
  {LeBouquin}, {Liu}, {Loreggia}, {Makarov}, {Marseille}, {Martayan},
  {Martinez-Rubi}, {Massart}, {Meynadier}, {Mignot}, {Munari}, {Nguyen},
  {Nordlander}, {Ocvirk}, {O'Flaherty}, {Olias Sanz}, {Ortiz}, {Osorio},
  {Oszkiewicz}, {Ouzounis}, {Palmer}, {Park}, {Pasquato}, {Peltzer}, {Peralta},
  {P{\'e}turaud}, {Pieniluoma}, {Pigozzi}, {Poels}, {Prat}, {Prod'homme},
  {Raison}, {Rebordao}, {Risquez}, {Rocca-Volmerange}, {Rosen}, {Ruiz-Fuertes},
  {Russo}, {Sembay}, {Serraller Vizcaino}, {Short}, {Siebert}, {Silva},
  {Sinachopoulos}, {Slezak}, {Soffel}, {Sosnowska}, {Strai{\v{z}}ys}, {ter
  Linden}, {Terrell}, {Theil}, {Tiede}, {Troisi}, {Tsalmantza}, {Tur},
  {Vaccari}, {Vachier}, {Valles}, {Van Hamme}, {Veltz}, {Virtanen}, {Wallut},
  {Wichmann}, {Wilkinson}, {Ziaeepour}, \& {Zschocke}}]{gaia_2016}
{Gaia Collaboration}, {Prusti}, T., {de Bruijne}, J.~H.~J., {et~al.} 2016,
  \aap, 595, A1

\bibitem[{{Gaia Collaboration} {et~al.}(2023){Gaia Collaboration}, {Vallenari,
  A.}, {Brown, A. G. A.}, {Prusti, T.}, {et~al.}}]{GaiaColl2023}
{Gaia Collaboration}, {Vallenari, A.}, {Brown, A. G. A.}, {Prusti, T.},
  {et~al.} 2023, A\&A, 674, A1

\bibitem[{Ginzburg {et~al.}(2018)Ginzburg, Schlichting, \& Sari}]{ginzburg2018}
Ginzburg, S., Schlichting, H.~E., \& Sari, R. 2018, Monthly Notices of the
  Royal Astronomical Society, 476, 759

\bibitem[{{Gomes da Silva} {et~al.}(2018){Gomes da Silva}, {Figueira},
  {Santos}, \& {Faria}}]{Gomes2018}
{Gomes da Silva}, J., {Figueira}, P., {Santos}, N., \& {Faria}, J. 2018, The
  Journal of Open Source Software, 3, 667

\bibitem[{{Gomes da Silva} {et~al.}(2021){Gomes da Silva}, {Santos},
  {Adibekyan}, {Sousa}, {Campante}, {Figueira}, {Bossini}, {Delgado-Mena},
  {Monteiro}, {de Laverny}, {Recio-Blanco}, \& {Lovis}}]{Gomes2021}
{Gomes da Silva}, J., {Santos}, N.~C., {Adibekyan}, V., {et~al.} 2021, \aap,
  646, A77

\bibitem[{{Grunblatt} {et~al.}(2015){Grunblatt}, {Howard}, \&
  {Haywood}}]{Grunblatt2015}
{Grunblatt}, S.~K., {Howard}, A.~W., \& {Haywood}, R.~D. 2015, \apj, 808, 127

\bibitem[{{G{\"u}nther} {et~al.}(2019){G{\"u}nther}, {Pozuelos}, {Dittmann},
  {Dragomir}, {Kane}, {Daylan}, {Feinstein}, {Huang}, {Morton}, {Bonfanti},
  {Bouma}, {Burt}, {Collins}, {Lissauer}, {Matthews}, {Montet}, {Vanderburg},
  {Wang}, {Winters}, {Ricker}, {Vanderspek}, {Latham}, {Seager}, {Winn},
  {Jenkins}, {Armstrong}, {Barkaoui}, {Batalha}, {Bean}, {Caldwell}, {Ciardi},
  {Collins}, {Crossfield}, {Fausnaugh}, {Furesz}, {Gan}, {Gillon}, {Guerrero},
  {Horne}, {Howell}, {Ireland}, {Isopi}, {Jehin}, {Kielkopf}, {Lepine},
  {Mallia}, {Matson}, {Myers}, {Palle}, {Quinn}, {Relles}, {Rojas-Ayala},
  {Schlieder}, {Sefako}, {Shporer}, {Su{\'a}rez}, {Tan}, {Ting}, {Twicken}, \&
  {Waite}}]{Gunther2019}
{G{\"u}nther}, M.~N., {Pozuelos}, F.~J., {Dittmann}, J.~A., {et~al.} 2019,
  Nature Astronomy, 3, 1099

\bibitem[{Gupta \& Schlichting(2019)}]{gupta2019}
Gupta, A. \& Schlichting, H.~E. 2019, Monthly Notices of the Royal Astronomical
  Society, 487, 24

\bibitem[{{Hara} {et~al.}(2017){Hara}, {Bou{\'e}}, {Laskar}, \&
  {Correia}}]{hara2017}
{Hara}, N.~C., {Bou{\'e}}, G., {Laskar}, J., \& {Correia}, A.~C.~M. 2017,
  \mnras, 464, 1220

\bibitem[{{Hart} {et~al.}(2023){Hart}, {Shappee}, {Hey}, {Kochanek}, {Stanek},
  {Lim}, {Dobbs}, {Tucker}, {Jayasinghe}, {Beacom}, {Boright}, {Holoien},
  {Ong}, {Prieto}, {Thompson}, \& {Will}}]{hart2023}
{Hart}, K., {Shappee}, B.~J., {Hey}, D., {et~al.} 2023, arXiv e-prints,
  arXiv:2304.03791

\bibitem[{{Hawthorn} {et~al.}(2023){Hawthorn}, {Bayliss}, {Wilson}, {Bonfanti},
  {Adibekyan}, {Alibert}, {Sousa}, {Collins}, {Bryant}, {Osborn}, {Armstrong},
  {Abe}, {Acton}, {Addison}, {Agabi}, {Alonso}, {Alves}, {Anglada-Escud{\'e}},
  {B{\'a}rczy}, {Barclay}, {Barrado}, {Barros}, {Baumjohann}, {Bendjoya},
  {Benz}, {Bieryla}, {Bonfils}, {Bouchy}, {Brandeker}, {Broeg}, {Brown},
  {Burleigh}, {Buttu}, {Cabrera}, {Caldwell}, {Casewell}, {Charbonneau},
  {Charnoz}, {Cloutier}, {Collier Cameron}, {Collins}, {Conti}, {Crouzet},
  {Czismadia}, {Davies}, {Deleuil}, {Delgado-Mena}, {Delrez}, {Demangeon},
  {Demory}, {Dransfield}, {Dumusque}, {Egger}, {Ehrenreich}, {Eigm{\"u}ller},
  {Erickson}, {Essack}, {Fortier}, {Fossati}, {Fridlund}, {G{\"u}nther},
  {G{\"u}del}, {Gandolfi}, {Gillard}, {Gillon}, {Gnilka}, {Goad}, {Goeke},
  {Guillot}, {Hadjigeorghiou}, {Hellier}, {Henderson}, {Heng}, {Hooton},
  {Horne}, {Howell}, {Hoyer}, {Irwin}, {Jenkins}, {Jenkins}, {Jensen}, {Kane},
  {Kendall}, {Kielkopf}, {Kiss}, {Lacedelli}, {Laskar}, {Latham}, {Etangs},
  {Leleu}, {Lendl}, {Lillo-Box}, {Lovis}, {M{\'e}karnia}, {Massey}, {Masters},
  {Maxted}, {Nascimbeni}, {Nielsen}, {O'Brien}, {Olofsson}, {Osborn}, {Pagano},
  {Pall{\'e}}, {Persson}, {Piotto}, {Plavchan}, {Pollacco}, {Queloz},
  {Ragazzoni}, {Rauer}, {Ribas}, {Ricker}, {S{\'e}gransan}, {Salmon},
  {Santerne}, {Santos}, {Scandariato}, {Schmider}, {Schwarz}, {Seager},
  {Shporer}, {Simon}, {Smith}, {Srdoc}, {Steller}, {Suarez}, {Szab{\'o}},
  {Teske}, {Thomas}, {Tilbrook}, {Triaud}, {Udry}, {Van Grootel}, {Walton},
  {Wang}, {Wheatley}, {Winn}, {Wittenmyer}, \& {Zhang}}]{Hawthorn2023}
{Hawthorn}, F., {Bayliss}, D., {Wilson}, T.~G., {et~al.} 2023, \mnras, 520,
  3649

\bibitem[{{Henden} \& {Munari}(2014)}]{henden_2014}
{Henden}, A. \& {Munari}, U. 2014, Contributions of the Astronomical
  Observatory Skalnate Pleso, 43, 518

\bibitem[{{Hidalgo} {et~al.}(2020){Hidalgo}, {Pall{\'e}}, {Alonso}, {Gandolfi},
  {Fridlund}, {Nowak}, {Luque}, {Hirano}, {Justesen}, {Cochran},
  {Barrag{\'a}n}, {Spina}, {Rodler}, {Albrecht}, {Anderson}, {Amado}, {Bryant},
  {Caballero}, {Cabrera}, {Csizmadia}, {Dai}, {De Leon}, {Deeg}, {Eigmuller},
  {Endl}, {Erikson}, {Esposito}, {Figueira}, {Georgieva}, {Grziwa}, {Guenther},
  {Hatzes}, {Hjorth}, {Hoeijmakers}, {Kabath}, {Korth}, {Kuzuhara}, {Lafarga},
  {Lampon}, {Le{\~a}o}, {Livingston}, {Mathur}, {Monta{\~n}es-Rodriguez},
  {Morales}, {Murgas}, {Nagel}, {Narita}, {Nielsen}, {Patzold}, {Persson},
  {Prieto-Arranz}, {Quirrenbach}, {Rauer}, {Redfield}, {Reiners}, {Ribas},
  {Smith}, {{\v{S}}ubjak}, {Van Eylen}, \& {Wilson}}]{hidalgo2020}
{Hidalgo}, D., {Pall{\'e}}, E., {Alonso}, R., {et~al.} 2020, \aap, 636, A89

\bibitem[{{Hippke} \& {Heller}(2019)}]{hippke2019_TLS}
{Hippke}, M. \& {Heller}, R. 2019, \aap, 623, A39

\bibitem[{{Hirano} {et~al.}(2018){Hirano}, {Dai}, {Gandolfi}, {Fukui},
  {Livingston}, {Miyakawa}, {Endl}, {Cochran}, {Alonso-Floriano}, {Kuzuhara},
  {Montes}, {Ryu}, {Albrecht}, {Barragan}, {Cabrera}, {Csizmadia}, {Deeg},
  {Eigm{\"u}ller}, {Erikson}, {Fridlund}, {Grziwa}, {Guenther}, {Hatzes},
  {Korth}, {Kudo}, {Kusakabe}, {Narita}, {Nespral}, {Nowak}, {P{\"a}tzold},
  {Palle}, {Persson}, {Prieto-Arranz}, {Rauer}, {Ribas}, {Sato}, {Smith},
  {Tamura}, {Tanaka}, {Van Eylen}, \& {Winn}}]{hirano2018}
{Hirano}, T., {Dai}, F., {Gandolfi}, D., {et~al.} 2018, \aj, 155, 127

\bibitem[{{Holmberg} \& {Madhusudhan}(2024)}]{Holmberg24}
{Holmberg}, M. \& {Madhusudhan}, N. 2024, \aap, 683, L2

\bibitem[{{Hord} {et~al.}(2024){Hord}, {Kempton}, {Evans-Soma}, {Latham},
  {Ciardi}, {Dragomir}, {Col{\'o}n}, {Ross}, {Vanderburg}, {de Beurs},
  {Collins}, {Watkins}, {Bean}, {Cowan}, {Daylan}, {Morley}, {Ih}, {Baker},
  {Barkaoui}, {Batalha}, {Behmard}, {Belinski}, {Benkhaldoun}, {Benni},
  {Bernacki}, {Bieryla}, {Binnenfeld}, {Bosch-Cabot}, {Bouchy}, {Bozza},
  {Brahm}, {Buchhave}, {Calkins}, {Chontos}, {Clark}, {Cloutier}, {Cointepas},
  {Collins}, {Conti}, {Crossfield}, {Dai}, {de Leon}, {Dransfield}, {Dressing},
  {Dustor}, {Esquerdo}, {Evans}, {Fajardo-Acosta}, {Fio{\l}ka},
  {For{\'e}s-Toribio}, {Frasca}, {Fukui}, {Fulton}, {Furlan}, {Gan},
  {Gandolfi}, {Ghachoui}, {Giacalone}, {Gilbert}, {Gillon}, {Girardin},
  {Gonzales}, {Grau Horta}, {Gregorio}, {Greklek-McKeon}, {Guerra}, {Hartman},
  {Hellier}, {Helm}, {He{\l}miniak}, {Henning}, {Hill}, {Horne}, {Howard},
  {Howell}, {Huber}, {Isopi}, {Jehin}, {Jenkins}, {Jensen}, {Johnson},
  {Jord{\'a}n}, {Kane}, {Kielkopf}, {Krushinsky}, {Lasota}, {Lee}, {Lewin},
  {Livingston}, {Lubin}, {Lund}, {Mallia}, {Mann}, {Marino}, {Maslennikova},
  {Massey}, {Matson}, {Matthews}, {Mayo}, {Mazeh}, {McLeod}, {Michaels},
  {Mo{\v{c}}nik}, {Mori}, {Mraz}, {Mu{\~n}oz}, {Narita}, {Natarajan},
  {Dyregaard Nielsen}, {Osborn}, {Palle}, {Panahi}, {Papini}, {Plavchan},
  {Polanski}, {Popowicz}, {Pozuelos}, {Quinn}, {Radford}, {Reed}, {Relles},
  {Rice}, {Robertson}, {Rodriguez}, {Rosenthal}, {Rubenzahl}, {Schanche},
  {Schlieder}, {Schwarz}, {Sefako}, {Shporer}, {Sozzetti}, {Srdoc},
  {Stockdale}, {Tarasenkov}, {Tan}, {Timmermans}, {Ting}, {Van Zandt},
  {Vignes}, {Waite}, {Watanabe}, {Weiss}, {Wittrock}, {Zhou}, {Ziegler}, \&
  {Zucker}}]{hord2024}
{Hord}, B.~J., {Kempton}, E. M.~R., {Evans-Soma}, T.~M., {et~al.} 2024, \aj,
  167, 233

\bibitem[{{Hu} {et~al.}(2024){Hu}, {Bello-Arufe}, {Zhang}, {Paragas},
  {Zilinskas}, {van Buchem}, {Bess}, {Patel}, {Ito}, {Damiano}, {Scheucher},
  {Oza}, {Knutson}, {Miguel}, {Dragomir}, {Brandeker}, \& {Demory}}]{Hu2024}
{Hu}, R., {Bello-Arufe}, A., {Zhang}, M., {et~al.} 2024, \nat, 630, 609

\bibitem[{Hu {et~al.}(2021)Hu, Damiano, Scheucher, Kite, Seager, \&
  Rauer}]{Hu_2021}
Hu, R., Damiano, M., Scheucher, M., {et~al.} 2021, The Astrophysical Journal
  Letters, 921, L8

\bibitem[{{Husser} {et~al.}(2013){Husser}, {Wende-von Berg}, {Dreizler},
  {Homeier}, {Reiners}, {Barman}, \& {Hauschildt}}]{husser2013new}
{Husser}, T.~O., {Wende-von Berg}, S., {Dreizler}, S., {et~al.} 2013, \aap,
  553, A6

\bibitem[{Husser {et~al.}(2013)Husser, {Wende-von Berg}, Dreizler, Homeier,
  Reiners, Barman, \& Hauschildt}]{Husser2013}
Husser, T.-O., {Wende-von Berg}, S., Dreizler, S., {et~al.} 2013, A{\&}A, 553,
  A6

\bibitem[{{Ikoma} \& {Hori}(2012)}]{IkomaHori12}
{Ikoma}, M. \& {Hori}, Y. 2012, \apj, 753, 66

\bibitem[{{Izidoro} {et~al.}(2021){Izidoro}, {Bitsch}, {Raymond}, {Johansen},
  {Morbidelli}, {Lambrechts}, \& {Jacobson}}]{izidoro2021}
{Izidoro}, A., {Bitsch}, B., {Raymond}, S.~N., {et~al.} 2021, \aap, 650, A152

\bibitem[{{Izidoro} {et~al.}(2022){Izidoro}, {Schlichting}, {Isella},
  {Dasgupta}, {Zimmermann}, \& {Bitsch}}]{Izidoro22}
{Izidoro}, A., {Schlichting}, H.~E., {Isella}, A., {et~al.} 2022, \apjl, 939,
  L19

\bibitem[{{Jahandar} {et~al.}(2024){Jahandar}, {Doyon}, {Artigau}, {Cook},
  {Cadieux}, {Lafreni{\`e}re}, {Forveille}, {Donati}, {Fouqu{\'e}}, {Carmona},
  {Cloutier}, {Cristofari}, {Gaidos}, {Gomes da Silva}, {Malo}, {Martioli}, {do
  Nascimento}, {Pelletier}, {Vandal}, \& {Venn}}]{Jahandar_2024}
{Jahandar}, F., {Doyon}, R., {Artigau}, {\'E}., {et~al.} 2024, \apj, 966, 56

\bibitem[{{Jeffers} {et~al.}(2022){Jeffers}, {Barnes}, {Sch{\"o}fer},
  {Quirrenbach}, {Zechmeister}, {Amado}, {Caballero}, {Fern{\'a}ndez},
  {Rodr{\'\i}guez}, {Ribas}, {Reiners}, {Cardona Guill{\'e}n}, {Cifuentes},
  {Czesla}, {Hatzes}, {K{\"u}rster}, {Montes}, {Morales}, {Pedraz}, \&
  {Sadegi}}]{jeffers2022}
{Jeffers}, S.~V., {Barnes}, J.~R., {Sch{\"o}fer}, P., {et~al.} 2022, \aap, 663,
  A27

\bibitem[{{Jenkins}(2020)}]{jenkins2020}
{Jenkins}, J.~M. 2020, {Kepler Data Processing Handbook}, Kepler Science
  Document KSCI-19081-003

\bibitem[{Jenkins {et~al.}(2016)Jenkins, Twicken, McCauliff, Campbell,
  Sanderfer, Lung, Mansouri-Samani, Girouard, Tenenbaum, Klaus, Smith,
  Caldwell, Chacon, Henze, Heiges, Latham, Morgan, Swade, Rinehart, \&
  Vanderspek}]{jenkins2016}
Jenkins, J.~M., Twicken, J.~D., McCauliff, S., {et~al.} 2016, in Society of
  Photo-Optical Instrumentation Engineers (SPIE) Conference Series, Vol. 9913,
  Software and Cyberinfrastructure for Astronomy IV, ed. G.~Chiozzi \& J.~C.
  Guzman, International Society for Optics and Photonics (SPIE), 1232 -- 1251

\bibitem[{{Jensen}(2013)}]{jensen2013}
{Jensen}, E. 2013, {Tapir: A web interface for transit/eclipse observability},
  Astrophysics Source Code Library, record ascl:1306.007

\bibitem[{{Kaltenegger} {et~al.}(2019){Kaltenegger}, {Pepper}, {Stassun}, \&
  {Oelkers}}]{kaltenegger2019}
{Kaltenegger}, L., {Pepper}, J., {Stassun}, K., \& {Oelkers}, R. 2019, \apjl,
  874, L8

\bibitem[{{Keles} {et~al.}(2022){Keles}, {Mallonn}, {Kitzmann}, {Poppenhaeger},
  {Hoeijmakers}, {Ilyin}, {Alexoudi}, {Carroll}, {Alvarado-Gomez}, {Ketzer},
  {Bonomo}, {Borsa}, {Gaudi}, {Henning}, {Malavolta}, {Molaverdikhani},
  {Nascimbeni}, {Patience}, {Pino}, {Scandariato}, {Schlawin}, {Shkolnik},
  {Sicilia}, {Sozzetti}, {Foster}, {Veillet}, {Wang}, {Yan}, \&
  {Strassmeier}}]{keles2022}
{Keles}, E., {Mallonn}, M., {Kitzmann}, D., {et~al.} 2022, \mnras, 513, 1544

\bibitem[{{Kempton} {et~al.}(2018){Kempton}, {Bean}, {Louie}, {Deming}, {Koll},
  {Mansfield}, {Christiansen}, {L{\'o}pez-Morales}, {Swain}, {Zellem},
  {Ballard}, {Barclay}, {Barstow}, {Batalha}, {Beatty}, {Berta-Thompson},
  {Birkby}, {Buchhave}, {Charbonneau}, {Cowan}, {Crossfield}, {de Val-Borro},
  {Doyon}, {Dragomir}, {Gaidos}, {Heng}, {Hu}, {Kane}, {Kreidberg}, {Mallonn},
  {Morley}, {Narita}, {Nascimbeni}, {Pall{\'e}}, {Quintana}, {Rauscher},
  {Seager}, {Shkolnik}, {Sing}, {Sozzetti}, {Stassun}, {Valenti}, \& {von
  Essen}}]{kempton2018}
{Kempton}, E. M.~R., {Bean}, J.~L., {Louie}, D.~R., {et~al.} 2018, \pasp, 130,
  114401

\bibitem[{{Kipping}(2013)}]{Kipping2013}
{Kipping}, D.~M. 2013, \mnras, 435, 2152

\bibitem[{{Kite} {et~al.}(2019){Kite}, {Fegley}, {Schaefer}, \&
  {Ford}}]{Kite19}
{Kite}, E.~S., {Fegley}, Bruce, J., {Schaefer}, L., \& {Ford}, E.~B. 2019,
  \apjl, 887, L33

\bibitem[{{Kochanek} {et~al.}(2017){Kochanek}, {Shappee}, {Stanek}, {Holoien},
  {Thompson}, {Prieto}, {Dong}, {Shields}, {Will}, {Britt}, {Perzanowski}, \&
  {Pojma{\'n}ski}}]{Kochanek2017}
{Kochanek}, C.~S., {Shappee}, B.~J., {Stanek}, K.~Z., {et~al.} 2017, \pasp,
  129, 104502

\bibitem[{{Korth} {et~al.}(2019){Korth}, {Csizmadia}, {Gandolfi}, {Fridlund},
  {P{\"a}tzold}, {Hirano}, {Livingston}, {Persson}, {Deeg}, {Justesen},
  {Barrag{\'a}n}, {Grziwa}, {Endl}, {Tronsgaard}, {Dai}, {Cochran}, {Albrecht},
  {Alonso}, {Cabrera}, {Cauley}, {Cusano}, {Eigm{\"u}ller}, {Erikson},
  {Esposito}, {Guenther}, {Hatzes}, {Hidalgo}, {Kuzuhara}, {Monta{\~n}es},
  {Napolitano}, {Narita}, {Niraula}, {Nespral}, {Nowak}, {Palle}, {Petrillo},
  {Redfield}, {Prieto-Arranz}, {Rauer}, {Smith}, {Tortora}, {Van Eylen}, \&
  {Winn}}]{Korth2019}
{Korth}, J., {Csizmadia}, S., {Gandolfi}, D., {et~al.} 2019, \mnras, 482, 1807

\bibitem[{{Kreidberg}(2015)}]{Kreidberg2015}
{Kreidberg}, L. 2015, \pasp, 127, 1161

\bibitem[{{Lacedelli} {et~al.}(2022){Lacedelli}, {Wilson}, {Malavolta},
  {Hooton}, {Collier Cameron}, {Alibert}, {Mortier}, {Bonfanti}, {Haywood},
  {Hoyer}, {Piotto}, {Bekkelien}, {Vanderburg}, {Benz}, {Dumusque}, {Deline},
  {L{\'o}pez-Morales}, {Borsato}, {Rice}, {Fossati}, {Latham}, {Brandeker},
  {Poretti}, {Sousa}, {Sozzetti}, {Salmon}, {Burke}, {Van Grootel},
  {Fausnaugh}, {Adibekyan}, {Huang}, {Osborn}, {Mustill}, {Pall{\'e}},
  {Bourrier}, {Nascimbeni}, {Alonso}, {Anglada}, {B{\'a}rczy}, {Barrado y
  Navascues}, {Barros}, {Baumjohann}, {Beck}, {Beck}, {Billot}, {Bonfils},
  {Broeg}, {Buchhave}, {Cabrera}, {Charnoz}, {Cosentino}, {Csizmadia},
  {Davies}, {Deleuil}, {Delrez}, {Demangeon}, {Demory}, {Ehrenreich},
  {Erikson}, {Esparza-Borges}, {Flor{\'e}n}, {Fortier}, {Fridlund}, {Futyan},
  {Gandolfi}, {Ghedina}, {Gillon}, {G{\"u}del}, {Guterman}, {Harutyunyan},
  {Heng}, {Isaak}, {Jenkins}, {Kiss}, {Laskar}, {Lecavelier des Etangs},
  {Lendl}, {Lovis}, {Magrin}, {Marafatto}, {Martinez Fiorenzano}, {Maxted},
  {Mayor}, {Micela}, {Molinari}, {Murgas}, {Narita}, {Olofsson}, {Ottensamer},
  {Pagano}, {Pasetti}, {Pedani}, {Pepe}, {Peter}, {Phillips}, {Pollacco},
  {Queloz}, {Ragazzoni}, {Rando}, {Ratti}, {Rauer}, {Ribas}, {Santos},
  {Sasselov}, {Scandariato}, {Seager}, {S{\'e}gransan}, {Serrano}, {Simon},
  {Smith}, {Steinberger}, {Steller}, {Szab{\'o}}, {Thomas}, {Twicken}, {Udry},
  {Walton}, \& {Winn}}]{lacedelli2022}
{Lacedelli}, G., {Wilson}, T.~G., {Malavolta}, L., {et~al.} 2022, \mnras, 511,
  4551

\bibitem[{{Langellier} {et~al.}(2021){Langellier}, {Milbourne}, {Phillips},
  {Haywood}, {Saar}, {Mortier}, {Malavolta}, {Thompson}, {Collier Cameron},
  {Dumusque}, {Cegla}, {Latham}, {Maldonado}, {Watson}, {Buchschacher},
  {Cecconi}, {Charbonneau}, {Cosentino}, {Ghedina}, {Gonzalez}, {Li}, {Lodi},
  {L{\'o}pez-Morales}, {Micela}, {Molinari}, {Pepe}, {Poretti}, {Rice},
  {Sasselov}, {Sozzetti}, {Udry}, \& {Walsworth}}]{langellier2021}
{Langellier}, N., {Milbourne}, T.~W., {Phillips}, D.~F., {et~al.} 2021, \aj,
  161, 287

\bibitem[{{Lee} \& {Chiang}(2016)}]{LeeChiang16}
{Lee}, E.~J. \& {Chiang}, E. 2016, \apj, 817, 90

\bibitem[{{Lee} {et~al.}(2014){Lee}, {Chiang}, \& {Ormel}}]{Lee14}
{Lee}, E.~J., {Chiang}, E., \& {Ormel}, C.~W. 2014, \apj, 797, 95

\bibitem[{{Lee} {et~al.}(2022){Lee}, {Karalis}, \& {Thorngren}}]{Lee22}
{Lee}, E.~J., {Karalis}, A., \& {Thorngren}, D.~P. 2022, \apj, 941, 186

\bibitem[{{L{\'e}ger} {et~al.}(2004){L{\'e}ger}, {Selsis}, {Sotin}, {Guillot},
  {Despois}, {Mawet}, {Ollivier}, {Lab{\`e}que}, {Valette}, {Brachet},
  {Chazelas}, \& {Lammer}}]{Leger04}
{L{\'e}ger}, A., {Selsis}, F., {Sotin}, C., {et~al.} 2004, \icarus, 169, 499

\bibitem[{{Lichtenberg} \& {Miguel}(2025)}]{Lichetnberg24}
{Lichtenberg}, T. \& {Miguel}, Y. 2025, Treatise on Geochemistry, 7, 51

\bibitem[{{Lissauer} {et~al.}(2011){Lissauer}, {Ragozzine}, {Fabrycky},
  {Steffen}, {Ford}, {Jenkins}, {Shporer}, {Holman}, {Rowe}, {Quintana},
  {Batalha}, {Borucki}, {Bryson}, {Caldwell}, {Carter}, {Ciardi}, {Dunham},
  {Fortney}, {Gautier}, {Howell}, {Koch}, {Latham}, {Marcy}, {Morehead}, \&
  {Sasselov}}]{Lissauer11}
{Lissauer}, J.~J., {Ragozzine}, D., {Fabrycky}, D.~C., {et~al.} 2011, \apjs,
  197, 8

\bibitem[{{Liu} {et~al.}(2017){Liu}, {Ormel}, \& {Lin}}]{Liu17}
{Liu}, B., {Ormel}, C.~W., \& {Lin}, D. N.~C. 2017, \aap, 601, A15

\bibitem[{{Livingston} {et~al.}(2018){Livingston}, {Endl}, {Dai}, {Cochran},
  {Barragan}, {Gandolfi}, {Hirano}, {Grziwa}, {Smith}, {Albrecht}, {Cabrera},
  {Csizmadia}, {de Leon}, {Deeg}, {Eigm{\"u}ller}, {Erikson}, {Everett},
  {Fridlund}, {Fukui}, {Guenther}, {Hatzes}, {Howell}, {Korth}, {Narita},
  {Nespral}, {Nowak}, {Palle}, {P{\"a}tzold}, {Persson}, {Prieto-Arranz},
  {Rauer}, {Tamura}, {Van Eylen}, \& {Winn}}]{livingstone2018}
{Livingston}, J.~H., {Endl}, M., {Dai}, F., {et~al.} 2018, \aj, 156, 78

\bibitem[{{Lodders}(2003)}]{Lodders03}
{Lodders}, K. 2003, \apj, 591, 1220

\bibitem[{{Lopez} \& {Fortney}(2013)}]{LopezFortney13}
{Lopez}, E.~D. \& {Fortney}, J.~J. 2013, \apj, 776, 2

\bibitem[{{Lovis} \& {Pepe}(2007)}]{Lovis_2007}
{Lovis}, C. \& {Pepe}, F. 2007, \aap, 468, 1115

\bibitem[{Luo {et~al.}(2024)Luo, Dorn, \& Deng}]{Luo24}
Luo, H., Dorn, C., \& Deng, J. 2024, Nature Astronomy, 8, 1399

\bibitem[{{Luque} {et~al.}(2022){Luque}, {Nowak}, {Hirano}, {Kossakowski},
  {Pall{\'e}}, {Nixon}, {Morello}, {Amado}, {Albrecht}, {Caballero},
  {Cifuentes}, {Cochran}, {Deeg}, {Dreizler}, {Esparza-Borges}, {Fukui},
  {Gandolfi}, {Goffo}, {Guenther}, {Hatzes}, {Henning}, {Kabath}, {Kawauchi},
  {Korth}, {Kotani}, {Kudo}, {Kuzuhara}, {Lafarga}, {Lam}, {Livingston},
  {Morales}, {Muresan}, {Murgas}, {Narita}, {Osborne}, {Parviainen},
  {Passegger}, {Persson}, {Quirrenbach}, {Redfield}, {Reffert}, {Reiners},
  {Ribas}, {Serrano}, {Tamura}, {Van Eylen}, {Watanabe}, \& {Zapatero
  Osorio}}]{Luque22}
{Luque}, R., {Nowak}, G., {Hirano}, T., {et~al.} 2022, \aap, 666, A154

\bibitem[{{Luque} {et~al.}(2019){Luque}, {Nowak}, {Pall{\'e}}, {Dai},
  {Kaminski}, {Nagel}, {Hidalgo}, {Bauer}, {Lafarga}, {Livingston},
  {Barrag{\'a}n}, {Hirano}, {Fridlund}, {Gandolfi}, {Justesen}, {Hjorth}, {Van
  Eylen}, {Winn}, {Esposito}, {Morales}, {Albrecht}, {Alonso}, {Amado}, {Beck},
  {Caballero}, {Cabrera}, {Cochran}, {Csizmadia}, {Deeg}, {Eigm{\"u}ller},
  {Endl}, {Erikson}, {Fukui}, {Grziwa}, {Guenther}, {Hatzes}, {Knudstrup},
  {Korth}, {Lam}, {Lund}, {Mathur}, {Monta{\~n}es-Rodr{\'\i}guez}, {Narita},
  {Nespral}, {Niraula}, {P{\"a}tzold}, {Persson}, {Prieto-Arranz},
  {Quirrenbach}, {Rauer}, {Redfield}, {Reiners}, {Ribas}, \&
  {Smith}}]{luque2019}
{Luque}, R., {Nowak}, G., {Pall{\'e}}, E., {et~al.} 2019, \aap, 623, A114

\bibitem[{{Luque} \& {Pall{\'e}}(2022)}]{luquepalle2022}
{Luque}, R. \& {Pall{\'e}}, E. 2022, Science, 377, 1211

\bibitem[{{Luque} {et~al.}(2021){Luque}, {Serrano}, {Molaverdikhani}, {Nixon},
  {Livingston}, {Guenther}, {Pall{\'e}}, {Madhusudhan}, {Nowak}, {Korth},
  {Cochran}, {Hirano}, {Chaturvedi}, {Goffo}, {Albrecht}, {Barrag{\'a}n},
  {Brice{\~n}o}, {Cabrera}, {Charbonneau}, {Cloutier}, {Collins}, {Collins},
  {Col{\'o}n}, {Crossfield}, {Csizmadia}, {Dai}, {Deeg}, {Esposito},
  {Fridlund}, {Gandolfi}, {Georgieva}, {Glidden}, {Goeke}, {Grziwa}, {Hatzes},
  {Henze}, {Howell}, {Irwin}, {Jenkins}, {Jensen}, {K{\'a}bath}, {Kidwell},
  {Kielkopf}, {Knudstrup}, {Lam}, {Latham}, {Lissauer}, {Mann}, {Matthews},
  {Mireles}, {Narita}, {Paegert}, {Persson}, {Redfield}, {Ricker}, {Rodler},
  {Schlieder}, {Scott}, {Seager}, {{\v{S}}ubjak}, {Tan}, {Ting}, {Vanderspek},
  {Van Eylen}, {Winn}, \& {Ziegler}}]{Luque21}
{Luque}, R., {Serrano}, L.~M., {Molaverdikhani}, K., {et~al.} 2021, \aap, 645,
  A41

\bibitem[{{MacDougall} {et~al.}(2023){MacDougall}, {Petigura}, {Gilbert},
  {Angelo}, {Batalha}, {Beard}, {Behmard}, {Blunt}, {Brinkman}, {Chontos},
  {Crossfield}, {Dai}, {Dalba}, {Dressing}, {Fetherolf}, {Fulton}, {Giacalone},
  {Hill}, {Holcomb}, {Howard}, {Huber}, {Isaacson}, {Kane}, {Kosiarek},
  {Lubin}, {Mayo}, {Mo{\v{c}}nik}, {Akana Murphy}, {Pidhorodetska}, {Polanski},
  {Rice}, {Robertson}, {Rosenthal}, {Roy}, {Rubenzahl}, {Scarsdale},
  {Turtelboom}, {Tyler}, {Van Zandt}, {Weiss}, \& {Yee}}]{MacDougall2023}
{MacDougall}, M.~G., {Petigura}, E.~A., {Gilbert}, G.~J., {et~al.} 2023, \aj,
  166, 33

\bibitem[{Madhusudhan {et~al.}(2023)Madhusudhan, Sarkar, Constantinou,
  Holmberg, Piette, \& Moses}]{Madhusudhan_2023}
Madhusudhan, N., Sarkar, S., Constantinou, S., {et~al.} 2023, The Astrophysical
  Journal Letters, 956, L13

\bibitem[{{Malavolta} {et~al.}(2018){Malavolta}, {Mayo}, {Louden}, {Rajpaul},
  {Bonomo}, {Buchhave}, {Kreidberg}, {Kristiansen}, {Lopez-Morales}, {Mortier},
  {Vand erburg}, {Coffinet}, {Ehrenreich}, {Lovis}, {Bouchy}, {Charbonneau},
  {Ciardi}, {Collier Cameron}, {Cosentino}, {Crossfield}, {Damasso},
  {Dressing}, {Dumusque}, {Everett}, {Figueira}, {Fiorenzano}, {Gonzales},
  {Haywood}, {Harutyunyan}, {Hirsch}, {Howell}, {Johnson}, {Latham}, {Lopez},
  {Mayor}, {Micela}, {Molinari}, {Nascimbeni}, {Pepe}, {Phillips}, {Piotto},
  {Rice}, {Sasselov}, {S{\'e}gransan}, {Sozzetti}, {Udry}, \&
  {Watson}}]{Malavolta2018}
{Malavolta}, L., {Mayo}, A.~W., {Louden}, T., {et~al.} 2018, \aj, 155, 107

\bibitem[{{Malavolta} {et~al.}(2016){Malavolta}, {Nascimbeni}, {Piotto},
  {Quinn}, {Borsato}, {Granata}, {Bonomo}, {Marzari}, {Bedin}, {Rainer},
  {Desidera}, {Lanza}, {Poretti}, {Sozzetti}, {White}, {Latham}, {Cunial},
  {Libralato}, {Nardiello}, {Boccato}, {Claudi}, {Cosentino}, {Covino},
  {Gratton}, {Maggio}, {Micela}, {Molinari}, {Pagano}, {Smareglia}, {Affer},
  {Andreuzzi}, {Aparicio}, {Benatti}, {Bignamini}, {Borsa}, {Damasso}, {Di
  Fabrizio}, {Harutyunyan}, {Esposito}, {Fiorenzano}, {Gandolfi}, {Giacobbe},
  {Gonz{\'a}lez Hern{\'a}ndez}, {Maldonado}, {Masiero}, {Molinaro}, {Pedani},
  \& {Scandariato}}]{Malavolta2016}
{Malavolta}, L., {Nascimbeni}, V., {Piotto}, G., {et~al.} 2016, \aap, 588, A118

\bibitem[{{Marboeuf} {et~al.}(2014){Marboeuf}, {Thiabaud}, {Alibert}, {Cabral},
  \& {Benz}}]{marboeuf2014}
{Marboeuf}, U., {Thiabaud}, A., {Alibert}, Y., {Cabral}, N., \& {Benz}, W.
  2014, \aap, 570, A36

\bibitem[{{Marcy} {et~al.}(2014{\natexlab{a}}){Marcy}, {Isaacson}, {Howard},
  {Rowe}, {Jenkins}, {Bryson}, {Latham}, {Howell}, {Gautier}, {Batalha},
  {Rogers}, {Ciardi}, {Fischer}, {Gilliland}, {Kjeldsen},
  {Christensen-Dalsgaard}, {Huber}, {Chaplin}, {Basu}, {Buchhave}, {Quinn},
  {Borucki}, {Koch}, {Hunter}, {Caldwell}, {Van Cleve}, {Kolbl}, {Weiss},
  {Petigura}, {Seager}, {Morton}, {Johnson}, {Ballard}, {Burke}, {Cochran},
  {Endl}, {MacQueen}, {Everett}, {Lissauer}, {Ford}, {Torres}, {Fressin},
  {Brown}, {Steffen}, {Charbonneau}, {Basri}, {Sasselov}, {Winn},
  {Sanchis-Ojeda}, {Christiansen}, {Adams}, {Henze}, {Dupree}, {Fabrycky},
  {Fortney}, {Tarter}, {Holman}, {Tenenbaum}, {Shporer}, {Lucas}, {Welsh},
  {Orosz}, {Bedding}, {Campante}, {Davies}, {Elsworth}, {Handberg}, {Hekker},
  {Karoff}, {Kawaler}, {Lund}, {Lundkvist}, {Metcalfe}, {Miglio}, {Silva
  Aguirre}, {Stello}, {White}, {Boss}, {Devore}, {Gould}, {Prsa}, {Agol},
  {Barclay}, {Coughlin}, {Brugamyer}, {Mullally}, {Quintana}, {Still},
  {Thompson}, {Morrison}, {Twicken}, {D{\'e}sert}, {Carter}, {Crepp},
  {H{\'e}brard}, {Santerne}, {Moutou}, {Sobeck}, {Hudgins}, {Haas},
  {Robertson}, {Lillo-Box}, \& {Barrado}}]{marcy2014b}
{Marcy}, G.~W., {Isaacson}, H., {Howard}, A.~W., {et~al.} 2014{\natexlab{a}},
  \apjs, 210, 20

\bibitem[{{Marcy} {et~al.}(2014{\natexlab{b}}){Marcy}, {Weiss}, {Petigura},
  {Isaacson}, {Howard}, \& {Buchhave}}]{marcy2014}
{Marcy}, G.~W., {Weiss}, L.~M., {Petigura}, E.~A., {et~al.} 2014{\natexlab{b}},
  Proceedings of the National Academy of Science, 111, 12655

\bibitem[{{Marfil} {et~al.}(2021){Marfil}, {Tabernero}, {Montes}, {Caballero},
  {L{\'a}zaro}, {Gonz{\'a}lez Hern{\'a}ndez}, {Nagel}, {Passegger},
  {Schweitzer}, {Ribas}, {Reiners}, {Quirrenbach}, {Amado}, {Cifuentes},
  {Cort{\'e}s-Contreras}, {Dreizler}, {Duque-Arribas},
  {Galad{\'\i}-Enr{\'\i}quez}, {Henning}, {Jeffers}, {Kaminski}, {K{\"u}rster},
  {Lafarga}, {L{\'o}pez-Gallifa}, {Morales}, {Shan}, \& {Zechmeister}}]{mar21}
{Marfil}, E., {Tabernero}, H.~M., {Montes}, D., {et~al.} 2021, \aap, 656, A162

\bibitem[{{Mayor} {et~al.}(2011){Mayor}, {Marmier}, {Lovis}, {Udry},
  {S{\'e}gransan}, {Pepe}, {Benz}, {Bertaux}, {Bouchy}, {Dumusque}, {Lo Curto},
  {Mordasini}, {Queloz}, \& {Santos}}]{Mayor11}
{Mayor}, M., {Marmier}, M., {Lovis}, C., {et~al.} 2011, arXiv e-prints,
  arXiv:1109.2497

\bibitem[{{McCully} {et~al.}(2018){McCully}, {Volgenau}, {Harbeck}, {Lister},
  {Saunders}, {Turner}, {Siiverd}, \& {Bowman}}]{McCully2018}
{McCully}, C., {Volgenau}, N.~H., {Harbeck}, D.-R., {et~al.} 2018, in Society
  of Photo-Optical Instrumentation Engineers (SPIE) Conference Series, Vol.
  10707, Software and Cyberinfrastructure for Astronomy V, ed. J.~C. {Guzman}
  \& J.~{Ibsen}, 107070K

\bibitem[{{Mikal-Evans} {et~al.}(2023){Mikal-Evans}, {Madhusudhan}, {Dittmann},
  {G{\"u}nther}, {Welbanks}, {Van Eylen}, {Crossfield}, {Daylan}, \&
  {Kreidberg}}]{Mikal_evans23}
{Mikal-Evans}, T., {Madhusudhan}, N., {Dittmann}, J., {et~al.} 2023, \aj, 165,
  84

\bibitem[{{Mistry} {et~al.}(2024){Mistry}, {Prasad}, {Maity}, {Pathak},
  {Gharat}, {Lekkas}, {Bhattarai}, {Kumar}, {Lissauer}, {Twicken}, {Soubkiou},
  {Pozuelos}, {Jenkins}, {Horne}, {Giacalone}, {Barkaoui}, {Timmermans},
  {Watkins}, {Sefako}, {Collins}, {Ciardi}, {Clark}, {Safonov}, {Shporer},
  {Schlieder}, {Benkhaldoun}, {Stockdale}, {Ziegler}, {Gilbert},
  {Emmanu{\"e}l}, {Murgas}, {Crossfield}, {Paegert}, {Lund}, {Narita},
  {Schwarz}, {Goeke}, {Fajardo-Acosta}, {Howell}, {Tan}, {Barclay}, \&
  {Kawai}}]{mistry2024}
{Mistry}, P., {Prasad}, A., {Maity}, M., {et~al.} 2024, \pasa, 41, e030

\bibitem[{{Mousis} {et~al.}(2020){Mousis}, {Deleuil}, {Aguichine}, {Marcq},
  {Naar}, {Aguirre}, {Brugger}, \& {Gon{\c{c}}alves}}]{Mousis20}
{Mousis}, O., {Deleuil}, M., {Aguichine}, A., {et~al.} 2020, \apjl, 896, L22

\bibitem[{{Mulders} {et~al.}(2018){Mulders}, {Pascucci}, {Apai}, \&
  {Ciesla}}]{Mulders18}
{Mulders}, G.~D., {Pascucci}, I., {Apai}, D., \& {Ciesla}, F.~J. 2018, \aj,
  156, 24

\bibitem[{{Ogihara} {et~al.}(2020){Ogihara}, {Kunitomo}, \& {Hori}}]{Ogihara20}
{Ogihara}, M., {Kunitomo}, M., \& {Hori}, Y. 2020, \apj, 899, 91

\bibitem[{{Osborn} {et~al.}(2021){Osborn}, {Armstrong}, {Cale}, {Brahm},
  {Wittenmyer}, {Dai}, {Crossfield}, {Bryant}, {Adibekyan}, {Cloutier},
  {Collins}, {Delgado Mena}, {Fridlund}, {Hellier}, {Howell}, {King},
  {Lillo-Box}, {Otegi}, {Sousa}, {Stassun}, {Matthews}, {Ziegler}, {Ricker},
  {Vanderspek}, {Latham}, {Seager}, {Winn}, {Jenkins}, {Acton}, {Addison},
  {Anderson}, {Ballard}, {Barrado}, {Barros}, {Batalha}, {Bayliss}, {Barclay},
  {Benneke}, {Berberian}, {Bouchy}, {Bowler}, {Brice{\~n}o}, {Burke},
  {Burleigh}, {Casewell}, {Ciardi}, {Collins}, {Cooke}, {Demangeon},
  {D{\'\i}az}, {Dorn}, {Dragomir}, {Dressing}, {Dumusque}, {Espinoza},
  {Figueira}, {Fulton}, {Furlan}, {Gaidos}, {Geneser}, {Gill}, {Goad},
  {Gonzales}, {Gorjian}, {G{\"u}nther}, {Helled}, {Henderson}, {Henning},
  {Hogan}, {Hojjatpanah}, {Horner}, {Howard}, {Hoyer}, {Huber}, {Isaacson},
  {Jenkins}, {Jensen}, {Jord{\'a}n}, {Kane}, {Kidwell}, {Kielkopf}, {Law},
  {Lendl}, {Lund}, {Matson}, {Mann}, {McCormac}, {Mengel}, {Morales},
  {Nielsen}, {Okumura}, {Osborn}, {Petigura}, {Plavchan}, {Pollacco},
  {Quintana}, {Raynard}, {Robertson}, {Rose}, {Roy}, {Reefe}, {Santerne},
  {Santos}, {Sarkis}, {Schlieder}, {Schwarz}, {Scott}, {Shporer}, {Smith},
  {Stibbard}, {Stockdale}, {Str{\o}m}, {Twicken}, {Tan}, {Tanner}, {Teske},
  {Tilbrook}, {Tinney}, {Udry}, {Villase{\~n}or}, {Vines}, {Wang}, {Weiss},
  {West}, {Wheatley}, {Wright}, {Zhang}, \& {Zohrabi}}]{Osborn2021}
{Osborn}, A., {Armstrong}, D.~J., {Cale}, B., {et~al.} 2021, \mnras, 507, 2782

\bibitem[{Osborne {et~al.}(2023)Osborne, Van Eylen, Goffo, Gandolfi, Nowak,
  Persson, Livingston, Weeks, Pallé, Luque, Hellier, Carleo, Redfield, Hirano,
  Garbaccio-Gili, Alarcon, Barragán, Casasayas-Barris, Díaz, Esposito,
  Knudstrup, Jenkins, Murgas, Orell-Miquel, Rodler, Serrano, Stangret,
  Albrecht, Alqasim, Cochran, Deeg, Fridlund, Hatzes, Korth, \&
  Lam}]{osborne2023}
Osborne, H. L.~M., Van Eylen, V., Goffo, E., {et~al.} 2023, Monthly Notices of
  the Royal Astronomical Society, 527, 11138

\bibitem[{{Otegi} {et~al.}(2020){Otegi}, {Bouchy}, \& {Helled}}]{Otegi20}
{Otegi}, J.~F., {Bouchy}, F., \& {Helled}, R. 2020, \aap, 634, A43

\bibitem[{{Owen} \& {Wu}(2013)}]{OwenWu13}
{Owen}, J.~E. \& {Wu}, Y. 2013, \apj, 775, 105

\bibitem[{{Owen} \& {Wu}(2016)}]{OwenWu16}
{Owen}, J.~E. \& {Wu}, Y. 2016, \apj, 817, 107

\bibitem[{{Panichi} {et~al.}(2017){Panichi}, {Go{\'z}dziewski}, \&
  {Turchetti}}]{Panichi_2017}
{Panichi}, F., {Go{\'z}dziewski}, K., \& {Turchetti}, G. 2017, \mnras, 468, 469

\bibitem[{{Parc} {et~al.}(2024){Parc}, {Bouchy}, {Venturini}, {Dorn}, \&
  {Helled}}]{parc2024}
{Parc}, L., {Bouchy}, F., {Venturini}, J., {Dorn}, C., \& {Helled}, R. 2024,
  \aap, 688, A59

\bibitem[{Parviainen(2015)}]{Parviainen2015_Pytransit}
Parviainen, H. 2015, MNRAS, 450, 3233

\bibitem[{Parviainen \& Aigrain(2015)}]{Parviainen2015}
Parviainen, H. \& Aigrain, S. 2015, MNRAS, 453, 3821

\bibitem[{{Parviainen} {et~al.}(2024{\natexlab{a}}){Parviainen}, {Luque}, \&
  {Palle}}]{spright}
{Parviainen}, H., {Luque}, R., \& {Palle}, E. 2024{\natexlab{a}}, \mnras, 527,
  5693

\bibitem[{{Parviainen} {et~al.}(2024{\natexlab{b}}){Parviainen}, {Luque}, \&
  {Palle}}]{parviainen2024}
{Parviainen}, H., {Luque}, R., \& {Palle}, E. 2024{\natexlab{b}}, \mnras, 527,
  5693

\bibitem[{{Patel} {et~al.}(2023){Patel}, {Egger}, {Wilson}, {Bourrier},
  {Carone}, {Beck}, {Ehrenreich}, {Sousa}, {Benz}, {Brandeker}, {Deline},
  {Alibert}, {Lam}, {Lendl}, {Alonso}, {Anglada}, {B{\'a}rczy}, {Barrado},
  {Barros}, {Baumjohann}, {Beck}, {Billot}, {Bonfils}, {Broeg}, {Busch},
  {Cabrera}, {Charnoz}, {Collier Cameron}, {Csizmadia}, {Davies}, {Deleuil},
  {Delrez}, {Demangeon}, {Demory}, {Erikson}, {Fortier}, {Fossati}, {Fridlund},
  {Gandolfi}, {Gillon}, {G{\"u}del}, {Heng}, {Hoyer}, {Isaak}, {Kiss}, {Kopp},
  {Laskar}, {Lecavelier des Etangs}, {Lovis}, {Magrin}, {Maxted}, {Nascimbeni},
  {Olofsson}, {Ottensamer}, {Pagano}, {Pall{\'e}}, {Peter}, {Piotto},
  {Pollacco}, {Queloz}, {Ragazzoni}, {Rando}, {Ratti}, {Rauer}, {Ribas},
  {Santos}, {Scandariato}, {S{\'e}gransan}, {Simon}, {Smith}, {Steller},
  {Szab{\'o}}, {Thomas}, {Udry}, {Ulmer}, {Van Grootel}, {Viotto}, \&
  {Walton}}]{patel2023}
{Patel}, J.~A., {Egger}, J.~A., {Wilson}, T.~G., {et~al.} 2023, \aap, 679, A92

\bibitem[{{Pepe} {et~al.}(2021){Pepe}, {Cristiani}, {Rebolo}, {Santos},
  {Dekker}, {Cabral}, {Di Marcantonio}, {Figueira}, {Lo Curto}, {Lovis},
  {Mayor}, {M{\'e}gevand}, {Molaro}, {Riva}, {Zapatero Osorio}, {Amate},
  {Manescau}, {Pasquini}, {Zerbi}, {Adibekyan}, {Abreu}, {Affolter}, {Alibert},
  {Aliverti}, {Allart}, {Allende Prieto}, {{\'A}lvarez}, {Alves}, {Avila},
  {Baldini}, {Bandy}, {Barros}, {Benz}, {Bianco}, {Borsa}, {Bourrier},
  {Bouchy}, {Broeg}, {Calderone}, {Cirami}, {Coelho}, {Conconi}, {Coretti},
  {Cumani}, {Cupani}, {D'Odorico}, {Damasso}, {Deiries}, {Delabre},
  {Demangeon}, {Dumusque}, {Ehrenreich}, {Faria}, {Fragoso}, {Genolet},
  {Genoni}, {G{\'e}nova Santos}, {Gonz{\'a}lez Hern{\'a}ndez}, {Hughes},
  {Iwert}, {Kerber}, {Knudstrup}, {Landoni}, {Lavie}, {Lillo-Box}, {Lizon},
  {Maire}, {Martins}, {Mehner}, {Micela}, {Modigliani}, {Monteiro}, {Monteiro},
  {Moschetti}, {Murphy}, {Nunes}, {Oggioni}, {Oliveira}, {Oshagh}, {Pall{\'e}},
  {Pariani}, {Poretti}, {Rasilla}, {Rebord{\~a}o}, {Redaelli}, {Santana
  Tschudi}, {Santin}, {Santos}, {S{\'e}gransan}, {Schmidt}, {Segovia},
  {Sosnowska}, {Sozzetti}, {Sousa}, {Span{\`o}}, {Su{\'a}rez Mascare{\~n}o},
  {Tabernero}, {Tenegi}, {Udry}, \& {Zanutta}}]{pepe2021}
{Pepe}, F., {Cristiani}, S., {Rebolo}, R., {et~al.} 2021, \aap, 645, A96

\bibitem[{{Pepe} {et~al.}(2002){Pepe}, {Mayor}, {Galland}, {Naef}, {Queloz},
  {Santos}, {Udry}, \& {Burnet}}]{Pepe2002}
{Pepe}, F., {Mayor}, M., {Galland}, F., {et~al.} 2002, \aap, 388, 632

\bibitem[{Pepe {et~al.}(2002)Pepe, Mayor, Rupprecht, Avila, Ballester, Beckers,
  Benz, Bertaux, Bouchy, Buzzoni, Cavadore, Deiries, Dekker, Delabre,
  D'Odorico, Eckert, Fischer, Fleury, George, Gilliotte, Gojak, Guzman, Koch,
  Kohler, Kotzlowski, Lacroix, Le~Merrer, Lizon, Lo~Curto, Longinotti,
  Megevand, Pasquini, Petitpas, Pichard, Queloz, Reyes, Richaud, Sivan,
  Sosnowska, Soto, Udry, Ureta, van Kesteren, Weber, Weilenmann, Wicenec,
  Wieland, Christensen-Dalsgaard, Dravins, Hatzes, Kürster, Paresce, \&
  Penny}]{Pepe_2002}
Pepe, F., Mayor, M., Rupprecht, G., {et~al.} 2002, The Messenger, 110, 9

\bibitem[{{Petigura} {et~al.}(2013){Petigura}, {Howard}, \&
  {Marcy}}]{Petigura13PNAS}
{Petigura}, E.~A., {Howard}, A.~W., \& {Marcy}, G.~W. 2013, Proceedings of the
  National Academy of Science, 110, 19273

\bibitem[{{Petigura} {et~al.}(2022){Petigura}, {Rogers}, {Isaacson}, {Owen},
  {Kraus}, {Winn}, {MacDougall}, {Howard}, {Fulton}, {Kosiarek}, {Weiss},
  {Behmard}, \& {Blunt}}]{Petigura22}
{Petigura}, E.~A., {Rogers}, J.~G., {Isaacson}, H., {et~al.} 2022, \aj, 163,
  179

\bibitem[{{Piaulet} {et~al.}(2023){Piaulet}, {Benneke}, {Almenara}, {Dragomir},
  {Knutson}, {Thorngren}, {Peterson}, {Crossfield}, {Kempton}, {Kubyshkina},
  {Howard}, {Angus}, {Isaacson}, {Weiss}, {Beichman}, {Fortney}, {Fossati},
  {Lammer}, {McCullough}, {Morley}, \& {Wong}}]{Piaulet22}
{Piaulet}, C., {Benneke}, B., {Almenara}, J.~M., {et~al.} 2023, Nature
  Astronomy, 7, 206

\bibitem[{{Piaulet-Ghorayeb} {et~al.}(2024){Piaulet-Ghorayeb}, {Benneke},
  {Radica}, {Raul}, {Coulombe}, {Ahrer}, {Kubyshkina}, {Howard},
  {Krissansen-Totton}, {MacDonald}, {Roy}, {Louca}, {Christie},
  {Fournier-Tondreau}, {Allart}, {Miguel}, {Schlichting}, {Welbanks},
  {Cadieux}, {Dorn}, {Evans-Soma}, {Fortney}, {Pierrehumbert},
  {Lafreni{\`e}re}, {Acu{\~n}a}, {Komacek}, {Innes}, {Beatty}, {Cloutier},
  {Doyon}, {Gagnebin}, {Gapp}, \& {Knutson}}]{Piaulet2024}
{Piaulet-Ghorayeb}, C., {Benneke}, B., {Radica}, M., {et~al.} 2024, \apjl, 974,
  L10

\bibitem[{{Plez}(2012)}]{ple12}
{Plez}, B. 2012, {Turbospectrum: Code for spectral synthesis}, Astrophysics
  Source Code Library

\bibitem[{{Polanski} {et~al.}(2024){Polanski}, {Lubin}, {Beard}, {Akana
  Murphy}, {Rubenzahl}, {Hill}, {Crossfield}, {Chontos}, {Robertson},
  {Isaacson}, {Kane}, {Ciardi}, {Batalha}, {Dressing}, {Fulton}, {Howard},
  {Huber}, {Petigura}, {Weiss}, {Angelo}, {Behmard}, {Blunt}, {Brinkman},
  {Dai}, {Dalba}, {Fetherolf}, {Giacalone}, {Hirsch}, {Holcomb}, {Kosiarek},
  {Mayo}, {MacDougall}, {Mo{\v{c}}nik}, {Pidhorodetska}, {Rice}, {Rosenthal},
  {Scarsdale}, {Turtelboom}, {Tyler}, {Van Zandt}, {Yee}, {Coria}, {Dulz},
  {Hartman}, {Householder}, {Lange}, {Langford}, {Louden}, {Siegel}, {Gilbert},
  {Gonzales}, {Schlieder}, {Boyle}, {Christiansen}, {Clark}, {Fernandes},
  {Lund}, {Savel}, {Gill}, {Beichman}, {Matson}, {Matthews}, {Furlan},
  {Howell}, {Scott}, {Everett}, {Livingston}, {Ershova}, {Cheryasov},
  {Safonov}, {Lillo-Box}, {Barrado}, \& {Morales-Calder{\'o}n}}]{polanski2024}
{Polanski}, A.~S., {Lubin}, J., {Beard}, C., {et~al.} 2024, \apjs, 272, 32

\bibitem[{{Queloz} {et~al.}(2001){Queloz}, {Henry}, {Sivan}, {Baliunas},
  {Beuzit}, {Donahue}, {Mayor}, {Naef}, {Perrier}, \& {Udry}}]{queloz2001}
{Queloz}, D., {Henry}, G.~W., {Sivan}, J.~P., {et~al.} 2001, \aap, 379, 279

\bibitem[{Rajpaul {et~al.}(2021)Rajpaul, Buchhave, Lacedelli, Rice, Mortier,
  Malavolta, Aigrain, Borsato, Mayo, Charbonneau, Damasso, Dumusque, Ghedina,
  Latham, López-Morales, Magazzù, Micela, Molinari, Pepe, Piotto, Poretti,
  Rowther, Sozzetti, Udry, \& Watson}]{rajpaul2021}
Rajpaul, V.~M., Buchhave, L.~A., Lacedelli, G., {et~al.} 2021, Monthly Notices
  of the Royal Astronomical Society, 507, 1847

\bibitem[{{Raymond} {et~al.}(2018{\natexlab{a}}){Raymond}, {Boulet}, {Izidoro},
  {Esteves}, \& {Bitsch}}]{Raymond18}
{Raymond}, S.~N., {Boulet}, T., {Izidoro}, A., {Esteves}, L., \& {Bitsch}, B.
  2018{\natexlab{a}}, \mnras, 479, L81

\bibitem[{{Raymond} {et~al.}(2018{\natexlab{b}}){Raymond}, {Boulet}, {Izidoro},
  {Esteves}, \& {Bitsch}}]{raymond2018}
{Raymond}, S.~N., {Boulet}, T., {Izidoro}, A., {Esteves}, L., \& {Bitsch}, B.
  2018{\natexlab{b}}, \mnras, 479, L81

\bibitem[{{Rein} \& {Liu}(2012)}]{rein2012}
{Rein}, H. \& {Liu}, S.~F. 2012, \aap, 537, A128

\bibitem[{{Rein} \& {Spiegel}(2015)}]{rein2015}
{Rein}, H. \& {Spiegel}, D.~S. 2015, \mnras, 446, 1424

\bibitem[{{Rein} \& {Tamayo}(2016)}]{Rein2016MNRAS.459.2275R}
{Rein}, H. \& {Tamayo}, D. 2016, \mnras, 459, 2275

\bibitem[{Ricker {et~al.}(2014)Ricker, Winn, Vanderspek, Latham, Bakos, Bean,
  Berta-Thompson, Brown, Buchhave, Butler, Butler, Chaplin, Charbonneau,
  Christensen-Dalsgaard, Clampin, Deming, Doty, Lee, Dressing, Dunham, Endl,
  Fressin, Ge, Henning, Holman, Howard, Ida, Jenkins, Jernigan, Johnson,
  Kaltenegger, Kawai, Kjeldsen, Laughlin, Levine, Lin, Lissauer, MacQueen,
  Marcy, McCullough, Morton, Narita, Paegert, Palle, Pepe, Pepper, Quirrenbach,
  Rinehart, Sasselov, Sato, Seager, Sozzetti, Stassun, Sullivan, Szentgyorgyi,
  Torres, Udry, \& Villasenor}]{ricker2014}
Ricker, G.~R., Winn, J.~N., Vanderspek, R., {et~al.} 2014, Journal of
  Astronomical Telescopes, Instruments, and Systems, 1, 1

\bibitem[{{Rogers} {et~al.}(2023){Rogers}, {Schlichting}, \& {Owen}}]{Rogers23}
{Rogers}, J.~G., {Schlichting}, H.~E., \& {Owen}, J.~E. 2023, \apjl, 947, L19

\bibitem[{{Rogers} {et~al.}(2024){Rogers}, {Schlichting}, \&
  {Young}}]{Rogers2024}
{Rogers}, J.~G., {Schlichting}, H.~E., \& {Young}, E.~D. 2024, \apj, 970, 47

\bibitem[{{Rogers}(2015)}]{Rogers15}
{Rogers}, L.~A. 2015, \apj, 801, 41

\bibitem[{{Rogers} \& {Seager}(2010)}]{RogersSeager10}
{Rogers}, L.~A. \& {Seager}, S. 2010, \apj, 712, 974

\bibitem[{{Rowe} {et~al.}(2014){Rowe}, {Bryson}, {Marcy}, {Lissauer},
  {Jontof-Hutter}, {Mullally}, {Gilliland}, {Issacson}, {Ford}, {Howell},
  {Borucki}, {Haas}, {Huber}, {Steffen}, {Thompson}, {Quintana}, {Barclay},
  {Still}, {Fortney}, {Gautier}, {Hunter}, {Caldwell}, {Ciardi}, {Devore},
  {Cochran}, {Jenkins}, {Agol}, {Carter}, \& {Geary}}]{rowe2014}
{Rowe}, J.~F., {Bryson}, S.~T., {Marcy}, G.~W., {et~al.} 2014, \apj, 784, 45

\bibitem[{{Ryabchikova} {et~al.}(2015){Ryabchikova}, {Piskunov}, {Kurucz},
  {Stempels}, {Heiter}, {Pakhomov}, \& {Barklem}}]{rya15}
{Ryabchikova}, T., {Piskunov}, N., {Kurucz}, R.~L., {et~al.} 2015, \physscr,
  90, 054005

\bibitem[{Schlichting \& Young(2022)}]{Schlichting_2022}
Schlichting, H.~E. \& Young, E.~D. 2022, \psj, 3, 127

\bibitem[{{Schulze} {et~al.}(2024){Schulze}, {Wang}, {Johnson}, {Gaudi},
  {Rodriguez Martinez}, {Unterborn}, \& {Panero}}]{Schulze2024}
{Schulze}, J.~G., {Wang}, J., {Johnson}, J.~A., {et~al.} 2024, \psj, 5, 71

\bibitem[{{Schweitzer} {et~al.}(2019){Schweitzer}, {Passegger}, {Cifuentes},
  {B{\'e}jar}, {Cort{\'e}s-Contreras}, {Caballero}, {del Burgo}, {Czesla},
  {K{\"u}rster}, {Montes}, {Zapatero Osorio}, {Ribas}, {Reiners},
  {Quirrenbach}, {Amado}, {Aceituno}, {Anglada-Escud{\'e}}, {Bauer},
  {Dreizler}, {Jeffers}, {Guenther}, {Henning}, {Kaminski}, {Lafarga},
  {Marfil}, {Morales}, {Schmitt}, {Seifert}, {Solano}, {Tabernero}, \&
  {Zechmeister}}]{schweitzer2019}
{Schweitzer}, A., {Passegger}, V.~M., {Cifuentes}, C., {et~al.} 2019, \aap,
  625, A68

\bibitem[{{Shappee} {et~al.}(2014){Shappee}, {Prieto}, {Grupe}, {Kochanek},
  {Stanek}, {De Rosa}, {Mathur}, {Zu}, {Peterson}, {Pogge}, {Komossa}, {Im},
  {Jencson}, {Holoien}, {Basu}, {Beacom}, {Szczygie{\l}}, {Brimacombe},
  {Adams}, {Campillay}, {Choi}, {Contreras}, {Dietrich}, {Dubberley},
  {Elphick}, {Foale}, {Giustini}, {Gonzalez}, {Hawkins}, {Howell}, {Hsiao},
  {Koss}, {Leighly}, {Morrell}, {Mudd}, {Mullins}, {Nugent}, {Parrent},
  {Phillips}, {Pojmanski}, {Rosing}, {Ross}, {Sand}, {Terndrup}, {Valenti},
  {Walker}, \& {Yoon}}]{Shappee2014}
{Shappee}, B.~J., {Prieto}, J.~L., {Grupe}, D., {et~al.} 2014, \apj, 788, 48

\bibitem[{{Sinukoff} {et~al.}(2017){Sinukoff}, {Howard}, {Petigura}, {Fulton},
  {Crossfield}, {Isaacson}, {Gonzales}, {Crepp}, {Brewer}, {Hirsch}, {Weiss},
  {Ciardi}, {Schlieder}, {Benneke}, {Christiansen}, {Dressing}, {Hansen},
  {Knutson}, {Kosiarek}, {Livingston}, {Greene}, {Rogers}, \&
  {L{\'e}pine}}]{sinukoff2017}
{Sinukoff}, E., {Howard}, A.~W., {Petigura}, E.~A., {et~al.} 2017, \aj, 153,
  271

\bibitem[{{Skrutskie} {et~al.}(2006){Skrutskie}, {Cutri}, {Stiening},
  {Weinberg}, {Schneider}, {Carpenter}, {Beichman}, {Capps}, {Chester},
  {Elias}, {Huchra}, {Liebert}, {Lonsdale}, {Monet}, {Price}, {Seitzer},
  {Jarrett}, {Kirkpatrick}, {Gizis}, {Howard}, {Evans}, {Fowler}, {Fullmer},
  {Hurt}, {Light}, {Kopan}, {Marsh}, {McCallon}, {Tam}, {Van Dyk}, \&
  {Wheelock}}]{skrutskie_2006}
{Skrutskie}, M.~F., {Cutri}, R.~M., {Stiening}, R., {et~al.} 2006, \aj, 131,
  1163

\bibitem[{{Smith} {et~al.}(2012){Smith}, {Stumpe}, {Van Cleve}, {Jenkins},
  {Barclay}, {Fanelli}, {Girouard}, {Kolodziejczak}, {McCauliff}, {Morris}, \&
  {Twicken}}]{smith2012}
{Smith}, J.~C., {Stumpe}, M.~C., {Van Cleve}, J.~E., {et~al.} 2012, \pasp, 124,
  1000

\bibitem[{Stassun {et~al.}(2018)Stassun, Oelkers, Pepper, Paegert, Lee, Torres,
  Latham, Charpinet, Dressing, Huber, Kane, L{\'{e}}pine, Mann, Muirhead,
  Rojas-Ayala, Silvotti, Fleming, Levine, \& Plavchan}]{Stassun2018}
Stassun, K.~G., Oelkers, R.~J., Pepper, J., {et~al.} 2018, The Astronomical
  Journal, 156, 102

\bibitem[{{Stumpe} {et~al.}(2014){Stumpe}, {Smith}, {Catanzarite}, {Van Cleve},
  {Jenkins}, {Twicken}, \& {Girouard}}]{Stumpe2014}
{Stumpe}, M.~C., {Smith}, J.~C., {Catanzarite}, J.~H., {et~al.} 2014, \pasp,
  126, 100

\bibitem[{{Stumpe} {et~al.}(2012){Stumpe}, {Smith}, {Van Cleve}, {Twicken},
  {Barclay}, {Fanelli}, {Girouard}, {Jenkins}, {Kolodziejczak}, {McCauliff}, \&
  {Morris}}]{Stumpe2012}
{Stumpe}, M.~C., {Smith}, J.~C., {Van Cleve}, J.~E., {et~al.} 2012, \pasp, 124,
  985

\bibitem[{{Su{\'a}rez Mascare{\~n}o} {et~al.}(2016){Su{\'a}rez Mascare{\~n}o},
  {Rebolo}, \& {Gonz{\'a}lez Hern{\'a}ndez}}]{suarez_mascareno2016}
{Su{\'a}rez Mascare{\~n}o}, A., {Rebolo}, R., \& {Gonz{\'a}lez Hern{\'a}ndez},
  J.~I. 2016, \aap, 595, A12

\bibitem[{{Su{\'a}rez Mascare{\~n}o} {et~al.}(2015){Su{\'a}rez Mascare{\~n}o},
  {Rebolo}, {Gonz{\'a}lez Hern{\'a}ndez}, \& {Esposito}}]{suarez_mascareno2015}
{Su{\'a}rez Mascare{\~n}o}, A., {Rebolo}, R., {Gonz{\'a}lez Hern{\'a}ndez},
  J.~I., \& {Esposito}, M. 2015, \mnras, 452, 2745

\bibitem[{{Tabernero} {et~al.}(2022){Tabernero}, {Marfil}, {Montes}, \&
  {Gonz{\'a}lez Hern{\'a}ndez}}]{tab22}
{Tabernero}, H.~M., {Marfil}, E., {Montes}, D., \& {Gonz{\'a}lez
  Hern{\'a}ndez}, J.~I. 2022, \aap, 657, A66

\bibitem[{{Tayar} {et~al.}(2022){Tayar}, {Claytor}, {Huber}, \& {van
  Saders}}]{tayar2022}
{Tayar}, J., {Claytor}, Z.~R., {Huber}, D., \& {van Saders}, J. 2022, \apj,
  927, 31

\bibitem[{{Teske} {et~al.}(2021){Teske}, {Wang}, {Wolfgang}, {Gan},
  {Plotnykov}, {Armstrong}, {Butler}, {Cale}, {Crane}, {Howard}, {Jensen},
  {Law}, {Shectman}, {Plavchan}, {Valencia}, {Vanderburg}, {Ricker},
  {Vanderspek}, {Latham}, {Seager}, {Winn}, {Jenkins}, {Adibekyan}, {Barrado},
  {Barros}, {Benkhaldoun}, {Brown}, {Bryant}, {Burt}, {Caldwell},
  {Charbonneau}, {Cloutier}, {Collins}, {Collins}, {Colon}, {Conti},
  {Demangeon}, {Eastman}, {Elmufti}, {Feng}, {Flowers}, {Guerrero},
  {Hojjatpanah}, {Irwin}, {Isopi}, {Lillo-Box}, {Mallia}, {Massey}, {Mori},
  {Mullally}, {Narita}, {Nishiumi}, {Osborn}, {Paegert}, {de Leon}, {Quinn},
  {Reefe}, {Schwarz}, {Shporer}, {Soubkiou}, {Sousa}, {Stockdale}, {Str{\o}m},
  {Tan}, {Tang}, {Tenenbaum}, {Wheatley}, {Wittrock}, {Yahalomi}, \&
  {Zohrabi}}]{Teske21}
{Teske}, J., {Wang}, S.~X., {Wolfgang}, A., {et~al.} 2021, \apjs, 256, 33

\bibitem[{{Thiabaud} {et~al.}(2014){Thiabaud}, {Marboeuf}, {Alibert}, {Cabral},
  {Leya}, \& {Mezger}}]{Thiabaud2014}
{Thiabaud}, A., {Marboeuf}, U., {Alibert}, Y., {et~al.} 2014, \aap, 562, A27

\bibitem[{{Tinetti} {et~al.}(2018){Tinetti}, {Drossart}, {Eccleston},
  {Hartogh}, {Heske}, {Leconte}, {Micela}, {Ollivier}, {Pilbratt}, {Puig},
  {Turrini}, {Vandenbussche}, {Wolkenberg}, {Beaulieu}, {Buchave}, {Ferus},
  {Griffin}, {Guedel}, {Justtanont}, {Lagage}, {Machado}, {Malaguti}, {Min},
  {N{\o}rgaard-Nielsen}, {Rataj}, {Ray}, {Ribas}, {Swain}, {Szabo}, {Werner},
  {Barstow}, {Burleigh}, {Cho}, {Coud{\'e} du Foresto}, {Coustenis}, {Decin},
  {Encrenaz}, {Galand}, {Gillon}, {Helled}, {Morales}, {Garc{\'\i}a Mu{\~n}oz},
  {Moneti}, {Pagano}, {Pascale}, {Piccioni}, {Pinfield}, {Sarkar}, {Selsis},
  {Tennyson}, {Triaud}, {Venot}, {Waldmann}, {Waltham}, {Wright}, {Amiaux},
  {Augu{\`e}res}, {Berth{\'e}}, {Bezawada}, {Bishop}, {Bowles}, {Coffey},
  {Colom{\'e}}, {Crook}, {Crouzet}, {Da Peppo}, {Sanz}, {Focardi}, {Frericks},
  {Hunt}, {Kohley}, {Middleton}, {Morgante}, {Ottensamer}, {Pace}, {Pearson},
  {Stamper}, {Symonds}, {Rengel}, {Renotte}, {Ade}, {Affer}, {Alard}, {Allard},
  {Altieri}, {Andr{\'e}}, {Arena}, {Argyriou}, {Aylward}, {Baccani}, {Bakos},
  {Banaszkiewicz}, {Barlow}, {Batista}, {Bellucci}, {Benatti}, {Bernardi},
  {B{\'e}zard}, {Blecka}, {Bolmont}, {Bonfond}, {Bonito}, {Bonomo}, {Brucato},
  {Brun}, {Bryson}, {Bujwan}, {Casewell}, {Charnay}, {Pestellini}, {Chen},
  {Ciaravella}, {Claudi}, {Cl{\'e}dassou}, {Damasso}, {Damiano}, {Danielski},
  {Deroo}, {Di Giorgio}, {Dominik}, {Doublier}, {Doyle}, {Doyon}, {Drummond},
  {Duong}, {Eales}, {Edwards}, {Farina}, {Flaccomio}, {Fletcher}, {Forget},
  {Fossey}, {Fr{\"a}nz}, {Fujii}, {Garc{\'\i}a-Piquer}, {Gear}, {Geoffray},
  {G{\'e}rard}, {Gesa}, {Gomez}, {Graczyk}, {Griffith}, {Grodent}, {Guarcello},
  {Gustin}, {Hamano}, {Hargrave}, {Hello}, {Heng}, {Herrero}, {Hornstrup},
  {Hubert}, {Ida}, {Ikoma}, {Iro}, {Irwin}, {Jarchow}, {Jaubert}, {Jones},
  {Julien}, {Kameda}, {Kerschbaum}, {Kervella}, {Koskinen}, {Krijger}, {Krupp},
  {Lafarga}, {Landini}, {Lellouch}, {Leto}, {Luntzer}, {Rank-L{\"u}ftinger},
  {Maggio}, {Maldonado}, {Maillard}, {Mall}, {Marquette}, {Mathis}, {Maxted},
  {Matsuo}, {Medvedev}, {Miguel}, {Minier}, {Morello}, {Mura}, {Narita},
  {Nascimbeni}, {Nguyen Tong}, {Noce}, {Oliva}, {Palle}, {Palmer}, {Pancrazzi},
  {Papageorgiou}, {Parmentier}, {Perger}, {Petralia}, {Pezzuto},
  {Pierrehumbert}, {Pillitteri}, {Piotto}, {Pisano}, {Prisinzano}, {Radioti},
  {R{\'e}ess}, {Rezac}, {Rocchetto}, {Rosich}, {Sanna}, {Santerne}, {Savini},
  {Scandariato}, {Sicardy}, {Sierra}, {Sindoni}, {Skup}, {Snellen}, {Sobiecki},
  {Soret}, {Sozzetti}, {Stiepen}, {Strugarek}, {Taylor}, {Taylor}, {Terenzi},
  {Tessenyi}, {Tsiaras}, {Tucker}, {Valencia}, {Vasisht}, {Vazan}, {Vilardell},
  {Vinatier}, {Viti}, {Waters}, {Wawer}, {Wawrzaszek}, {Whitworth}, {Yung},
  {Yurchenko}, {Zapatero Osorio}, {Zellem}, {Zingales}, \& {Zwart}}]{ARIEL}
{Tinetti}, G., {Drossart}, P., {Eccleston}, P., {et~al.} 2018, Experimental
  Astronomy, 46, 135

\bibitem[{{Tsai} {et~al.}(2021){Tsai}, {Innes}, {Lichtenberg}, {Taylor},
  {Malik}, {Chubb}, \& {Pierrehumbert}}]{tsai2021}
{Tsai}, S.-M., {Innes}, H., {Lichtenberg}, T., {et~al.} 2021, \apjl, 922, L27

\bibitem[{{Tsiaras} {et~al.}(2016){Tsiaras}, {Rocchetto}, {Waldmann}, {Venot},
  {Varley}, {Morello}, {Damiano}, {Tinetti}, {Barton}, {Yurchenko}, \&
  {Tennyson}}]{tsiaras2016}
{Tsiaras}, A., {Rocchetto}, M., {Waldmann}, I.~P., {et~al.} 2016, \apj, 820, 99

\bibitem[{{Turbet} {et~al.}(2020){Turbet}, {Bolmont}, {Ehrenreich}, {Gratier},
  {Leconte}, {Selsis}, {Hara}, \& {Lovis}}]{turbet2020}
{Turbet}, M., {Bolmont}, E., {Ehrenreich}, D., {et~al.} 2020, \aap, 638, A41

\bibitem[{{Valencia} {et~al.}(2013){Valencia}, {Guillot}, {Parmentier}, \&
  {Freedman}}]{valencia2013}
{Valencia}, D., {Guillot}, T., {Parmentier}, V., \& {Freedman}, R.~S. 2013,
  \apj, 775, 10

\bibitem[{{Van Eylen} {et~al.}(2018){Van Eylen}, {Agentoft}, {Lundkvist},
  {Kjeldsen}, {Owen}, {Fulton}, {Petigura}, \& {Snellen}}]{vanEylen2018}
{Van Eylen}, V., {Agentoft}, C., {Lundkvist}, M.~S., {et~al.} 2018, \mnras,
  479, 4786

\bibitem[{{Van Eylen} {et~al.}(2019){Van Eylen}, {Albrecht}, {Huang},
  {MacDonald}, {Dawson}, {Cai}, {Foreman-Mackey}, {Lundkvist}, {Silva Aguirre},
  {Snellen}, \& {Winn}}]{vanEylen2019}
{Van Eylen}, V., {Albrecht}, S., {Huang}, X., {et~al.} 2019, \aj, 157, 61

\bibitem[{{Vazan} {et~al.}(2022){Vazan}, {Sari}, \& {Kessel}}]{Vazan22}
{Vazan}, A., {Sari}, R., \& {Kessel}, R. 2022, \apj, 926, 150

\bibitem[{{Venturini} {et~al.}(2020){Venturini}, {Guilera}, {Haldemann},
  {Ronco}, \& {Mordasini}}]{Venturini20}
{Venturini}, J., {Guilera}, O.~M., {Haldemann}, J., {Ronco}, M.~P., \&
  {Mordasini}, C. 2020, \aap, 643, L1

\bibitem[{{Venturini} {et~al.}(2024){Venturini}, {Ronco}, {Guilera},
  {Haldemann}, {Mordasini}, \& {Miller Bertolami}}]{venturini2024}
{Venturini}, J., {Ronco}, M.~P., {Guilera}, O.~M., {et~al.} 2024, \aap, 686, L9

\bibitem[{{Weiss} \& {Marcy}(2014)}]{Weiss14}
{Weiss}, L.~M. \& {Marcy}, G.~W. 2014, \apjl, 783, L6

\bibitem[{{Wildi} {et~al.}(2022){Wildi}, {Bouchy}, {Doyon}, {Blind}, {Genolet},
  {Sordet}, {Segovia}, {Grieves}, {Malo}, {Artigau}, {St-Antoine},
  {Vall{\'e}e}, {Rasilla}, {Gracia Temich}, {Poulin-Girard}, {Brousseau},
  {Sosnowska}, {Reshetov}, {Baron}, {Thibault}, {Bovay}, {Frensch}, {Lo Curto},
  {Hubin}, {Zins}, {Peroux}, \& {Cabral}}]{Wildi_2022}
{Wildi}, F., {Bouchy}, F., {Doyon}, R., {et~al.} 2022, in Society of
  Photo-Optical Instrumentation Engineers (SPIE) Conference Series, Vol. 12184,
  Ground-based and Airborne Instrumentation for Astronomy IX, ed. C.~J.
  {Evans}, J.~J. {Bryant}, \& K.~{Motohara}, 121841H

\bibitem[{{Wright} {et~al.}(2010){Wright}, {Eisenhardt}, {Mainzer}, {Ressler},
  {Cutri}, {Jarrett}, {Kirkpatrick}, {Padgett}, {McMillan}, {Skrutskie},
  {Stanford}, {Cohen}, {Walker}, {Mather}, {Leisawitz}, {Gautier}, {McLean},
  {Benford}, {Lonsdale}, {Blain}, {Mendez}, {Irace}, {Duval}, {Liu}, {Royer},
  {Heinrichsen}, {Howard}, {Shannon}, {Kendall}, {Walsh}, {Larsen}, {Cardon},
  {Schick}, {Schwalm}, {Abid}, {Fabinsky}, {Naes}, \& {Tsai}}]{wright_2010}
{Wright}, E.~L., {Eisenhardt}, P. R.~M., {Mainzer}, A.~K., {et~al.} 2010, \aj,
  140, 1868

\bibitem[{Wu(2019)}]{Wu_2019}
Wu, Y. 2019, The Astrophysical Journal, 874, 91

\bibitem[{{Wyatt} {et~al.}(2020){Wyatt}, {Kral}, \& {Sinclair}}]{Wyatt20}
{Wyatt}, M.~C., {Kral}, Q., \& {Sinclair}, C.~A. 2020, \mnras, 491, 782

\bibitem[{{Xie}(2014)}]{xie2014}
{Xie}, J.-W. 2014, \apjs, 210, 25

\bibitem[{{York} {et~al.}(2000){York}, {Adelman}, {Anderson}, {Anderson},
  {Annis}, {Bahcall}, {Bakken}, {Barkhouser}, {Bastian}, {Berman}, {Boroski},
  {Bracker}, {Briegel}, {Briggs}, {Brinkmann}, {Brunner}, {Burles}, {Carey},
  {Carr}, {Castander}, {Chen}, {Colestock}, {Connolly}, {Crocker}, {Csabai},
  {Czarapata}, {Davis}, {Doi}, {Dombeck}, {Eisenstein}, {Ellman}, {Elms},
  {Evans}, {Fan}, {Federwitz}, {Fiscelli}, {Friedman}, {Frieman}, {Fukugita},
  {Gillespie}, {Gunn}, {Gurbani}, {de Haas}, {Haldeman}, {Harris}, {Hayes},
  {Heckman}, {Hennessy}, {Hindsley}, {Holm}, {Holmgren}, {Huang}, {Hull},
  {Husby}, {Ichikawa}, {Ichikawa}, {Ivezi{\'c}}, {Kent}, {Kim}, {Kinney},
  {Klaene}, {Kleinman}, {Kleinman}, {Knapp}, {Korienek}, {Kron}, {Kunszt},
  {Lamb}, {Lee}, {Leger}, {Limmongkol}, {Lindenmeyer}, {Long}, {Loomis},
  {Loveday}, {Lucinio}, {Lupton}, {MacKinnon}, {Mannery}, {Mantsch}, {Margon},
  {McGehee}, {McKay}, {Meiksin}, {Merelli}, {Monet}, {Munn}, {Narayanan},
  {Nash}, {Neilsen}, {Neswold}, {Newberg}, {Nichol}, {Nicinski}, {Nonino},
  {Okada}, {Okamura}, {Ostriker}, {Owen}, {Pauls}, {Peoples}, {Peterson},
  {Petravick}, {Pier}, {Pope}, {Pordes}, {Prosapio}, {Rechenmacher}, {Quinn},
  {Richards}, {Richmond}, {Rivetta}, {Rockosi}, {Ruthmansdorfer}, {Sandford},
  {Schlegel}, {Schneider}, {Sekiguchi}, {Sergey}, {Shimasaku}, {Siegmund},
  {Smee}, {Smith}, {Snedden}, {Stone}, {Stoughton}, {Strauss}, {Stubbs},
  {SubbaRao}, {Szalay}, {Szapudi}, {Szokoly}, {Thakar}, {Tremonti}, {Tucker},
  {Uomoto}, {Vanden Berk}, {Vogeley}, {Waddell}, {Wang}, {Watanabe},
  {Weinberg}, {Yanny}, {Yasuda}, \& {SDSS Collaboration}}]{york_2000}
{York}, D.~G., {Adelman}, J., {Anderson}, John~E., J., {et~al.} 2000, \aj, 120,
  1579

\bibitem[{{Yu} {et~al.}(2021){Yu}, {He}, {Zhang}, {H{\"o}rst}, {Dymont},
  {McGuiggan}, {Moses}, {Lewis}, {Fortney}, {Gao}, {Kempton}, {Moran},
  {Morley}, {Powell}, {Valenti}, \& {Vuitton}}]{yu2021}
{Yu}, X., {He}, C., {Zhang}, X., {et~al.} 2021, Nature Astronomy, 5, 822

\bibitem[{{Zechmeister} \& {K{\"u}rster}(2009)}]{Zechmeister2009}
{Zechmeister}, M. \& {K{\"u}rster}, M. 2009, \aap, 496, 577

\bibitem[{{Zechmeister} {et~al.}(2018){Zechmeister}, {Reiners}, {Amado},
  {Azzaro}, {Bauer}, {B{\'e}jar}, {Caballero}, {Guenther}, {Hagen}, {Jeffers},
  {Kaminski}, {K{\"u}rster}, {Launhardt}, {Montes}, {Morales}, {Quirrenbach},
  {Reffert}, {Ribas}, {Seifert}, {Tal-Or}, \& {Wolthoff}}]{zechmeister2018}
{Zechmeister}, M., {Reiners}, A., {Amado}, P.~J., {et~al.} 2018, \aap, 609, A12

\bibitem[{{Zeng} {et~al.}(2019){Zeng}, {Jacobsen}, {Sasselov}, {Petaev},
  {Vanderburg}, {Lopez-Morales}, {Perez-Mercader}, {Mattsson}, {Li}, {Heising},
  {Bonomo}, {Damasso}, {Berger}, {Cao}, {Levi}, \& {Wordsworth}}]{Zeng2019}
{Zeng}, L., {Jacobsen}, S.~B., {Sasselov}, D.~D., {et~al.} 2019, Proceedings of
  the National Academy of Science, 116, 9723

\bibitem[{{Zhang} {et~al.}(2021){Zhang}, {Weiss}, {Huber}, {Blunt}, {Chontos},
  {Fulton}, {Grunblatt}, {Howard}, {Isaacson}, {Kosiarek}, {Petigura},
  {Rosenthal}, \& {Rubenzahl}}]{zhang2021}
{Zhang}, J., {Weiss}, L.~M., {Huber}, D., {et~al.} 2021, \aj, 162, 89

\end{thebibliography}

\begin{appendix} 
\section{\tess\ target pixel files}\label{appendix:tpf}
\begin{figure}[h]
\centering
  \includegraphics[width=0.8\linewidth]{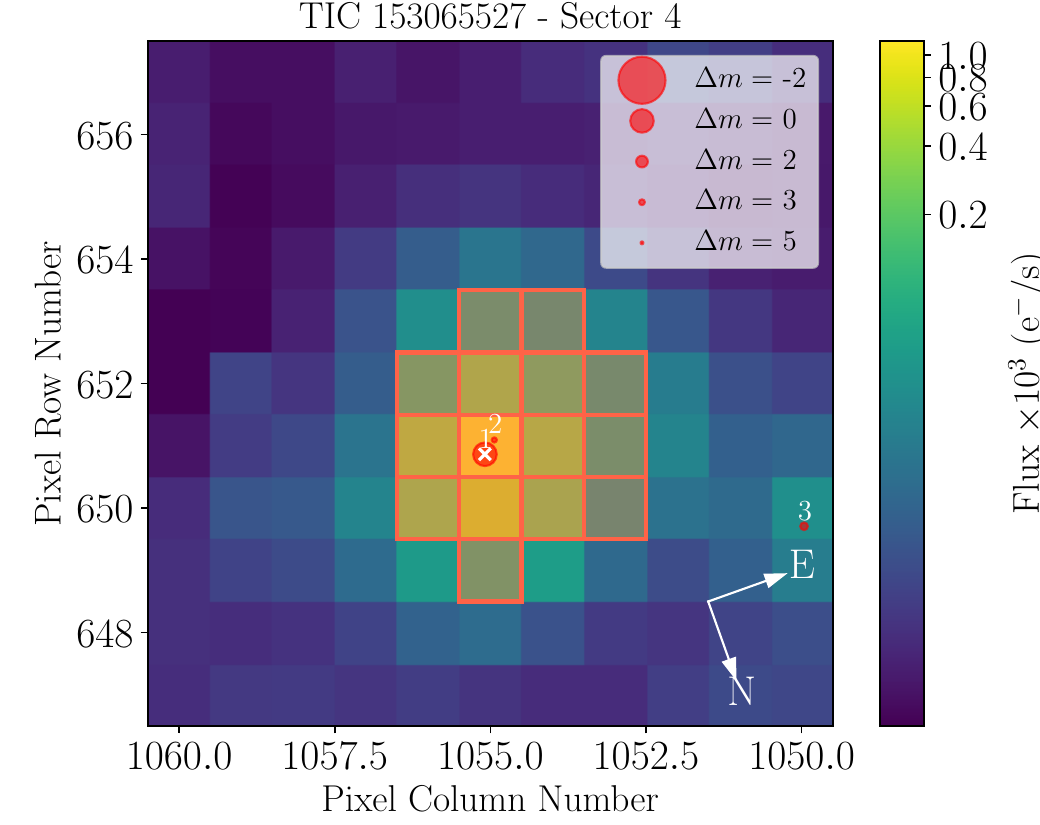}  
  \includegraphics[width=0.8\linewidth]{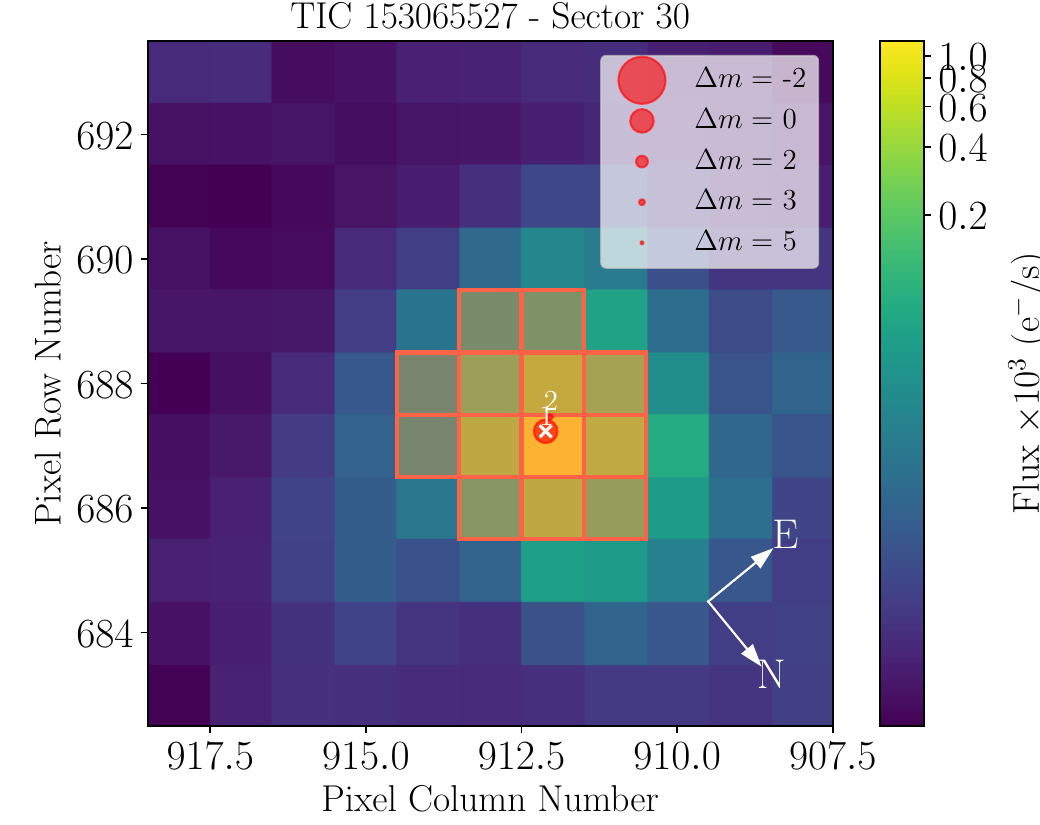}
  \includegraphics[width=0.8\linewidth]{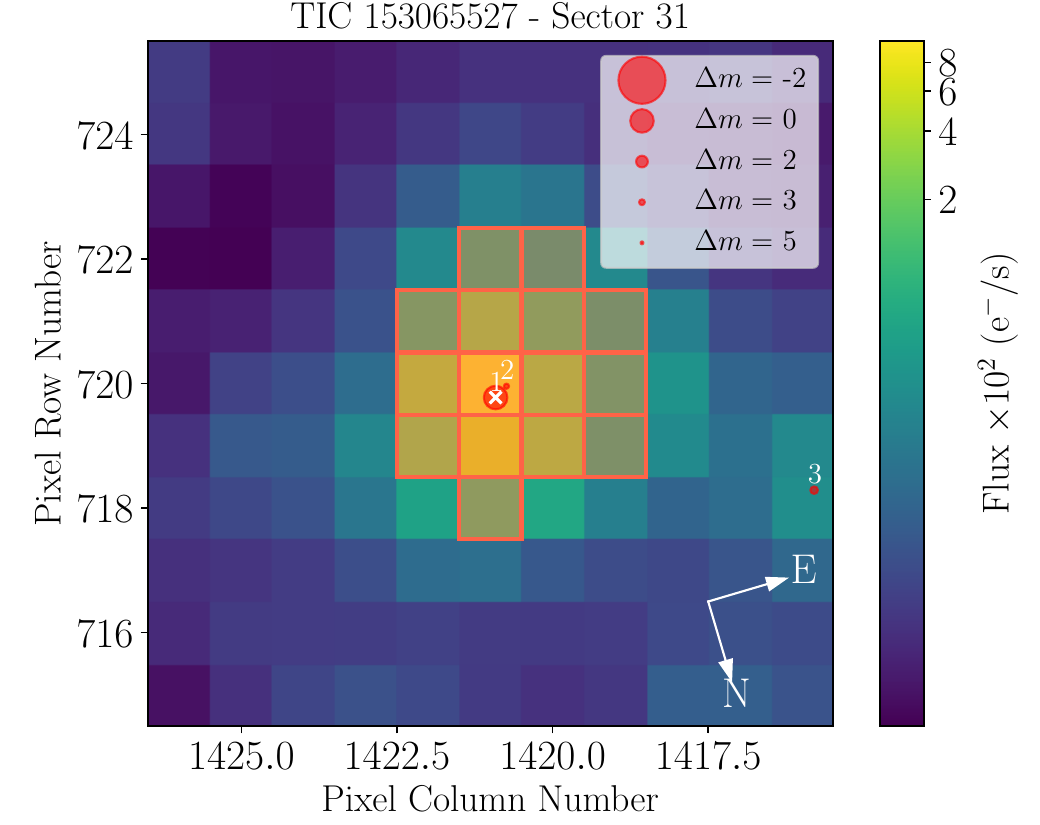}
  \caption{\texttt{tpfplotter} TPF image of \starname, as in Fig.~\ref{fig:tess_tpf}, but for sectors 4, 30, 31.
  }
    \label{fig:tess_tpf_all}
\end{figure}

We show in Fig.~\ref{fig:tess_tpf_all} the TPF plots for TOI-406
for all observed sectors save sector 3, which is shown in Fig.~\ref{fig:tess_tpf}.

\section{{\it TESS} light curves}\label{appendix:light_curves}

\begin{figure*}
\centering
  \includegraphics[width=\linewidth]{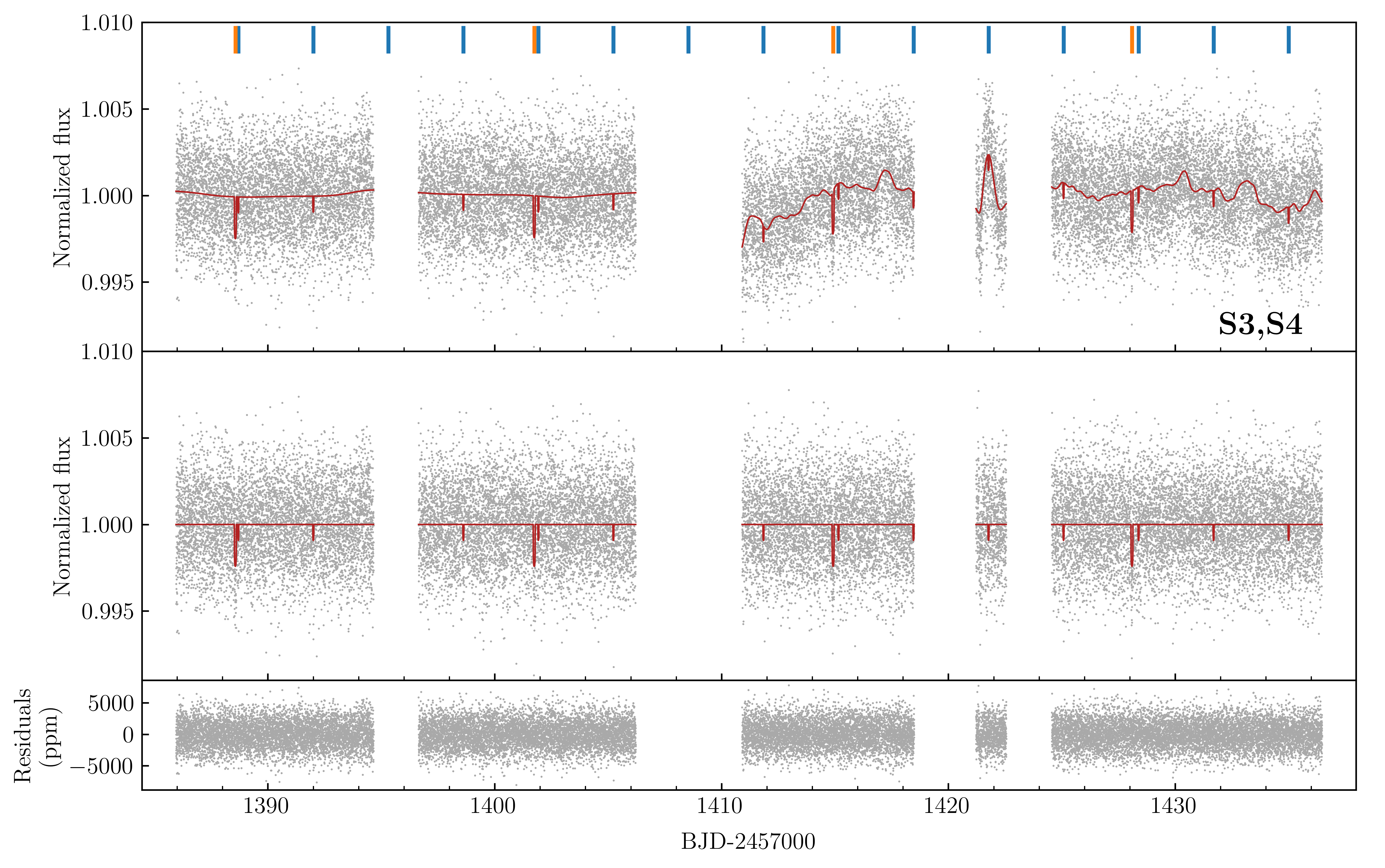}
  \includegraphics[width=\linewidth]{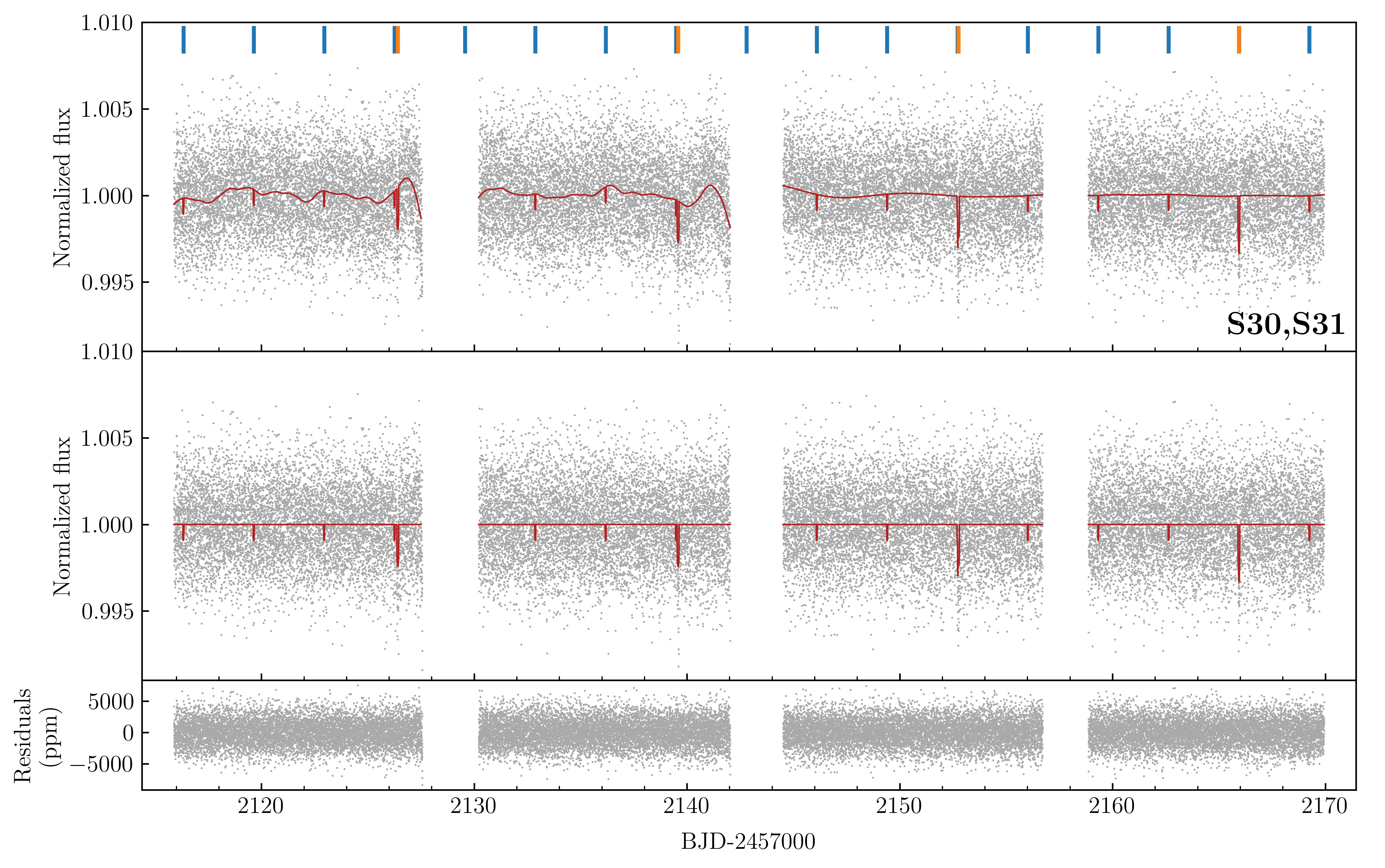}
  \caption{{\it TESS} sectors $3$, $4$ (top), and $30$, $31$ (bottom) PDCSAP light curves of TOI-406. In each figure, we show the best-fitting transit and Matérn-3/2 kernel GP model derived in Sect.~\ref{sec:global_analysis} with a solid dark red line (top panel). The flattened light curve is shown in the central panel, after removing the GP component, with the solid red line showing the best-fitting model. The bottom panel shows the light curves residuals. The transits of planets b and c are highlighted with vertical orange and blue lines, respectively.
  }
    \label{fig:tess_lc}
\end{figure*}

We show in Fig.~\ref{fig:tess_lc} the original and detrended {\it TESS} light curves of TOI-406 with our best-fit model. 

\section{RV data}

\begin{table*}
\tiny
  \caption{ESPRESSO RV measurements and activity indicators of TOI-406.}
\label{table:RV_table_ESPRESSO}      
\centering                                      
\resizebox{\linewidth}{!}{\begin{tabular}{c c c c c c c c c c}          
\hline\hline                        
  BJD$_{\rm TDB}$ & RV & dLW  & CRX & Na \textsc{i} D$_1$ & Na \textsc{i} D$_2$ & H$\alpha$ & FWHM & BIS & $S$-index\\
 $($d$)$  & (\ms ) & (m$^2$ s$^{-2}$) & (\ms) & (dex)  & (dex) & (dex) & (\ms) & (\ms) & \\
\hline                                 
2460140.87040  & $7.68 \pm 0.52 $& $-1.8 \pm  0.3 $&  $-3.9 \pm  3.5 $  & $0.2350 \pm  0.0043 $ & $0.2210 \pm  0.0044 $  &   $0.9420 \pm  0.0028 $ & $2724.4 \pm 2.3 $ & $4060.1 \pm 2.3 $ & $0.966 \pm 0.033$\\
2460143.89259  & $ 5.02 \pm 0.51$  & $-0.7 \pm  0.3 $ & $-3.0 \pm  3.3 $ &  $0.2141 \pm  0.0035 $   & $0.2138 \pm  0.0037 $  &  $0.9588 \pm  0.0025 $ & $2727.2 \pm  2.0 $& $4065.8 \pm 1.9 $ & $1.029 \pm 0.025$\\
2460146.86288  &  $6.45  \pm  0.48$ & $-1.6 \pm  0.3 $ &  $2.8 \pm  3.2 $ &  $0.2358 \pm  0.0037 $   & $0.2221 \pm  0.0037 $  &  $0.9414 \pm  0.0025 $ & $2726.7 \pm 2.0 $& $4063.0 \pm 2.0 $ & $0.969 \pm 0.024$\\
... & ... & ... & ... & ... & ... & ... & ... & ... & ...\\
\hline
\end{tabular}}
\tablefoot{This table is available in its entirety in machine-readable form.}
\end{table*}

\begin{table}
\tiny
  \caption{NIRPS RV measurements and activity indicators of TOI-406.}
\label{table:RV_table_NIRPS}      
\centering                                      
\begin{tabular}{c c c c}          
\hline\hline                        
  BJD$_{\rm TDB}$ & RV& FWHM & Contrast \\
 $($d$)$  & (\ms ) & (\ms) &  \\
\hline                                 
2460200.88073  & $14995.93 \pm 5.22 $ &  $5501.6 \pm  9.9 $  & $0.9999519 \pm  9.0$E-07 \\
2460201.87678  & $14983.03 \pm 4.84 $ &  $5493.8 \pm  9.2 $  & $0.9999637 \pm  7.0$E-07 \\
2460201.88729  & $14985.89 \pm 5.10 $ &  $5483.2 \pm  9.7 $  & $0.9999610 \pm  8.0$E-06 \\
... & ... & ... & ...  \\
\hline
\end{tabular}
\tablefoot{This table is available in its entirety in machine-readable form.}
\end{table}

\begin{table}
\tiny
  \caption{HARPS RV measurements and activity indicators of TOI-406.}
\label{table:RV_table_HARPS}      
\centering                                      
\begin{tabular}{c c c c}          
\hline\hline                        
  BJD$_{\rm TDB}$ & RV & FWHM & Contrast \\
 $($d$)$  & (\ms ) & (\ms) &  \\
\hline                                 
2460181.91467  & $14741.56 \pm 3.20 $&  $3716.0 \pm  6.6 $  & $0.8397 \pm  0.0018$ \\
2460182.89296  & $14734.58 \pm 5.47 $&   $3736.1 \pm  11.2 $  &  $0.8257 \pm 0.0030$\\
2460185.80265  & $14728.73 \pm 4.85 $& $3765.6 \pm  10.9$ &  $0.8150 \pm  0.0027$ \\
... & ... & ... & ... \\
\hline
\end{tabular}
\tablefoot{This table is available in its entirety in machine-readable form.}
\end{table}

\end{appendix}

\end{document}